\documentclass{aa}  

\usepackage{graphicx}
\usepackage{lipsum}
\usepackage{caption}
\usepackage{subcaption}
\usepackage{float}
\usepackage[utf8]{inputenc}
\usepackage[T1]{fontenc}
\usepackage{ulem}
\usepackage{appendix}
\usepackage{sidecap}

\usepackage{txfonts}
\usepackage[colorlinks,citecolor=blue,hypertexnames=true,bookmarks=false,breaklinks=true]{hyperref}

\begin{document}

   \title{[CII] line intensity mapping the epoch of reionization with the 
   Prime-Cam on FYST}

   \subtitle{II. CO foreground masking based on an external catalog}

   \author{C. Karoumpis \thanks{\email{karoumpis@astro.uni-bonn.de}}
          \inst{1}
          \and
          B. Magnelli\inst{2}
          \and
          E. Romano-D{\'\i}az \inst{1}
          \and
          K. Garcia \inst{3}
          \and
          A. Dev \inst{1}
          \and
          J. Clarke\inst{4}
          \and
          T.-M. Wang \inst{1}
          \and
          T. B\u{a}descu \inst{1}
          \and
          D. Riechers \inst{4}
          \and
          F. Bertoldi \inst{1}} %\fnmsep

   \institute{ Argelander Institut für Astronomie, Universität Bonn,
            Auf dem Hügel 71, D-53121 Bonn, Germany
            \and
            Universit{\'e} Paris-Saclay, Universit{\'e} Paris Cit{\'e}, CEA, CNRS, AIM, 91191, Gif-sur-Yvette, France
            \and
            Department of Astronomy, University of Florida, 211 Bryant Space Science Center, Gainesville, FL 32611, USA
            \and
            I. Physikalisches Institut, Universität zu Köln, Zülpicher Straße 77, D-50937 Köln, Germany
           }

% \abstract{}{}{}{}{} 
% 5 {} token are mandatory
 
  \abstract
   { The Fred Young Submillimeter Telescope (FYST) line intensity mapping (LIM) survey will measure the power spectrum (PS) of the singly ionized carbon 158 $\rm \mu$m fine-structure line, [CII], to trace the appearance of the first galaxies that emerged during and right after the epoch of reionization (EoR, $6<z<9$).}
   {We aim to quantify the contamination of the (post-)EoR [CII] LIM signal by foreground carbon monoxide (CO) line emission ($3 < J_{ \rm up} < 12$) and assess the efficiency to retrieve this [CII] LIM signal by the targeted masking of bright CO emitters.}
   {Using the IllustrisTNG300 simulation, we produced mock CO intensity tomographies based on empirical star formation rate-to-CO luminosity relations. Combining these predictions with the [CII] PS predictions of the first paper of this series, we evaluated a masking technique where the interlopers are identified and masked using an external catalog whose properties are equivalent to those of a deep Euclid survey.}
  {Prior to masking, our [CII] PS forecast is an order of magnitude lower than the predicted CO contamination in the 225 GHz ([CII] emitted at $z=6.8-8.3$) band of the FYST LIM survey, at the same level in its 280 GHz ([CII] emitted at $z=5.3-6.3$) and 350 GHz ([CII] emitted at $z=4.1-4.8$) bands, and an order of magnitude higher in its 410 GHz ([CII] emitted at $z=3.4-3.9$) band. For our fiducial model, the optimal masking depth is reached when less than 10\% of the survey volume is masked at 350 and 410 GHz but around 40\% at 280 GHz and 60 \% at 225 GHz. At these masking depths we anticipate a detection of the [CII] PS at 350 and 410 GHz, a tentative detection at 280 GHz, whereas at 225 GHz the CO signal still dominates our model. In the last case, alternative decontamination techniques will be needed.}
{}

   \maketitle
%
%-------------------------------------------------------------------

\section{Introduction}
\label{sec:intro}

In its first billion years, the Universe underwent numerous changes. Following the Big Bang, the hot and ionized plasma cooled down, leading to the formation of neutral hydrogen atoms \citep[e.g.,][]{Peebles1968, Zeldovic1969, Seager2000}. This resulted in the Universe being in a state of darkness, known as the "Dark Ages." The advent of the first stars and galaxies, roughly 300 million years after the Big Bang, re-illuminated the Universe \citep[e.g.,][]{Harikane2022}. What makes this period in the history of the Universe even more special is that the appearance of the first light sources concurs with a phase transition in the intergalactic medium, from neutral to ionized \citep[e.g.,][]{Becker2001,Barkana2001,Becker2015,Fan2006,Veneman2013}, a period known as the epoch of reionization (EoR).

The connection between the reionization of the Universe and the emergence of the first light sources is well established in current theoretical models \citep[e.g.,][]{Zaroubi2013}. Early massive stars located in the first incipient galaxies were the primary drivers of reionization, powering it through their ultraviolet (UV) emission. Despite this understanding, the intricacies of the interaction between early galaxy formation, evolution, and the reionization process remain elusive. Key uncertainties include the energy emitted by the early stars, the fraction of this energy that escaped the early galaxies \citep[e.g.,][]{Bouwens2016}, as well as the distribution of those ionizing sources in space and time. Unfortunately, these galaxies are too faint to be detected in large samples with current telescopes, including the state-of-the-art \textit{James Webb Space Telescope} \citep[JWST,][]{Boylan2015}, which creates challenges for accurately estimating the properties of their population and does not allow for tracing the large-scale structure in which they reside.

A more effective approach than traditional galaxy surveys for tracing both star formation in early galaxies and the large-scale structure in which they reside is line intensity mapping \citep[LIM; see, e.g.,][]{Kovetz2019, Bernal2022}. This method measures the integrated emission of spectral lines from galaxies within a given spatial-spectral resolution element (voxel) using a spectro-imager. By focusing on the singly ionized carbon 158 $\rm \mu$m fine-structure line, [CII], which is the brightest line in typical star-forming galaxies \citep[e.g.,][]{Stacey1991} and an indicator of star formation rate \citep[SFR; e.g,][]{DeLooze2014,LeFevreALPINE,BetALP2020,Schaerer2020,Bowens2022b}, a three-dimensional (3D) tomography of galaxies during the early universe can be obtained \citep[e.g.,][]{Karoumpis}. Several LIM experiments, such as the EXperiment for Cryogenic Large-Aperture Intensity Mapping \citep[EXCLAIM,][]{Pullen2022}, the Terahertz Intensity Mapper \citep[TIM,][]{TIM2020}, the CarbON CII line in post-rEionization and ReionizaTiOn (CONCERTO) project \citep[][]{CONCERTO2020}, the Tomographic Ionized-carbon Mapping Experiment \citep[TIME,][]{TIMENEW}, and the Fred Young Submillimeter Telescope \citep[FYST,][]{FYST}, are underway to target [CII] emission coming from EoR ($6 < z < 9$) and post-EoR ($3 < z < 6$ ) redshifts. 

Many of these observatories will prioritize measuring the variance in the Fourier mode of these 3D tomographies through their spherically averaged power spectrum (PS). This statistical metric, although relatively simple, offers powerful limits on the underlying line luminosity function, given that the PS correlates with both the first moment (for scales $> 10 \ \rm{Mpc}$) and the second moment (at scales $< 10 \ \rm{Mpc}$) of the line luminosity function \citep{Karoumpis}. Such robust constrains of the luminosity functions of [CII] will provide rigorous statistical constraints on the SFR of (post-)EoR galaxies—a pivotal, yet elusive parameter in reionization models \citep[e.g.,][]{FYST, Karoumpis}.

\begin{figure}[ht]
    \centering
        \includegraphics[width=0.45 \textwidth]{{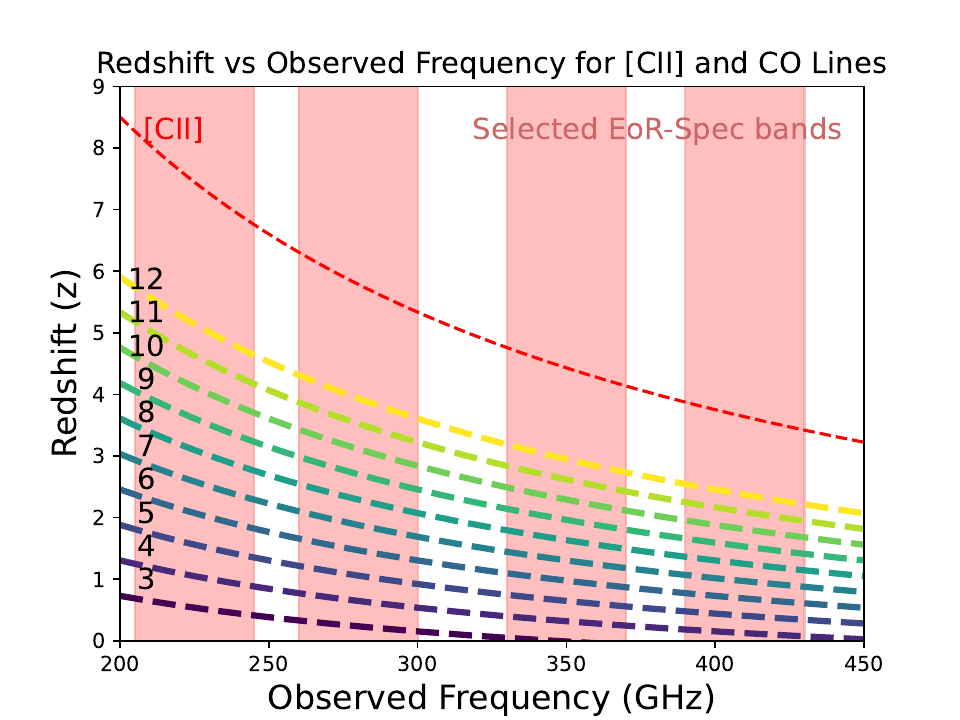}}
        \caption{Spectral lines observable within four selected EoR-Spec bands (illustrated by red-shaded regions), and originated from galaxies situated at different redshifts. The dashed numbered lines represent transitions within the CO rotational ladder, with the corresponding numbers indicating the $J_{\rm up}$ for each specific transition. These CO lines act as the foreground contaminant for the [CII] (red thin dashed line) LIM survey.}
        \label{fig:COREDSHIFT}
\end{figure}

In this context, where precise measurements of the [CII] PS at different scales during and after the EoR will be possible, it is crucial to study how various cosmic galaxy evolution models influence this PS and can thus be tested by comparison with future LIM measurements. To contribute to this effort, in the first article of this series \citep{Karoumpis}, we generated a span of predictions by post-processing the dark matter halo catalog from the IllustrisTNG300 hydrodynamical simulation \citep[TNG300;][]{Pillepich:Springel:Nelson2017}. We then used our predictions to assess the feasibility of detecting the PS of the [CII] line from galaxies at redshifts between 3 and 8 using the spectro-imager of FYST. Our results, which are consistent with empirically motivated predictions \citep{Clarke2024}, demonstrated promising potential for detecting the [CII] PS at the critical comoving length scale of 10 Mpc in four selected FYST bands centered at redshifts 3.7, 4.3, 5.8, and tentatively at redshift 7 (i.e., 410, 350, 280, and 225~GHz). However, those predictions did not account for a significant challenge faced by the LIM technique: the difficulty in distinguishing the contribution of (post-)EoR [CII] galaxies from the infrared (IR) continuum and carbon monoxide (CO) rotational line emission ($2<J_{\rm up}<12$) of galaxies located at the same solid angle but at lower redshifts ($z=0-5.7$; see Fig.~\ref{fig:COREDSHIFT} and Table ~\ref{tab:CORED}). While the IR continuum emission is not a major concern for the PS measurement as it is spectrally smooth and its frequency-coherent spectrum can be fitted and removed \citep{VanCuyck}, the contribution of CO emitters presents a real challenge since the variance of their emission is of the same order or even exceeds that of the [CII] galaxies \citep[e.g.,][]{Bethermin2022,Roy2023}.

In this paper, we utilized the halo catalog from our previous work \citep{Karoumpis}, and by implementing empirical relations between SFRs and CO line luminosities of galaxies, we modeled the foreground line emission corresponding to the same frequencies as the (post-)EoR [CII] emission. Drawing from these predictions, we evaluated a foreground removal method where voxels containing luminous CO emitters are identified and masked using mock external catalogs with realistic characteristics. To this end, we assumed that the FYST LIM survey will cover a region of the sky benefiting from an external catalog with similar stellar mass completeness limits as the Euclid deep fields \citep{Euclid2022}. This external catalog provides the angular position, redshift, stellar mass and SFR of the galaxies and it is used to pinpoint and mask voxels with a high likelihood of containing bright CO emitters. By carefully simulating this technique, we not only assess its effectiveness in mitigating CO contamination, but also examine its influence on the recovered [CII] PS. This includes an analysis of heightened measurement uncertainty resulting from diminished survey volume and the effects of a convolved PS attributable to the window function of the mask.

The paper is organized as follows. Section 2 introduces the formalism adopted for the modeling of the [CII] and CO emission. Section 3  describes the construction of the mock external catalog, and provides an overview of the masking technique. Section 4 examines the statistical properties of the individual CO lines, the impact of masking on the CO PS, the targeted [CII] PS, and the total line CO+[CII] PS.  Finally, in Section 5, we critically assess the limitations of our work and suggest potential improvements and additions to our models. Throughout the paper, we adopt a $\Lambda$CDM cosmology with the same parameters %as 
used in the TNG300 simulation \citep{Pillepich:Springel:Nelson2017}: $h=0.71$, $\Omega_{\rm b}=0.046$, $\Omega_{\rm m}=0.281$, $\sigma_{\rm 8}=0.8$, $\Omega_{\Lambda}=0.719$, and $n_{\rm s}=0.963$.

%--------------------------------------------------------------------
\section{Generating the mock FYST LIM survey}
\label{sec:Method}

The Fred Young Submillimeter Telescope \citep[FYST,][]{FYST} is set to be a state-of-the-art, 6-meter diameter telescope designed for submillimeter to millimeter observations. Placed at 5600 meters on Cerro Chajnantor, FYST will utilize a novel crossed-Dragone optical design \citep{Dragone1978,Parshley2018} for fast and efficient wide-field-of-view sky mapping. One of its key instruments, the Prime-Cam receiver, will offer impressive spectroscopic and broadband measurement capabilities, enabling a mapping speed over ten times faster than existing facilities \citep{Vava18}.
Prime-Cam features seven instrument modules tailored to specific scientific programs. Two of the modules, named Epoch of Reionization Spectrometer \citep[EoR-Spec,][]{Huber22}, integrate a Fabry-Perot interferometer \citep{FP1899,Zou2022} to observe the [CII] line emission at $z = 3.5-8.05$ (210 to 420 GHz) with a resolving power of approximately $\Delta \lambda/ \lambda=100$. EoR-Spec will target two 5 $\rm{deg} ^2$ regions in a survey dedicated to measuring the [CII] PS from (post-)EoR galaxies \citep[][]{FYST}.
The chosen survey fields require sensitive auxiliary observations with extensive wavelength coverage to help remove foreground sources (see Sect.~\ref{sec:maskingtechnique}). Consequently, the Extended-COSMOS \citep[E-COSMOS,][]{Aihara}, and Extended-Chandra Deep Field South \citep[E-CDFS,][]{Lehmer2005} contained in the Euclid Deep Field Fornax \citep[EDF-F,][]{Euclid2022} are the selected fields for the FYST LIM survey. 
Existing deep coverage in various bands for these fields will soon be augmented by forthcoming imaging and spectroscopy data from present and future state-of-the-art observatories like JWST, Roman, Euclid, and the Large Millimeter Telescope (LMT). For a comprehensive overview of both current and upcoming datasets, see \cite{FYST}.

In this context, realistic simulations of the FYST LIM survey are crucial. They allow for the optimization of the survey strategy, the examination of its ability to detect the [CII] LIM signal during the (post-)EoR, and the testing of various foreground mitigation strategies. To this end, we utilize the TNG300 simulation to create a $4^\circ \times 4^\circ$ cone of mock galaxies spanning from $z \approx 0-9$ (see Sect. ~\ref{Subsection:MockCone}). The [CII] line emission from these galaxies is then incorporated, focusing on the $z \approx 3.4-8.3$ range to which the FYST observed frequencies are sensitive \citep[][hereafter Paper I; see Sect.~\ref{Subsection:CII} for a summary]{Karoumpis}. Subsequently, we introduce the CO ($J_{\rm up}=3-12$) foregrounds which, in the context of FYST, are emitted by galaxies in the $z \approx 0-5.7$ range (see Sect.~\ref{Subsection:CO}). The constructed observable cone is then transformed into mock FYST LIM tomographies (see Sect.~\ref{Subsubsection:Tomographies}), containing both the foreground CO and background [CII] emission, that serves as the basis for evaluating our masking technique.

\subsection{The cone of mock galaxies}
\label{Subsection:MockCone}

In this study, we utilized the dark matter (DM) halo catalog presented in Paper I, which contains all the essential results of the TNG300 simulation needed to estimate the line luminosity of the mock galaxies (i.e., angular position, redshift, $M_{\ast}$, SFR) which are defined as the gravitationally bound substructures hosted in the DM halos. This catalog corresponds to an observational cone encompassing a sky area of $4^\circ \times 4^\circ$ and a redshift range of $z=0-9$, and it was constructed using the TNG300 halo catalogs as building blocks and corrected for resolution effects with the TNG100 DM halo catalogs as a reference \citep{Pillepich:Springel:Nelson2017}. However, for the redshifts of the cone where galaxies emit the brightest CO lines in the frequency range of FYST (i.e., $z=0-5.7$), we recalculated the resolution corrections, this time using the newly available TNG50 simulation \citep{Pille19}. The reason for this update was that the CO emission originates from redshifts where the TNG100 simulation does not accurately reproduce the observed evolution of the cosmic SFR density \citep{Madau2014}, with the peak epoch appearing at earlier cosmic times in the simulation ($z \approx 3$) than the observations ($z \approx 2$). Owing to its 17 times better mass resolution, TNG50 closely follows the cosmic SFR density evolution of \cite{Madau2014}. The same applies for our cone after the new mass resolution corrections.

To quantify the expected sample variance, we employed an analytical relation derived by \cite{Gkogou2022} to directly calculate the coefficient of variation of the PS (CV$_{\rm PS}$) of the LIM tomography. In their study, \cite{Gkogou2022} investigated the impact of field-to-field variance (which we refer to as sample variance) on CO and [CII] PS predictions for future LIM experiments.  This was achieved by combining a 117 deg$^2$ dark matter lightcone with a model of the infrared sky. They developed an analytical formula that estimates the expected CV$_{\rm PS}$ of the [CII] and CO LIM tomographies, given their dimensions. Upon inputting the survey parameters of a 40~GHz frequency coverage and a $4^\circ \times 4^\circ$ field into their formula, we find that the CV$_{\rm PS}$ for the clustering (shot noise) component of the CO PS peaks at 410~GHz, with a value of about 7\% (4\%), while for the [CII] PS, the CV$_{\rm PS}$ peaks at 225~GHz with a value of about 8\% (7\%).

Our predictions of line intensities are based on a single realization of this $4^\circ \times 4^\circ$ field and therefore may be subject to sample variance, due to the small number of galaxies traced. This is particularly true at low redshift (near $z=0$) where the comoving volume covered by our cone is small, and at high redshift (near $z=7$) where star formation occurs predominantly in overdense regions. To quantify the expected sample variance, we employed an analytical relation derived by \cite{Gkogou2022} to directly calculate the coefficient of variation of the PS (CV$_{\rm PS}$) of the LIM tomography. In their study, \cite{Gkogou2022} investigated the impact of field-to-field variance (which we refer to as sample variance) on CO and [CII] PS predictions for future LIM experiments.  This was achieved by combining a 117 deg$^2$ dark matter lightcone with a model of the infrared sky. They developed an analytical formula that estimates the expected CV$_{\rm PS}$ of the [CII] and CO LIM tomographies, given their dimensions. Upon inputting the survey parameters of a 40~GHz frequency coverage and a $4^\circ \times 4^\circ$ field into their formula, we find that the CV$_{\rm PS}$ for the clustering (shot noise) component of the CO PS peaks at 410~GHz, with a value of about 7\% (4\%), while for the [CII] PS, the CV$_{\rm PS}$ peaks at 225~GHz with a value of about 8\% (7\%). Nevertheless, in Sects.~\ref{Subsection:CII} and ~\ref{Subsection:CO} we demonstrate that uncertainties on the modeling of the CO and [CII] emission of galaxies introduced much larger uncertainties on the CO and [CII] PS. For example, in the cases of the CO PS at 350~GHz and the [CII] PS at 225~GHz,  the relative variability of the models in relation to their mean, $(P_{\rm max}-P_{\rm min})/(P_{\rm max}+P_{\rm min})$, is $\approx 85\%$ and $\approx 90\%$ respectively. Therefore, while sample variance affects LIM survey covering only $4^\circ \times 4^\circ$ field, it does not significantly affect the evaluation of the masking technique performed in this paper.

\subsection{The [CII] emission}
\label{Subsection:CII}

The mock galaxy catalog detailed in Sect.~\ref{Subsection:MockCone} underwent various post-processing methods to associate [CII] emission to its galaxies. We restricted these calculations to the redshift range of $z \approx 3.4-8.3$, where the [CII] emission of galaxies is redshifted into the observing bands of EoR-Spec. All the related methods and calculations are described in detail in Paper I and here we only summarize the most important steps.
Firstly, SFRs were attributed to mock galaxies in two ways. One approach used their intrinsic TNG300 SFR, corrected to account for the mass resolution  limitation of TNG300 (see Sect.~\ref{Subsection:MockCone}). The other approach matched the mock galaxy abundance with the observed, dust-corrected ultraviolet luminosity function of high-redshift galaxies \citep{Bouwens:Illingworth:2015}, assigning luminosities to the mock galaxies which are subsequently converted into SFRs following \cite{Kennicutt98}.
Secondly, the [CII] luminosities of the mock galaxies were estimated from the SFRs using three different SFR-to-$L_{\rm [CII]}$ 
relations: from a semi-analytic model of galaxy formation \citep{Lagache2018}, from a hydrodynamical simulation of a high-redshift galaxy \citep{Vallini2015}, and from a high-redshift [CII] galaxy survey \citep{ALPINE}.

The various galaxy-to-SFR and SFR-to-$L_{\rm [CII]}$ relations translate into large variations of the amplitude of the expected [CII] PS (Paper I). Such variations render the assessment of the detectability of the [CII] PS more problematic but at the same time demonstrate the utility of this signal in discriminating between all these models. In the following, our fiducial [CII] model refers to the combination of the recalibrated TNG300 SFR with the SFR-to-$L_{\rm[CII]}$ relation from \cite{Vallini2015}.  It was chosen as it produces results that closely align with the median value of our predictions (Paper I).

\subsection{The CO emission}
\label{Subsection:CO}

 \begin{figure}[ht]
    \centering
        \includegraphics[width=0.45 \textwidth]{{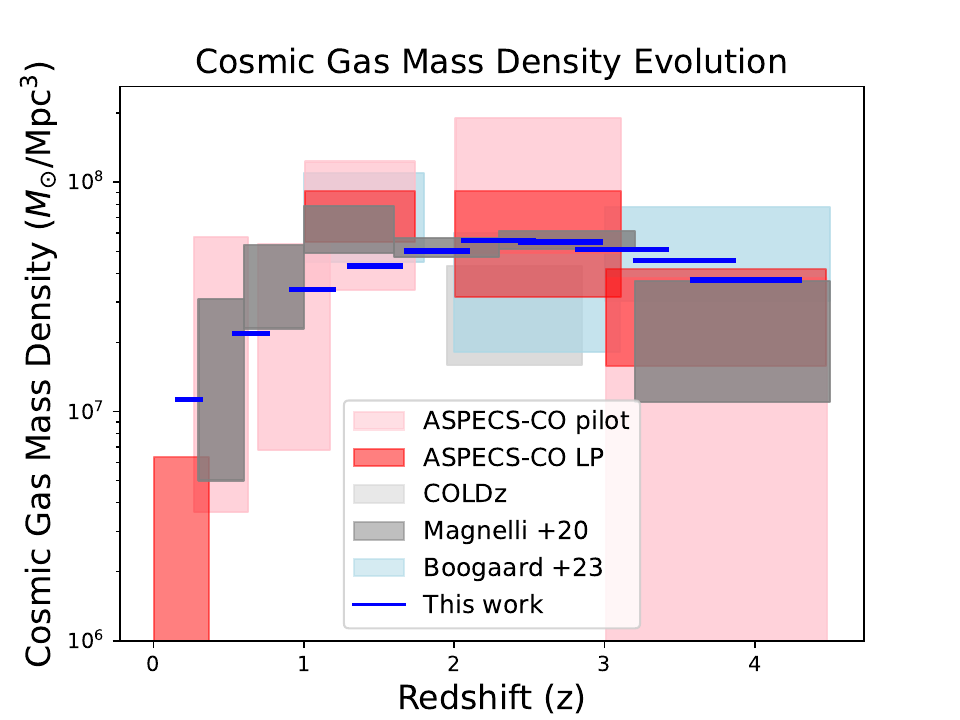}}
        \caption{Redshift evolution of the cosmic molecular gas mass density of galaxies. The blue horizontal line represents predictions from our cone of mock galaxies, while observational constraints are presented by rectangles. The gray rectangles are for the dust-based estimates of \cite{Magnelli2020}, the light gray rectangles are for the COLDz CO survey \citep{Riechers2019}, the light blue rectangle for the CO-based estimates of \cite{Boogaard2023}, the red-shaded rectangles are for the ASPECS CO pilot survey \citep{Decarli2020}, while the pink rectangles are for the ASPECS CO LP survey \citep{Decarli2020}. The width of the rectangles indicates the redshift bin size and the height the 1-$\sigma$ confidence region.}
        \label{fig:MOLG}
\end{figure}

The CO molecule is the second most abundant molecule in the Universe, surpassed only by molecular hydrogen ($\rm{H}_2$). In the typical conditions of giant molecular clouds, the rotational transitions of the CO that emit line emission are easily excited, making it the most widely used tracer of molecular gas in both the local \citep[e.g.,][]{Wang2020,denBrok} and high-redshift universe \citep[e.g.,][]{Solomon2005, Daddi15, Aravena2016, Riechers2019}. Its luminosity is proportional to its mass, assuming the number of clouds is small enough for them not to overshadow each other and that they are close to virial equilibrium \citep[e.g.,][]{CW13}. The strong correlation between the CO line luminosities of galaxies and their molecular gas content enables us to forecast these foregrounds using our cone of mock galaxies.

\subsubsection{The CO (1-0) luminosity}

We estimated the molecular gas content of a galaxy, relying on the well-established relationship between molecular gas and star formation rate. Several studies have shown that the depletion time ($t_{\rm dep}=M_{\rm gas}/\rm{SFR}$) does not depend significantly on the stellar mass of galaxies and only slightly increases from $z \approx 4$ to $z \approx 0$ \citep[e.g.,][]{Scoville2017, Tacconi2018, Kaasinen2019, Magnelli2020, Wang2022}. Therefore, changes in the SFR of galaxies with stellar mass and redshift are primarily driven by changes in their gas reservoir. This allowed us to make the simplifying assumption that all galaxies convert their molecular gas into stars at a constant universal rate. To estimate the molecular gas masses of our mock galaxies, we multiplied their SFRs (the corrected intrinsic TNG300 SFRs presented in Sect. \ref{Subsection:MockCone}) by $t_{\rm dep}=570 \ \rm{Myr}$, corresponding to the depletion time of $z \approx 2$ galaxies with a stellar mass of $M_{\ast} \approx 10^{10} M_{\odot}$ \citep{Tacconi2018}. We verified that even with this simple approach, the cosmic molecular gas mass density inferred from our cone of mock galaxies was compatible with the observations of \cite{Riechers2019},  \cite{Magnelli2020}, \cite{Decarli2020}, and \cite{Boogaard2023} (Fig.~\ref{fig:MOLG}). This agreement is particularly true around the peak of the molecular gas mass density at $z \approx 2$, from where the bulk of the CO foregrounds originate. A more complex model that accounts for changes in the depletion time with redshift and distance from the main sequence (MS; for a definition, see Sect.~\ref{Subsection:SLED}) produces a similar gas mass density evolution, but to avoid introducing additional parameters, we adopted the simpler model of a constant depletion time.

The calculated molecular gas masses were then used to predict the CO line emission of each mock galaxy. We adopted a constant light-to-mass ratio between the specific luminosity of the CO (1-0) line and the molecular gas mass, 
\begin{equation}
 \alpha _{\rm CO} = \frac{M_{\rm gas}}{L'_{\rm CO(1-0)}}=3.6 \ M _{\odot} / \rm{K \ km \ s ^{-1} \ pc ^2,}   
\end{equation}
as its value for the MS galaxies does not seem to change significantly up to $z \approx 3$ \citep{Decarli2016}, and is adopted by the majority of the molecular gas measurements \citep[][]{Daddi15, Riechers2019, Decarli2020, Lenkic2020, Chung2022}.

There is observational evidence that starburst galaxies might require adjustments to their $\alpha _{\rm CO}$ and $t_{\rm dep}$. Specifically, a lower $t_{\rm dep}$ \citep[$ \approx 150\ \rm{Myr}$ as suggested in][]{Tacconi2018} and a reduced $\alpha _{\rm CO}$ value of 0.8 
\citep[as suggested in][]{Downes1998}. Interestingly, the ratio of the proposed depletion times (570/150) is roughly equivalent to the ratio of the $\alpha _{\rm CO}$ values (3.6/0.8), meaning these changes may roughly offset each other. Therefore, while incorporating these adjustments would not significantly alter the $L_{\rm CO(1-0)}$ of our starbursts, it complicates the model further. For this reason, we have chosen to use the simpler model of a constant depletion time and $\alpha _{\rm CO}$. Nevertheless, the effect of starburst galaxies is further examined in Sect.~\ref{Subsection:SLED}. 

More generally, although we may underestimate the stochasticity in our scaling relations for the CO(1-0) luminosities due to the simple assumptions of constant $t_{\rm dep}$ and $\alpha {\rm CO}$, the significant uncertainties introduced in Sect.~\ref{Subsection:SLED} make any additional scatter in the CO(1-0) luminosities unnecessary for the scope of this study. There, by making some assumptions on the spectral line energy distribution (SLED) of our mock galaxies, we use these CO(1-0) line luminosities to predict the CO line luminosities in transitions that pose the most significant foreground challenges for our LIM survey (from $J_{\rm up}=3$ to $J_{\rm up}=12$).

\subsubsection{The CO SLED}
\label{Subsection:SLED}

 \begin{figure}[ht]
    \centering
        \includegraphics[width=0.45 \textwidth]{{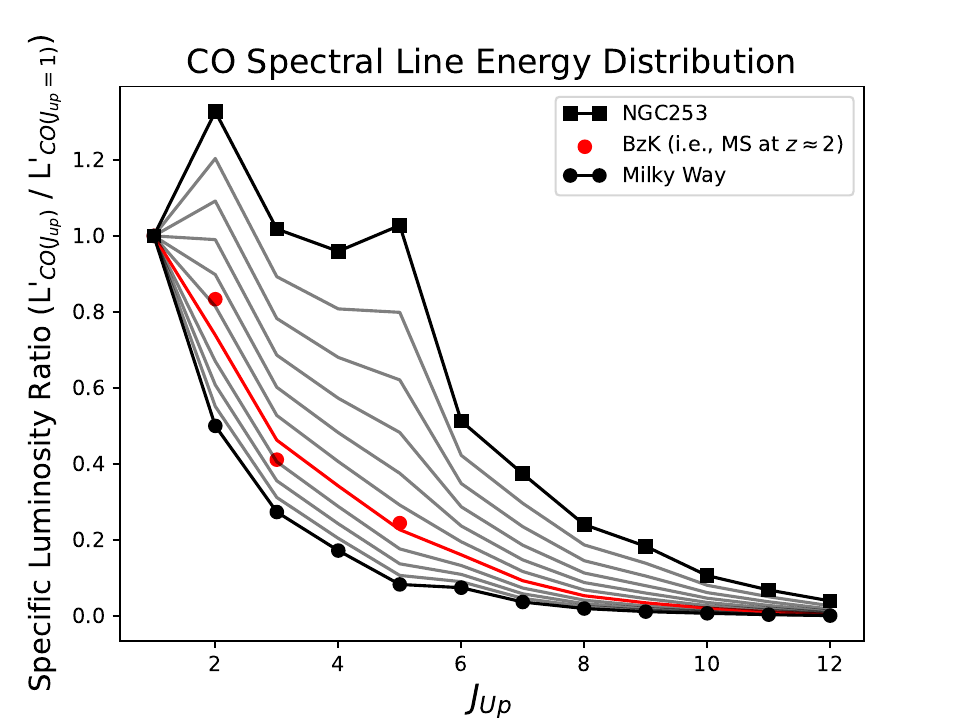}}
        \caption{CO SLED of the Milky Way (black circles) and NGC253 (black squares). The SLEDs resulting from the linear combination of the two, following Eq.~\ref{eq:SLED}, are shown by solid lines, with $\mu$ varying from 0 to 1 in steps of 0.1. The red line represents the combination (with $\mu=0.4$) that best fit the BzK  galaxies (red points) and is used here as reference for high-redshift MS galaxies.}
        \label{fig:SLED}
\end{figure}

The CO SLED of each galaxy is largely defined by the physical conditions (gas density and temperature) of its interstellar medium \citep[ISM; e.g.,][]{Narayanan2014}. It is established that due to the diverse range of conditions in the ISM, the CO SLED of a galaxy often requires more than one temperature component to accurately describe it \citep{Valentino2020}. However, simulating the complex interactions of gas physics is beyond the scope of this paper.

\begin{figure*}[ht]
    \centering
    %\sidecaption
    %\centering
    \begin{subfigure}[b]{7cm}  % Each subfigure is 6 cm wide
        \centering
        \includegraphics[width=\textwidth]{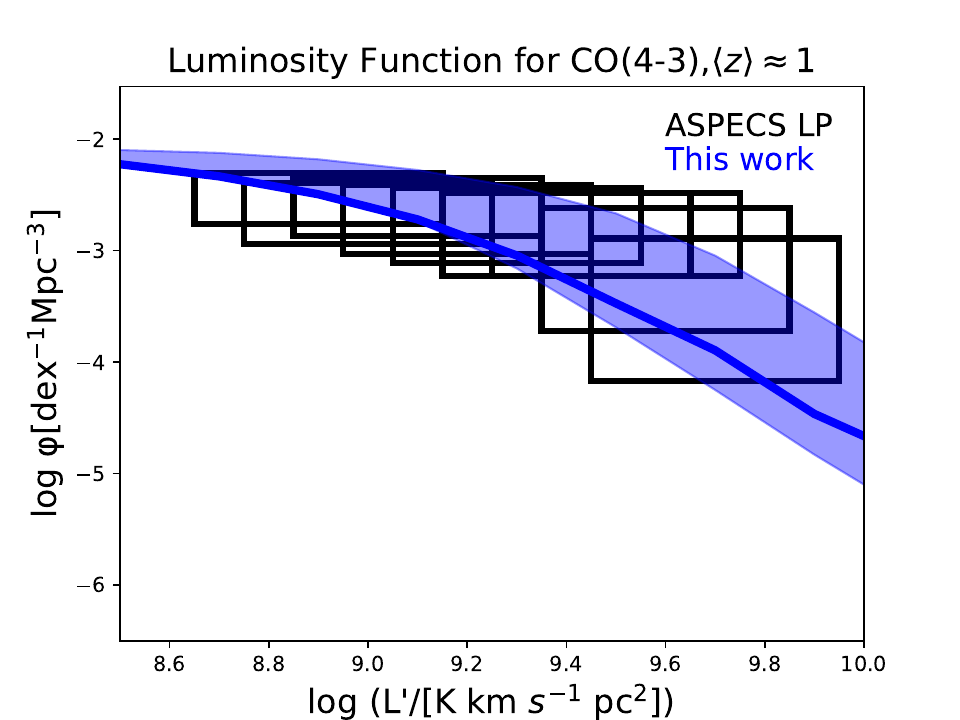}
        \label{fig:one}
    \end{subfigure}
    %\hfill
    \begin{subfigure}[b]{7cm}  % Each subfigure is 6 cm wide
        \centering
        \includegraphics[width=\textwidth]{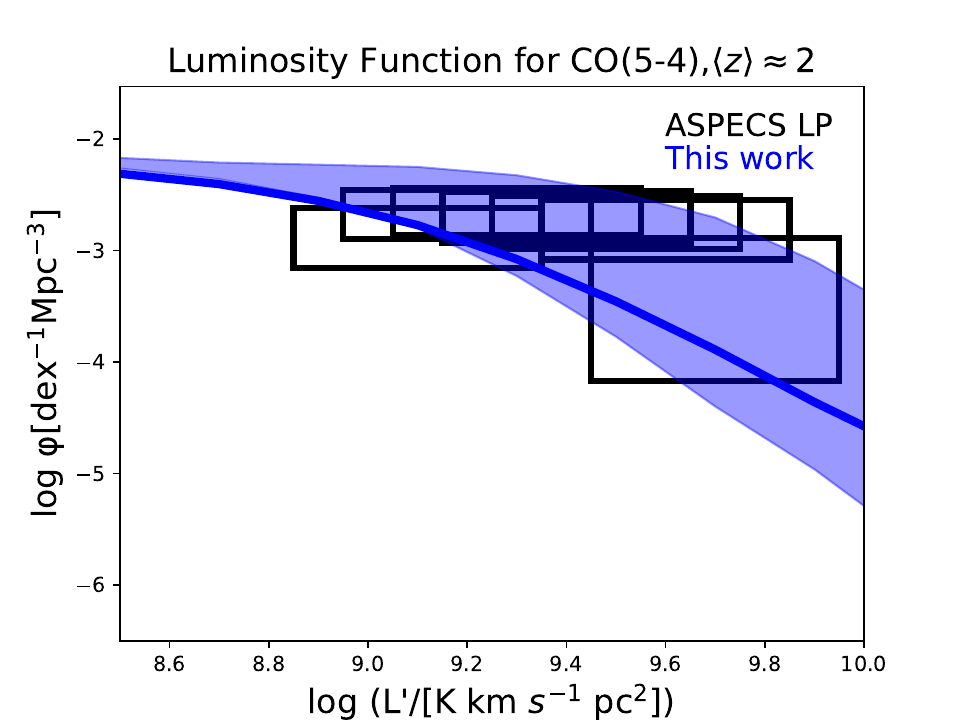}
        \label{fig:two}
    \end{subfigure}
    \vfill
    \begin{subfigure}[b]{7cm}  % Each subfigure is 6 cm wide
        \centering
        \includegraphics[width=\textwidth]{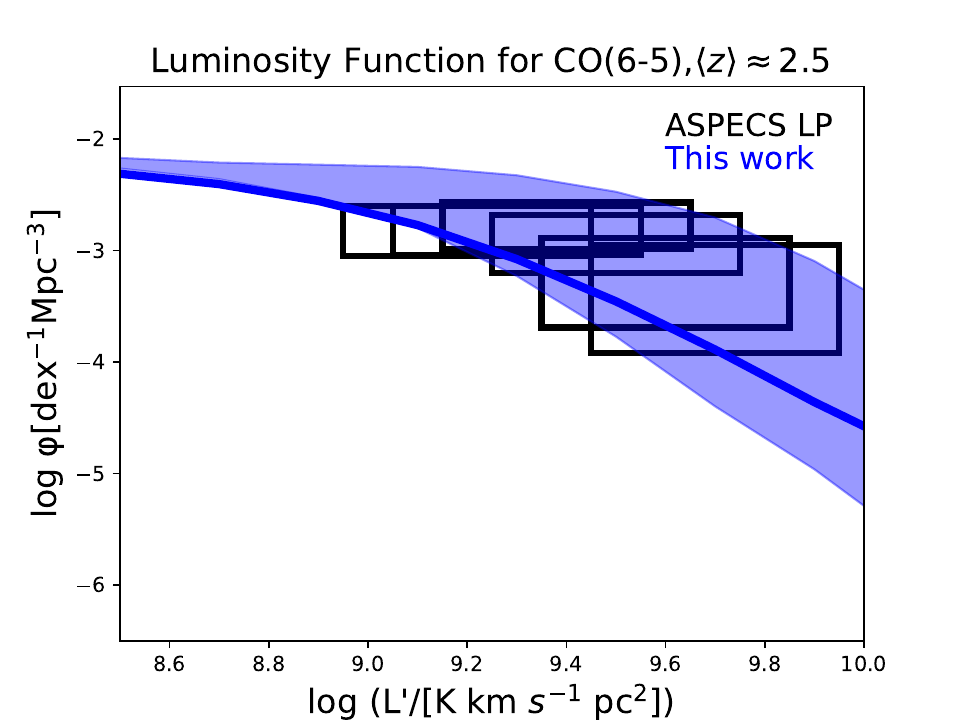}
        \label{fig:three}
    \end{subfigure}
    %\hfill
    \begin{subfigure}[b]{7cm}  % Each subfigure is 6 cm wide
        \centering
        \includegraphics[width=\textwidth]{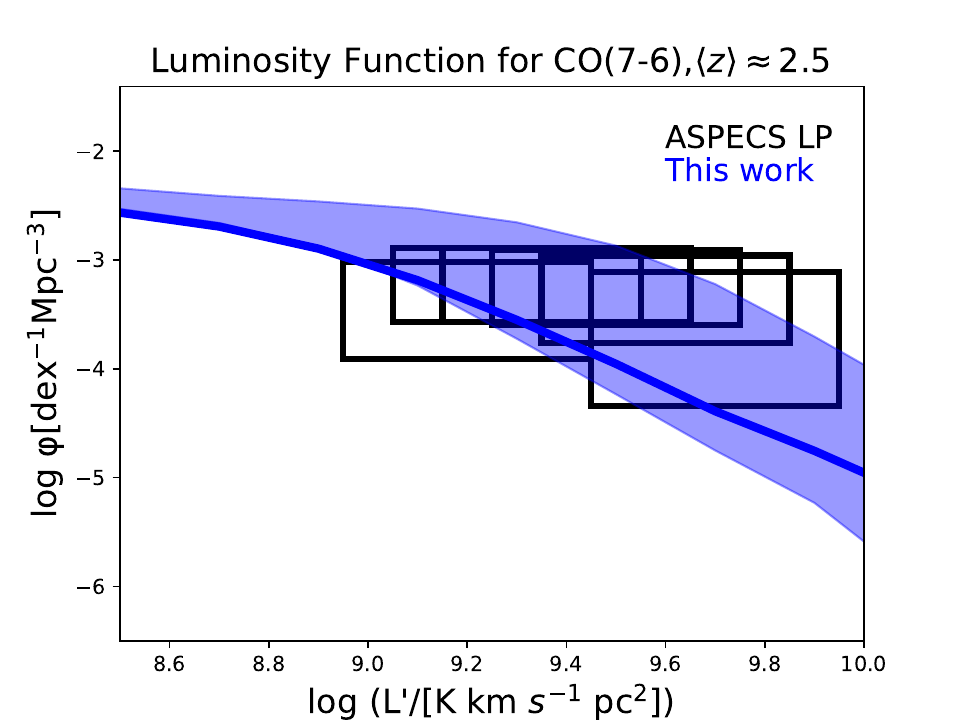}
        \label{fig:four}
    \end{subfigure}
    \caption{CO luminosity function for various transitions and redshifts. The blue line corresponds to our fiducial model, while the shaded region indicates the range resulting from our optimistic and pessimistic models. Rectangles represent observational data from the ASPECS LP survey \citep{Decarli2020}, with the width indicating the bin size and the height indicating the 1-$\sigma$ confidence region.}
    \label{fig:COLUM}
\end{figure*}

Here we employed a model with a small number of free parameters, yet sufficiently advanced to replicate the CO luminosity function in various transitions observed by the Atacama Large Millimeter/submillimeter Array (ALMA) Spectroscopic Survey \citep[ASPECS, ][]{Decarli2020,Boogaard2020,Riechers2020} at frequencies akin to our 225~GHz band (see Fig.~\ref{fig:COLUM} and Table~\ref{tab:CORED}). To achieve this, we adopted an empirical approach based on the galaxy position in the $M_{\ast}-\rm{SFR}$ diagram, which is a valuable tool for evaluating the properties of a galaxy. The majority of the galaxies on this diagram follows a so-called MS \citep{Elbaz07}, wherein more massive galaxies tend to have higher SFRs. However, there are distinct exceptions: starburst galaxies, for instance, exhibit exceptionally high SFRs for their stellar masses; or passive galaxies, that have low SFR for their stellar masses. Here, we defined our MSs by fitting a log-linear relation to the $M_{\ast}-\rm{SFR}$ sequence followed by our mock galaxies at different redshifts. Then, we parameterized the position of a galaxy in this diagram using their distance from the MS, $\Delta \text{MS} = \log (\text{SFR} / \text{SFR}_{\text{MS}})$, where $\text{SFR}_{\text{MS}}$ is the MS SFR at the redshift and stellar mass of the galaxy. Galaxies close to and below the MS have relatively cold ISM \citep[e.g.,][]{CW13,Magdis2021}, while those above the MS, such as starburst galaxies, have a warmer, denser ISM \citep[e.g.,][]{Silich2009, CW13, Kamenetzky2017}. Based on this assumption, we modeled the CO SLEDs of our mock galaxies by linearly combining the observed and well sampled CO SLEDs of a MS-like galaxy and a starburst-like galaxy. We used the CO SLED of the Milky Way (MW), as a characteristic star-forming galaxy with cold ISM conditions, slightly below the MS. On the other hand, we used the CO SLED of NGC253, a characteristic starburst galaxy with warm and dense ISM condition. This linear combination allowed us to create a continuum of conditions that encompassed a broad range of galaxies from MS to starburst. This approach is supported by observations, which report a correlation between the CO SLED shape of a galaxy and its distance from the MS \citep[e.g., ][]{Valentino2020,Cassata2020}. 

The CO SLED of the MW and NGC253 used in our study are displayed in Fig. \ref{fig:SLED}. The MW observations up to $J_{\text{up}} = 5$ were taken from \cite{CW13}, and the SLED was extended to $J_{\text{up}} = 6-10$ using the $\mathrm{CO}\left(J_{\text{up}} - J_{\text{low}}\right) / \mathrm{CO}(5-4)$ measurement from \cite{Wilson17}. The CO SLED of NGC253 was taken from \cite{Mashian15}. The difference between the MW and NGC253 CO SLEDs, which is greater at $J_{\rm up} = 5$, is due to the contrasting star formation environments in these galaxies. The MW features a cold, more diffuse molecular gas component, while the intense star formation of NGC253 creates a warm and dense gas environment, leading to higher excitation of CO molecules and an enhancement mid- to high-$J$ ($J_{\rm up} = 4-8$) transition emission.

We assigned each mock galaxy a SLED resulting from its distance from the MS and the linear combination of these two extreme CO SLEDs. The mixing parameter ($\mu$) was calibrated so that $\Delta \text{MS} = 0$ yields a SLED consistent with typical MS galaxies at $z = 1.5$ \citep[BzK galaxies;][]{Daddi15}, and $\Delta \text{MS} = 1.5$ produces the SLED of NGC253. In turn, this yields,
\begin{equation}
\label{eq:mu}
\mu=\frac{\Delta \text{MS}+1}{2.5}.
\end{equation}
The CO SLED resulting from the linear mixing is given by the equation,
\begin{align}
\label{eq:SLED}
\log L_{\text{CO}(J_{\text{up}} - J_{\text{low}})}^{\text{'galaxy}} &= (1 - \mu) \log L_{\text{CO}(J_{\text{up}} - J_{\text{low}})}^{\text{'MW}} \nonumber \\
&+ \mu \log L_{\text{CO}(J_{\text{up}} - J_{\text{low}})}^{\text{'NGC253}}\ ,
\end{align}
where $L_{\text{CO}(J_{\text{up}} - J_{\text{low}})}^{\text{'galaxy}} $, $L_{\text{CO}(J_{\text{up}} - J_{\text{low}})}^{\text{'MW}}$, and $L_{\text{CO}(J_{\text{up}} - J_{\text{low}})}^{\text{'NGC253}}$ are the CO$(J_{\text{up}} - J_{\text{low}})$ specific line luminosities of our mock galaxy, the MW, and NGC253, respectively. Figure~\ref{fig:SLED} displays an example of the individual CO SLEDs resulting from Eq.~\ref{eq:SLED} (for $\mu =0-1$ with steps of $\delta \mu=0.1$). The linear combination highlighted in red corresponds to $\Delta \text{MS} = 0$ (i.e. $\mu=0.4$), and aligns by design closely with the observed CO SLEDs from BzK galaxies.  

This approach where each galaxy has its individual CO SLED constitutes our fiducial model. However, we also considered two extreme models in addition to our fiducial, creating a larger range of possible values for the CO intensity tomographies. In the "low-contamination" model, all galaxies have the SLED of a typical MS galaxy \citep[i.e., $\mu=0.4$; BzK galaxies;][]{Daddi15}, while in the "high-contamination" model, all galaxies have the SLED of a typical starburst (i.e., $\mu=1$; NGC253).

To verify the accuracy of this approximation, we compared the CO luminosity function inferred from our fiducial, "low-" and "high-contamination"  models to that observed by ASPECS (Fig.~\ref{fig:COLUM}). Our model predictions cover a good range of the 1-$\sigma$ interval compatible with the ASPECS measurements, with, however, a slight over-prediction of the faint end and under-prediction of the bright end. This former discrepancy could be due to the fact that ASPECS is a line-flux-limited CO survey, which naturally becomes less sensitive and more incomplete at the faint end of the CO luminosity function. The difference at the bright end between our models and the 
survey has been reported previously in the semi-analytical approach of \cite{Popping2019, Bethermin2022, Gkogou2022}. As suggested by \cite{Dave2020}, this could be due to the assumption of a constant $\alpha _{\rm CO}$ for all of our galaxies. 
In any case, the galaxies of the brightest bin are few in number and therefore constitute a small percentage of the overall number of interlopers that should be masked. This slight underestimation should not therefore significantly impact our results.

\subsection{The [CII] and CO mock intensity tomographies}
\label{Subsubsection:Tomographies}

In order to translate the line luminosities of our mock galaxies into mock line intensity tomographies, we assumed that the intensity of a line (either CO $J_{\rm up}=3-12$ or [CII]) in a given voxel, positioned at a specific R.A., Dec, and $\nu$, is expressed as:
\begin{equation} \label{eq:map}
I_{\rm line} = \frac{1}{\Delta \nu_0( \Delta \theta_{\rm b})^2} \left (\sum_{j \in \text{voxel}} \frac{L_{\rm line}^j}{4 \pi \ r^2_j(1+z_j)^2} \right ) \\ * G(R.A., Dec, \nu),
\end{equation}
where $r_{j}$ is the comoving distance of the $j$-th galaxy, which resides at a redshift $z_{j}$ and has a line luminosity $L_{\rm line}^{j}$; $*$ is the symbol of the convolution; and $G(R.A., Dec, \nu)$ is the 3D Gaussian function representing the angular and spectral resolution element of the EoR-Spec. To adhere to the Nyquist-Shannon sampling theorem \citep{Shanon49}, we selected the grid spacing of the mock tomographies to be three times smaller than the full width at half maximum (FWHM) of this 3D Gaussian function. Since the summation of all galaxy luminosities within a voxel essentially represents the integration of the intensity field over the volume of the voxel, we divide twice by $\Delta \theta_{\rm b}=\rm{FWHM} _{\rm R.A.}/3= \rm{FWHM} _{ \rm Dec}/3= \rm{FWHM} _{ \rm ang}/3$ and once by $\Delta \nu_0=\rm{FWHM} _{ \rm \nu}/3$ to account for the spatial and frequency dimensions of the voxel. This ensures that the calculated intensity is properly normalized by the volume of the voxel.

We assumed that the FYST LIM survey will be divided into four tomographies for each of the four selected frequency bands of the EoR-Spec. These bands are centered at $225 ±\pm 20$, $280 \pm 20$, $350 \pm 20$, and $410 \pm 20 \ \rm{GHz}$. The average FWHM of the spectral channels for these bands are 2.1, 2.7, 3.6, and 4.4 GHz, respectively. Additionally, the angular beam FWHM for these bands are 0.88, 0.77, 0.65, and 0.62 arcmin, respectively. The intended survey will cover an area of 5 $\rm{deg}^2$. However, for our analysis, we focused on a 16 $\rm{deg}^2$ field, as it offers a testbed with reduced sample variance. It is important to note that when calculating uncertainties (Sect. \ref{sec:Results}), we considered the sample variance and the white noise of the expected (i.e. two fields of 5 $\rm{deg}^2$) size and observational time (2000 hours for each field) of the FYST LIM survey. This way we ensure that while the statistical parameters of the foregrounds and the targeted signal approximate the global values, the impact of any low number statistics is effectively captured within the uncertainty estimates.

\begin{table*}
\centering
\caption{Rest-frame frequencies and redshift ranges of the brightest CO lines and [CII].}
\begin{tabular}{lccccc}
\hline \hline 
Line & $v_r[\mathrm{GHz}]$ & $\Delta z (225 \pm 20$ GHz) & $\Delta z (280 \pm 20$ GHz) & $\Delta z (350 \pm 20$ GHz) & $\Delta z (410 \pm 20$ GHz) \\
\hline 
$\mathrm{C \ II}$ & $1901.0$ & $6.8-8.3$ & $5.3-6.3$ & $4.1-4.8$ & $3.4-3.9$ \\
$\mathrm{CO} \ J=3-2$ & $345.8$ & $0.4-0.7$ & $0.2-0.3$ & $0.0-0.05$ & $\cdots$ \\
$\mathrm{CO} \ J=4-3$ & $461.0$ & $0.9-1.2$ & $0.5-0.8$ & $0.2-0.4$ & $0.1-0.2$ \\
$\mathrm{CO} \ J=5-4$ & $576.3$ & $1.4-1.8$ & $0.9-1.2$ & $0.6-0.7$ & $0.3-0.5$ \\
$\mathrm{CO} \ J=6-5$ & $691.5$ & $1.8-2.4$ & $1.3-1.7$ & $0.9-1.1$ & $0.6-0.8$ \\
$\mathrm{CO} \ J=7-6$ & $806.7$ & $2.3-2.9$ & $1.7-2.1$ & $1.2-1.4$ & $0.9-1.1$ \\
$\mathrm{CO} \ J=8-7$ & $921.8$ & $2.8-3.5$ & $2.1-2.5$ & $1.5-1.8$ & $1.1-1.4$ \\
$\mathrm{CO} \ J=9-8$ & $1036.9$ & $3.2-4.0$ & $2.5-3.0$ & $1.8-2.1$ & $1.4-1.7$ \\
$\mathrm{CO} \ J=10-9$ & $1152.0$ & $3.7-4.6$ & $2.8-3.4$ & $2.1-2.5$ & $1.7-2.0$ \\
$\mathrm{CO} \ J=11-10$ & $1267.0$ & $4.2-5.2$ & $3.2-3.9$ & $2.4-2.8$ & $1.9-2.2$ \\
$\mathrm{CO} \ J=12-11$ & $1382.0$ & $4.6-5.7$ & $3.6-4.3$ & $2.7-3.2$ & $2.2-2.5$ \\
\hline
\end{tabular}
\label{tab:CORED}
\tablefoot{The rest-frame frequencies ($ \nu _r$) of the brightest CO lines present in the selected EoR-Spec frequency bands and [CII], together with the redshift range ($\Delta z$) from which their emission originate.}
\end{table*}

\section{The masking technique}
\label{sec:maskingtechnique}
 \begin{figure}[ht]
    \centering
        \includegraphics[width=0.45 \textwidth]{{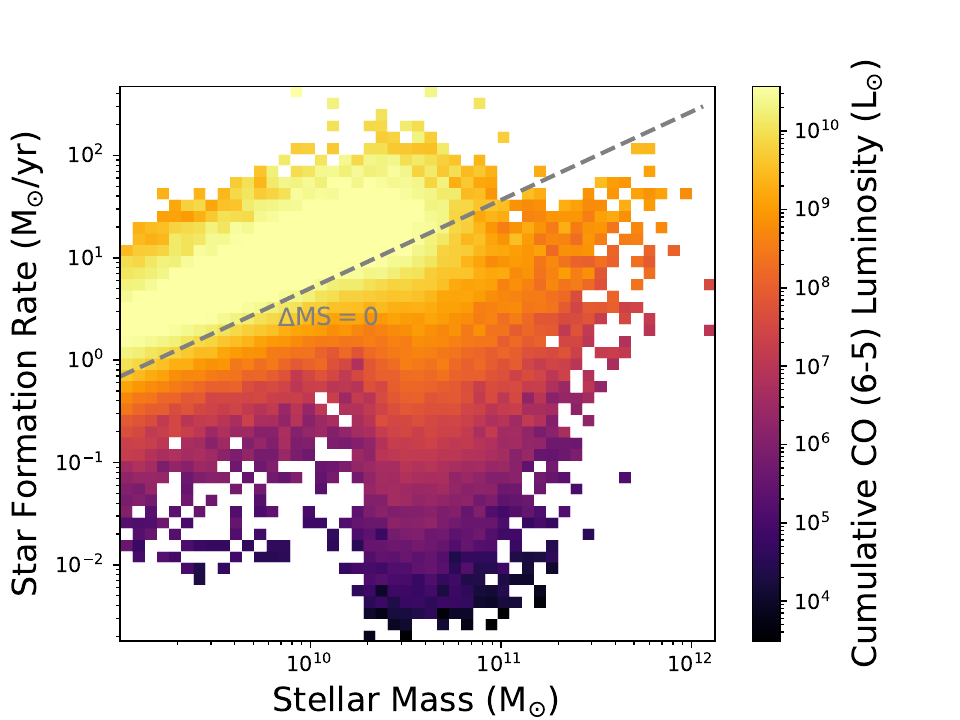}}
        \caption{SFR as a function of stellar mass with the bins color-coded to reflect the cumulative CO (6-5) luminosity of the 2D bin. The dashed line indicates the thresholds that define our "bright" subsample with $\Delta MS >0$ (i.e., galaxies on and above the MS) }
        \label{fig:MS}
  
\end{figure}

While various alternative methods for addressing submillimeter line foreground contamination have been proposed in the literature, including spectral template fitting \citep{Kogut2015, Cheng2020}, blind masking \citep{Breysse2015}, and machine learning techniques \citep{Moriwaki2020, Moriwaki2021, Zhou2023}, we focus on what appears to be the most effective approach—targeted masking \citep[e.g.,][]{Yue:Ferrara2015, Silva2015, Sun2018}. This technique employs external catalogs to identify the locations of the brightest foreground sources and mitigates their influence on the targeted PS by excluding the implicated voxels from further analysis. 

In our study, the masks were created based on a mock external stellar mass-selected catalog (Sect.~\ref{Subsection:External}). In order to explore several depths of masking, for each CO line transition in each FYST frequency band, we created two masks of different depth: the "bright" and the "complete" masking arrays. To generate the "complete" masking arrays, we masked every galaxy in our stellar mass-complete external catalog with redshift $z_{\rm J_{\rm up}}=(\nu_{\rm J_{\rm up}}/\nu_{\rm obs}\pm 20 \ \rm{GHz})-1$, where $J_{\rm up}$ refers to the upper level of the CO transition, and $\nu _{\rm obs}$ is the central observed frequency of one of the four selected EoR-Spec bands.

For the "bright" masking arrays, we employed the same method but limited it to galaxies above the MS, (i.e., $\Delta \rm{MS} > 0$). This way to identify bright CO contaminants solely based on $\Delta \rm{MS}$ stemmed from the observation that the cumulative CO line contamination from galaxies correlates with the MS boundary (see e.g., Fig.~\ref{fig:MS}). By exclusively masking galaxies above the MS, we effectively limited our masks to those galaxies that significantly contaminate the targeted [CII] signal. This approach optimizes the survey volume available for our PS analysis. The masking arrays, constructed using these two subsamples, are presented in Sect.~\ref{Subsubsection:MaskArray}, and the methodology behind this masking strategy is discussed in further detail in Sect.~\ref{Subsection:Evaluating}.

\subsection{The mock external catalog}
\label{Subsection:External}

\begin{figure} %[ht]
    \centering
        \includegraphics[width=0.45 \textwidth]{{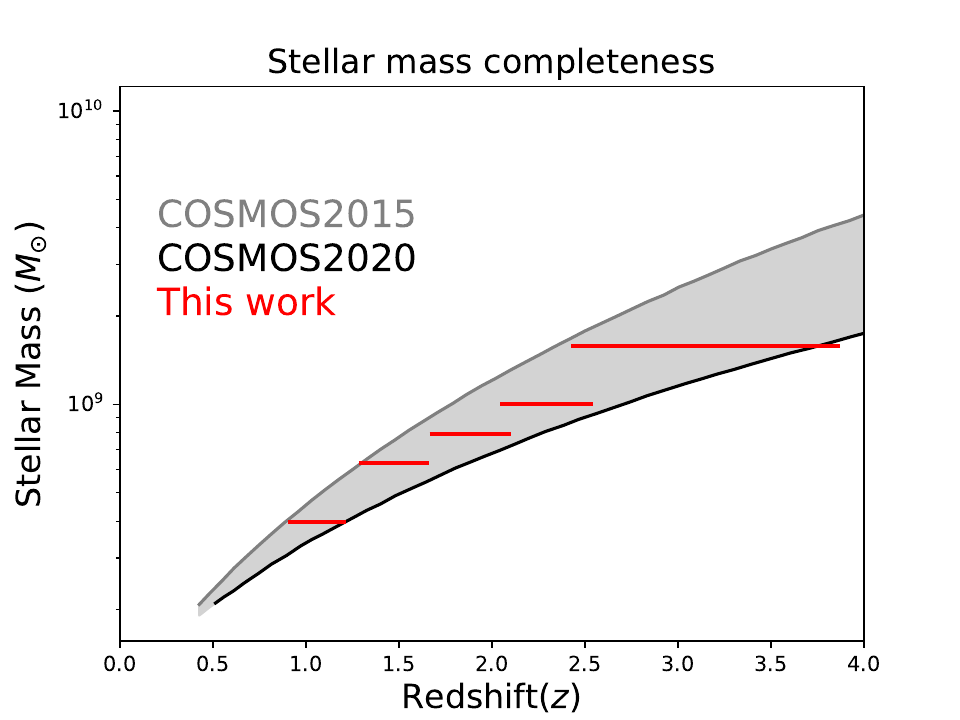}}
        \caption{ Comparison of the stellar mass completeness limits applied to our mock external catalog (red line) with those of the COSMOS2015 \citep[light gray line][]{Laigle2016} and COSMOS2020 \citep[dark gray line][]{Weaver2022} catalogs. The stellar mass completeness limits of the deep Euclid catalogs is expected to lie between these two lines.}
        \label{fig:Complit}
  
\end{figure}

The fields that will be targeted by the FYST LIM survey \citep{FYST} are the E-COSMOS \citep{Aihara} and the E-CDFS \citep{Lehmer2005}, with the latter potentially being extended to cover the whole EDF-F. To mimic the way these surveys would observe our cone of mock galaxies, we simply applied to our mock galaxy catalog redshift-dependent stellar mass limits akin to those affecting an Euclid-deep-like survey \citep{Euclid2020}. According to the latest forecasts, such a survey will have multi-wavelength photometric sensitivities that will be equivalent to those obtained today in the COSMOS field. By taking the stellar mass completeness limits from the COSMOS2015 \citep{Laigle2016} and COSMOS2020 \citep{Weaver2022} catalogs, we should thus obtain realistic redshift-dependent stellar mass limits to apply to our mock galaxy catalog. These redshift-dependent limits are shown in Fig.~\ref{fig:Complit}, starting at $\approx 0.5 \times 10^9 M_{\odot}$ at low redshifts ($z \approx 0-2$) and rising to $\approx 10^9 M_{\odot}$ at high-redshifts ($z \approx 2-4$).

In addition to the angular position, redshift information, and stellar mass, we also included in our mock external catalog the SFR for each galaxy. Indeed, we anticipated that the Euclid Deep Fields will provide SFR estimates for the majority of detected galaxies through the spectral energy distribution (SED) fit of all the available optical to near-infrared bands. 

We note that the limited angular resolution of the EoR-Spec renders insignificant any potential offset between the CO position and the optically-based position of galaxies in this external catalog, or any potential astrometry inaccuracies in the external catalog. Uncertainties on the angular position of our mock galaxies were thus not included in our mock external catalog. Moreover, given that the $\Delta z$ associated with our frequency channels largely exceeds the anticipated photometric redshift accuracy of 0.002 $\times$ (1+$z$) from Euclid \citep{Ilbert2021}, uncertainties on the line-of-sight distance were also not included in our mock external catalog.

\subsection{The masking array}
\label{Subsubsection:MaskArray}

The process of masking voxels containing bright CO contaminant is represented by a 3D array of zeros and ones, with zeros indicating voxels likely containing a bright CO emitter and thus requiring masking. This array is referred to as the binary masking array, $B$. Here, through empirical testing, we found that the most effective voxel size for masking is that of $(\rm{FWHM} _{ \rm \nu}/3) \times (\rm{FWHM}_{\rm ang}/3)^2$.
It is important to note that observed LIM tomographies undergo convolution with the beam of the telescope. This results in intensity from galaxies within a voxel spilling over to neighboring voxels. Additionally, sharp masks, like our binary masking arrays, produce high-frequency components in their Fourier transform that are not well captured when sampling this space discretely. This causes high-frequency details to alias into lower frequencies, distorting the Fourier-transformed data.

A remedy for both issues is to smooth the binary mask using a 3D Gaussian normalized such that its volume integral is equal to 1, a 3D equivalent of the 2D method used in CMB studies \citep{Ponthieu:Grain2011, Kim2011}. While \cite{Ponthieu:Grain2011} recommends a Gaussian size double the map resolution, we have opted for a size consistent with the 3D beam of the EoR-Spec. Indeed, due to our broader binning in Fourier space, we do not face the same level of aliasing as CMB studies. This choice aligns with the inherent resolution of the instrument and addresses potential aliasing without significantly reducing the survey volume. However, as highlighted by \cite{Kim2011}, Gaussian smoothing can risk foreground contamination, that is, pixels initially zero in the binary mask might become non-zero after smoothing. To counteract this, we multiplied the smoothed mask with the original binary mask.

With the above considerations, we generated our mask as follows: we subtracted $B$ from a unit array, then convolved the outcome with the normalized 3D beam of the EoR-Spec,  $G(R.A., Dec, \nu)$. By subtracting the convolution result from another unit array and multiplying it by the original binary array, we obtained the "applied masking array," $M$. Mathematically, this is expressed as:
\begin{equation}
    M = B \times [\mathbf{1} - (\mathbf{1} - B) *  G(R.A., Dec, \nu)],
\end{equation}
where $\mathbf{1}$ is an array of ones of identical dimensions to $B$. Through this approach, the impact $M$ has on the PS of a tomography mirrors the impact $B$ has on the PS of a perfectly deconvolved map. This method bypasses deconvolution, a procedure that is not only computationally intensive but that can also amplify the noise of the observed data.

\section{Results}
\label{sec:Results}

\subsection{Line intensities}
\label{Subsection:Intensities}

 \begin{figure}[ht]
    \centering
    \includegraphics[width=0.5 \textwidth]{{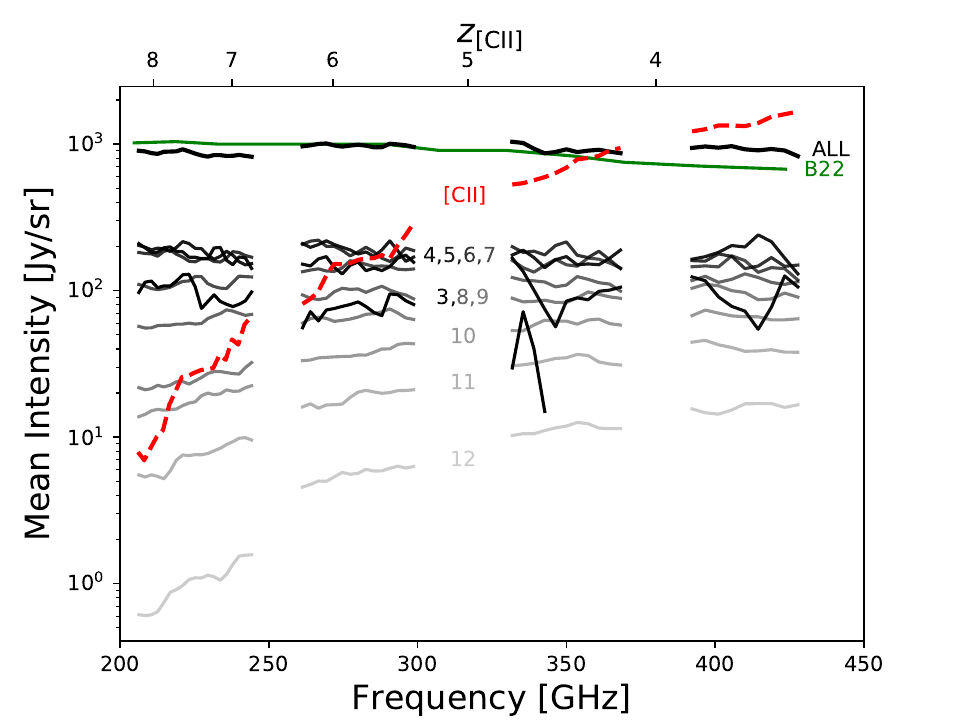}}
        \caption{Mean [CII] and CO line intensities as a function of the observed frequency (or emitted redshift for [CII]). The red dashed line depicts the [CII] fiducial model from \cite{Karoumpis}. Continuous gray lines represent the prediction for the CO emission, with line shade decreasing as $J_{\rm up}$ (gray numbers) increases. The thicker black line indicates the mean total CO emission, while the \cite{Bethermin2022} prediction for total CO mean intensity (green line) is included for comparison.}
        \label{fig:meanpix}
\end{figure}
\begin{figure} %[H]
    \centering
        \includegraphics[width=0.45 \textwidth]{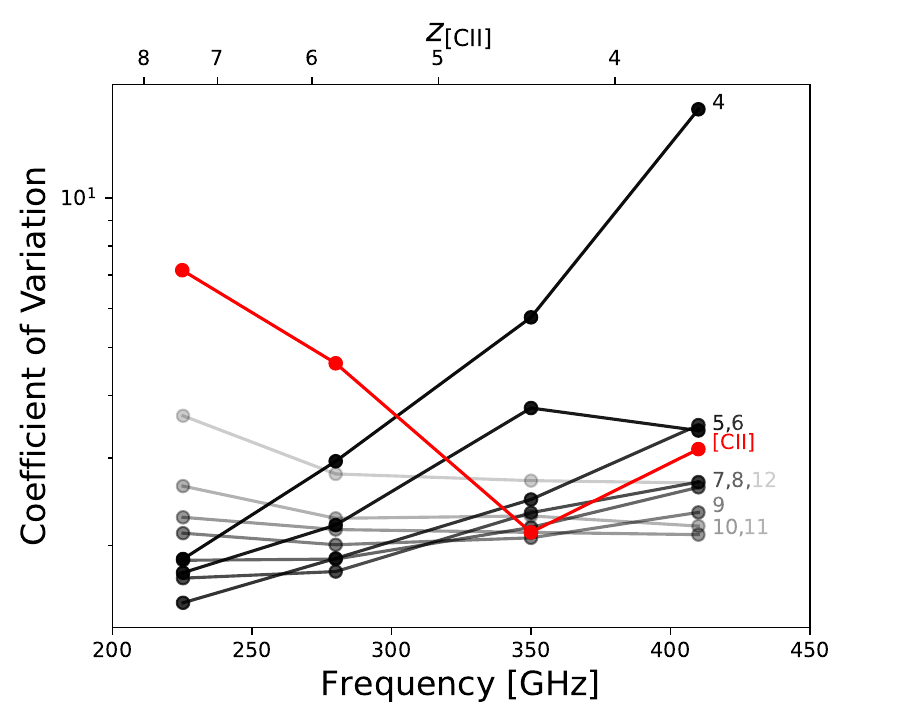}
        \caption{Coefficient of variation of the [CII] (red) and CO lines (gray) intensities for the four FYST frequency bands. The gray numbers of decreasing opacity denote the $J_{ \rm up}$ of the CO transition. The secondary $x$-axis indicating redshift applies solely to [CII] emitters.}
        \label{fig:VAR}
\end{figure}

To inform our masking strategy, we begin by analyzing the predicted intensities of the CO lines at the frequencies of the selected EoR-Spec bands ($225 \pm 20$ GHz, $280 \pm 20$ GHz, $350 \pm 20$ GHz, and $410 \pm 20$ GHz), that is, comparing the mean values (Fig.~\ref{fig:meanpix}) and CV (i.e., $ \sigma _{\rm I}/ \mu _{\rm I}$; Fig.~\ref{fig:VAR}) of their intensities to those of the targeted line, [CII].

Variations in the mean intensities of spectral lines at specific observed frequencies are primarily due to two factors: the different line luminosities of the galaxy populations emitting each line, and the different distances to each of these populations. In Fig.~\ref{fig:meanpix}, we see a clear trend in the evolution of line mean intensity with observed frequency. As we shift towards lines with higher rest-frame frequency which therefore originate from higher redshifts, the gradient of their mean intensity as a function of frequency becomes markedly steeper. This steepening is a consequence of more substantial changes between the luminosity distances of galaxies of subsequent frequency channels when dealing with higher redshifts.

Notably the mean intensities of CO(3-2) and CO(4-3) exhibit significant fluctuations at frequencies above 300 GHz. This variability can be attributed to their origin from redshifts $z<0.4$, where the volume encompassed by the observational cone is relatively small. Consequently, there is a limited number of emitting galaxies within this volume, leading to intensity fluctuations that are strongly driven by small number statistics. Furthermore, CO(3-2) cannot be observed at frequencies above 345.8 GHz, its rest frequency.

While the mean of the total CO intensity remains relatively constant at all frequencies ($ \approx 10^3 $  Jy/sr) and is always dominated by the low-$J$ CO transition from low redshift galaxies, the [CII] intensity shows a notable increase from about  $\approx 10 $ Jy/sr at 225~GHz to $\approx 2000$ Jy/sr at 410~GHz. We estimate that [CII] becomes the dominant line at $>364$ GHz, a prediction that matches well with the results of \cite{Bethermin2022} which quotes a crossover at 358 GHz.

The variability in intensity distribution, as indicated by the CV, is also critical for the efficiency of our masking method. A high CV implies that the line intensity is concentrated in fewer voxels, while a lower CV indicates a more uniform voxel-to-voxel distribution. This aspect becomes increasingly relevant as [CII] signal recovery becomes more challenging at lower frequencies (i.e. higher redshifts).

The differences in CV arise from variations in the size of the galaxy populations contributing to the intensity of each line. This size is contingent on the survey volume, which expands with increasing redshift, and the number density of star-forming galaxies, which peaks around $z \approx 3$ in the case of our mock observational cone \citep[in agreement with observations like][]{Conselice}. Consequently, the CV typically decreases from $z \approx 0$ to $z \approx 4$ and then increases for higher redshifts, as shown in Fig.~\ref{fig:VAR} in the case of the [CII] line. For the bright CO lines (i.e., $J_{ \rm up}=4-9$, see Fig.~\ref{fig:meanpix}), which are emitted by galaxies at $z<4$, only the decreasing trend appears. On the other hand, for the CO lines with $J_{\rm up}>10$, CV is constant with redshift at $>280$ GHz since they originate from $z \approx 2-3$, where the survey volume expands slowly. Additionally, for these $J_{ \rm up}$, an increasing trend appears at lower frequencies ($<280$ GHz) where these lines are emitted by galaxies at $z>4$ where the number density of galaxies decreases rapidly. 

Just by comparing the mean and the CV of the line intensities and before even applying any mask, we observed a clear dichotomy. On the one hand, for the recovery of the [CII] emission originating from $z<5$ galaxies, the situation is encouraging as the foregrounds have comparable intensity with the targeted signal and the dominating low-$J$ line emission is scarcely distributed in the surveyed volume (i.e., their CV is high). On the other hand, for the recovery of the [CII] emission from $z>5$ galaxies, the situation appears less favorable as the foregrounds are on average more than an order of magnitude brighter than the targeted signal and the dominating low-$J$ line emission is homogeneously distributed in the surveyed volume (i.e., CV is low).  In the following section, we explore and quantify how these differences impact the effectiveness of the masking technique in removing the CO signal. There, it becomes clear that the differences in the CV of the CO intensity result in high masking efficiency (significant CO reduction with a small fraction of the survey volume masked) for our low-redshift (high-frequency) tomographies, but low masking efficiency (poor CO reduction with a large fraction of the survey volume masked) for our high-redshift (low-frequency) tomographies.

\subsection{Evaluating the masking technique efficiency}
\label{Subsection:Evaluating}

The effectiveness of a foreground removal technique can only be evaluated by its impact on the statistic of interest. Our study utilizes the spherically averaged PS, computed within the comoving volume of the [CII] reference frame, thus standardizing the analysis frame to avoid constant conversions between different lines. Contrasting with the approach in Paper I, here we compute the PS for tomographies convolved with the 3D beam of the EoR-Spec, rather than incorporating the beam attenuation into the uncertainty of the PS. This adjustment allows us to simulate the effect that the interplay between the transfer functions of the beam and the mask has on the measured PS. We are focusing on the PS at $k=0.02-0.32 \ \rm{Mpc}^{-1}$, a scale that is less affected by sample variance and masking bias. However, our findings hold true at higher $k$-bins ($k=0.32-0.62\ \rm{Mpc}^{-1}$ and $k=0.62-0.92\ \rm{Mpc}^{-1}$), suggesting that the effectiveness of the masking strategy is relatively constant across a broad range of scales.

Mitigating CO contamination while preserving maximal survey volume is far from straightforward due to the variety of possible CO masks and their nuanced effects on both CO and [CII] PS. To unravel these complex behaviors, we have adopted a three-step approach: initially focusing on CO alone to deduce the optimal masking sequence and strategy (Sect. \ref{Subsubsection:MaskCO}), then examining the impact of masking on the [CII] signal in isolation (Sect. \ref{Subsubsection:MaskCII}), and finally analyzing the combined effects to fully understand the interplay between masking, CO contamination, and the [CII] signal (Sect. \ref{Subsubsection:MaskCOCII}). This comprehensive approach will not only enable us to identify the optimal masking depth, but also to assess the level of contamination at this depth.

\subsubsection{Impact of masking on the CO PS}
\label{Subsubsection:MaskCO}

\begin{figure*}
  \centering
  \includegraphics[width=12cm]{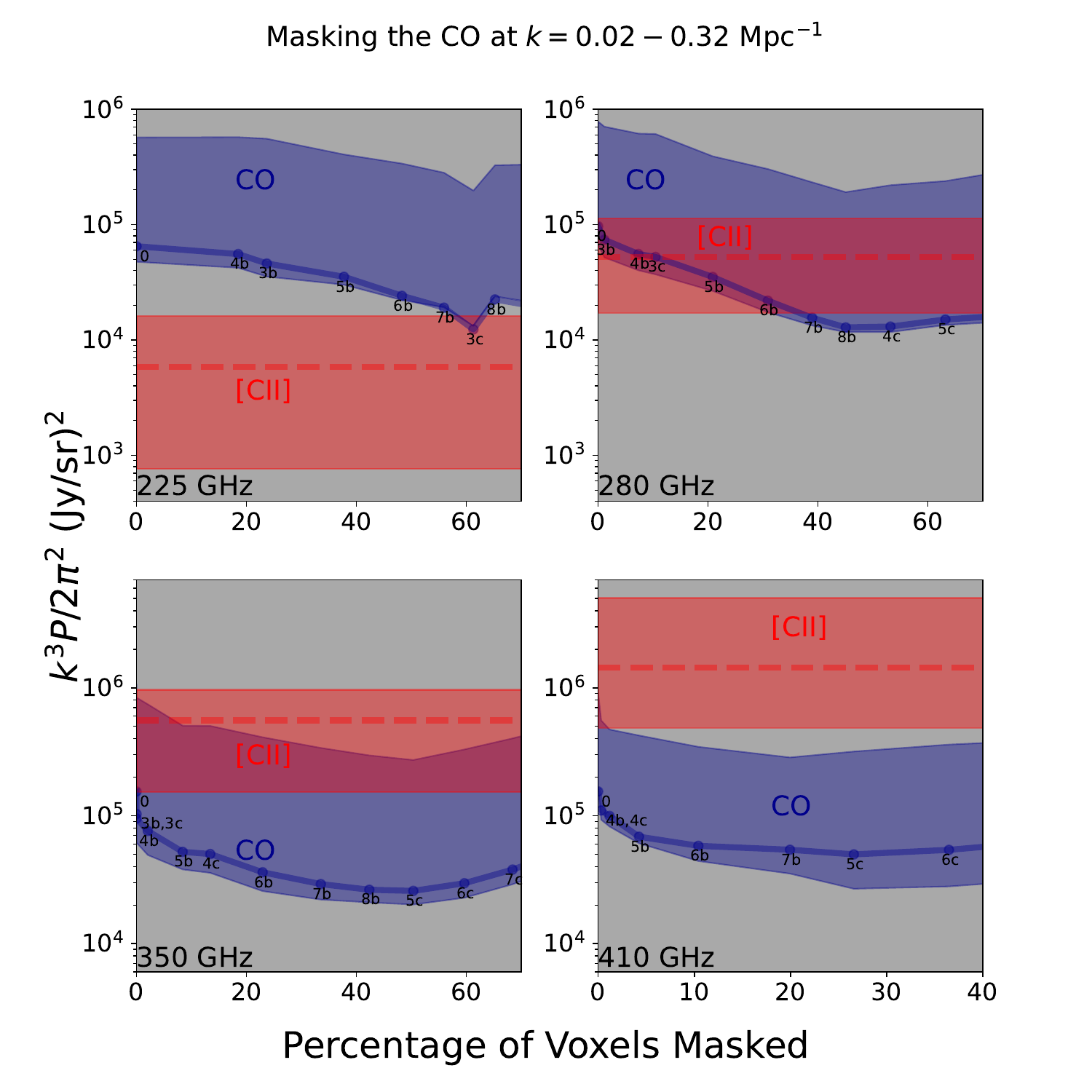}
  \caption{Effect of masking on the CO-only PS. The range of PS between the "low-contamination" and "high-contamination" models (Sect. \ref{Subsection:SLED}) at $k=0.02-0.32 \ \rm{Mpc}^{-1}$ for CO lines (blue) is shown in relation to the percentage of voxels masked within the LIM tomographies for the frequency ranges of $225 \pm 40$ GHz (associated with $z_{\rm [CII]}=7.4$), $280 \pm 40$ GHz ( $z_{\rm [CII]}=5.8$), $350 \pm 40$ GHz ( $z_{ \rm [CII]}=4.3$), and $410 \pm 40$ GHz ( $z_{ \rm [CII]}=3.7$). The lines represent the optimal CO masking sequence for our fiducial model, while each point and label indicate the $J_{\rm up}$ and subsample ("b" for "bright" and "c" for "complete") associated with the masking array applied. The same masking sequence is applied to the "low-contamination" and "high-contamination" models (Sect. \ref{Subsubsection:MaskCO}). As an illustration and assuming at this stage that the mask does not affect the [CII] PS (but see Sect. 4.2.2), we show the fiducial and range of [CII] PS predictions in red.}
  \label{fig:coimpact}
\end{figure*}

As already discussed in Sect.~\ref{sec:maskingtechnique}, for each frequency band and for each CO line we generated two distinct masking arrays: a "complete" masking array that masks all the galaxies in our external catalog with $ z_{ \rm J_{ \rm up}}=(\nu_{ \rm J_{ \rm  up}}/\nu_{ \rm obs} \pm 20 \ \rm{GHz}) - 1$  and a "bright" masking array that among the galaxies of the "complete" subsample masks only those with $ \Delta \rm{MS} > 0 $.

Given the multitude of available masks, each with a unique impact on the observed data, identifying the optimal masking sequence becomes a critical challenge in our analysis. Therefore, it is important to establish a reliable criterion for determining the most effective sequence of masks. The effectiveness, $E$($J_{\rm up}$,~bright~or~complete) of masking arrays, $ M $($J_{\rm up}$,~bright~or~complete), is quantified as, 
\begin{equation}
\label{eq:effectiveness}
E = \frac{\text{Standard deviation} _{\rm unmasked} - \text{Standard deviation}_{\rm masked}}{\text{Number of masked voxels}}.
\end{equation}
The reason for this definition of $E$ is that we need a scale independent value that correlates with the amplitude of the PS per number of masked voxels, like the standard deviation of the intensity normalized by the number of voxels does.

The masking array with the highest $E$ is applied iteratively to the CO-only tomography, continuing until all arrays have been utilized. With each iteration, we recalculate the effectiveness of subsequent masks on the updated data. This iterative process, performed on our CO-only simulations, identifies a sequence of masking arrays for each LIM tomography. It is important to note that this optimal sequence of CO masks remains unchanged at $350 \pm 20$ GHz and $410 \pm 20$ GHz if evaluated on a tomography including CO, [CII], and realistic white noise for the FYST LIM survey. In such tomographies, the masking efficiency ($E$) for each CO line is indeed still dominated by the CO signal. However, at $225 \pm 20$ GHz and $280 \pm 20$ GHz, where the [CII] signal and white noise become increasingly important, it is not practical to evaluate the optimal CO masking sequence directly from a realistic tomography including CO, [CII] and instrumental white noise. However, for all our tomographies, we observed that the optimal CO masking sequence does not vary much across the range of CO foreground models implemented in our work. This implies that the optimal CO masking sequences inferred here on our CO-only tomographies are applicable to first order to real observations. Any deviation from the "real" optimal sequence is unlikely to affect the result significantly.

The optimal CO masking sequence as well as their impact on the CO PS as a function of masking depth is depicted in Fig.~\ref{fig:coimpact} at $k=0.02-0.32 \ \rm{Mpc}^{-1}$. By construction of our optimal CO masking sequence, the slopes of the lines connecting the CO PS as a function of masking depth starts by having negative values that get progressively less negative as the masking depth increase and so does the masking effectiveness, $E$. This trend of progressively less negative slope is, however, not always observed in this initial phase due to fact that the definition of $E$ is scale independent and thus not tailored to the specific $k \approx 0.02-0.32 \ \rm{Mpc}^{-1}$ explored in Fig.~\ref{fig:coimpact}. Overall, this figure demonstrates that in this initial phase, the decrease of the CO PS with masking depth is mostly driven at the high frequency bands by the masking of the CO $J_{\rm up} =3$ (for $350 \pm 20$ GHz) and $J_{\rm up}=4$ line (for $350 \pm 20$ GHz and $410 \pm 20$ GHz) which results in a 50\% reduction of the CO PS. At the low frequency bands ($225 \pm 20$ GHz and $280 \pm 20$ GHz) the decrease is more gradual, yielding to a reduction of one order of magnitude by masking all $J_{\rm up}<8$ lines.

As the masking process intensifies, a critical threshold is reached where the amplitude of the CO PS reaches a minimum value (hereafter called the optimal masking depth) and after which it rises despite the increasing masking depth. This occurs when the CO line being masked contributes less to the total intensity of the tomography than the residual emissions from the brighter CO lines that have been masked previously. At this point, the mask begins to function akin to a random mask on the LIM tomography, as it is no longer targeting the brightest voxels. Moreover, since the intensity of the voxels follow a sparse, log-normal distribution (Sect. \ref{Subsubsection:MaskCOCII}), this random masking results in the removal of more lower-intensity than high-intensity voxels. However, the correction we apply to the PS for the survey volume lost due to masking assumes that the intensity is uniformly distributed throughout the survey volume. This assumption of uniformity, although simplistic, is necessary because the non-homogeneity of the maps is model-dependent and cannot be accurately predicted without specific models. Nevertheless, it results in an increase in the amplitude of the PS when the mask does not follow the dominant galaxy population.

From Fig.~\ref{fig:coimpact}, we note that the optimal masking depth as we move to higher frequency tomographies is reached at progressively lower masking depths ($60\%$, $40\%$, $40\%$ and $30\%$ for the 225 GHz, 280 GHz, 350 GHz and 410 GHz accordingly). Comparing the amplitude of the CO PS at these optimal masking depths with that of the CII PS--assuming, at this stage, that it remains unaffected by the mask (but see Sect. \ref{Subsubsection:MaskCII} for further discussion)--allows us to preliminarily evaluate the feasibility of using this targeted masking approach to extract the CII PS signal. We find that in all frequency bands except for the $225 \pm 40$ GHz, the minimum fiducial CO PS is lower than the corresponding [CII] PS.

\subsubsection{Impact of masking on the [CII] PS}
\label{Subsubsection:MaskCII}

\begin{figure}[ht] % Use the figure* environment for a two-column wide figure
    \centering
    \includegraphics[width=0.45 \textwidth]{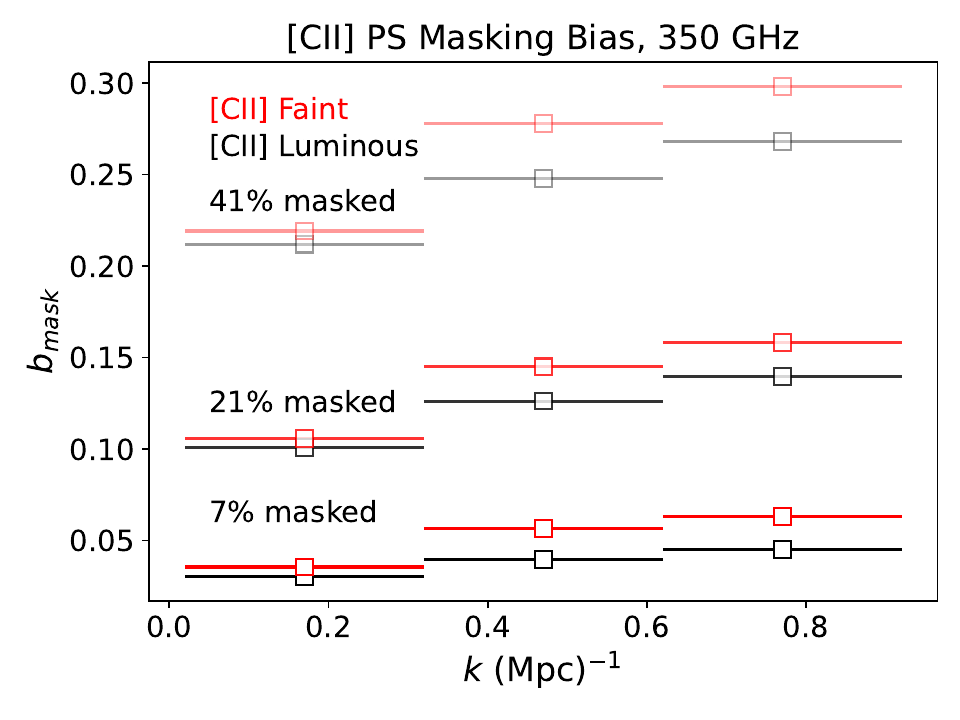}
    \caption{Masking bias representing the relative difference in the PS between masked and unmasked [CII] tomographies for the $350 \pm 20 $ GHz frequency band. Comparisons are made for the most pessimistic ("Faint," red lines) and optimistic ("Luminous," black lines) [CII] PS predictions for the masking depths of 7\%, 21\%, and 41\%.}
    \label{fig:masktransfer}
\end{figure}
In the analysis carried out so far, we have assumed that masking does not affect the PS of the [CII] signal. In this section, we assess the validity of this assumption across different scales. 

By examining the influence of the masking on the [CII] PS we deal with an issue already discussed in the context of the CO residuals (Sect.~\ref{Subsubsection:MaskCO}). Due to the sparsity of the signal, the masking is artificially boosting the [CII] PS. Unlike the case of the CO emission, where this effect becomes prominent only after a certain threshold, boosting of the [CII] PS commences from the onset of the masking procedure as the [CII] emission is invariably not correlated with the mask. This boosting of the [CII] PS can henceforth be seen as a bias introduced by the masking process, that contributes additional uncertainty to our analysis. We defined this masking bias, $b_{\rm mask}$, as the relative difference at a given scale between the PS amplitudes of the masked and unmasked [CII] tomographies, that is, 

\begin{equation}
    b_{\text{mask}} = \left|\frac{P_{\text{masked}} - P_{\text{unmasked}}}{P_{\text{unmasked}}}\right|.
\end{equation}
We evaluated this scale-dependent bias using two extreme models for the [CII] tomography, that is, our model with the highest (hereafter "Luminous") and our model with the lowest [CII] PS signal ("Faint"; Fig \ref{fig:masktransfer}). Then, we evaluated these biases for three masking scenarios, the "light" ($7\%$), "moderate" ($21\%$), and "deep" ($41\%$). In Fig.~\ref{fig:masktransfer}, we present these results for the $350 \pm 20$ GHz tomography. 

For the "light," "moderate," and "deep" scenarios, we find biases of $\approx 3\%$, $\approx 10\%$, and $\approx 22\%$ at $k=0.02-0.32 \ \text{Mpc}^{-1}$ which increase to $\approx 6\%$, $\approx 15\%$, and $\approx 28\%$ at $k=0.62-0.92 \ \text{Mpc}^{-1}$, respectively. The scale dependence of the masking bias depends thus on the level of masking, and it intensifies as the masking becomes more aggressive. At the limiting case of  of a masking depth of $41\%$,  \(b_{\text{mask}}\) remains below \(25\%\) at large scales, but it can escalate to \(30\%\) at small scales. A similar scale-dependent bias is observed by \cite{VanCuyck} in their angular [CII] PS predictions. This scale-dependent bias stems from the disparate impact of bright sources on the [CII] PS amplitude at different scales. On larger scales, the PS is predominantly influenced by the clustering signal of numerous star forming galaxies, which itself is proportional to the first moment of the luminosity function (Paper I). However, as we move to smaller scales, the PS is dominated by shot noise, which arises from the discrete nature of the luminous sources and is proportional to the second moment of the [CII] luminosity function (Paper I). At this scale the PS is thus strongly affected by the few, more luminous sources. The low probability of masking these rare bright sources leads to a significant masking bias due to volume corrections calculated assuming an homogeneous source distribution.

Inverting the masking process to correct for masking bias, that is, deconvolving the masking array with the masked intensity maps, would require uncertain modeling of clustering components in order to extrapolate the [CII] emission of the masked voxels. Furthermore, this approach could introduce artifacts when applied to a map with white noise, further complicating the PS analysis. However, it should be noted that this scale-dependent, and sometimes significant, masking bias does not vary greatly (<5\%) depending on the exact [CII] model used. This suggests that reliable scale-dependent corrections for this masking bias can be efficiently calculated using the specific masking strategies employed and [CII] models that are correct to first order. Of course, forward model fits to the observed [CII] PS will be able to account for this bias by construction.

\subsubsection{Impact of masking on the CO-Contaminated [CII] PS}
\label{Subsubsection:MaskCOCII}

%CII PS uncertainty correct it after referee report
\begin{figure*}
  \centering
  \includegraphics[width=12cm]{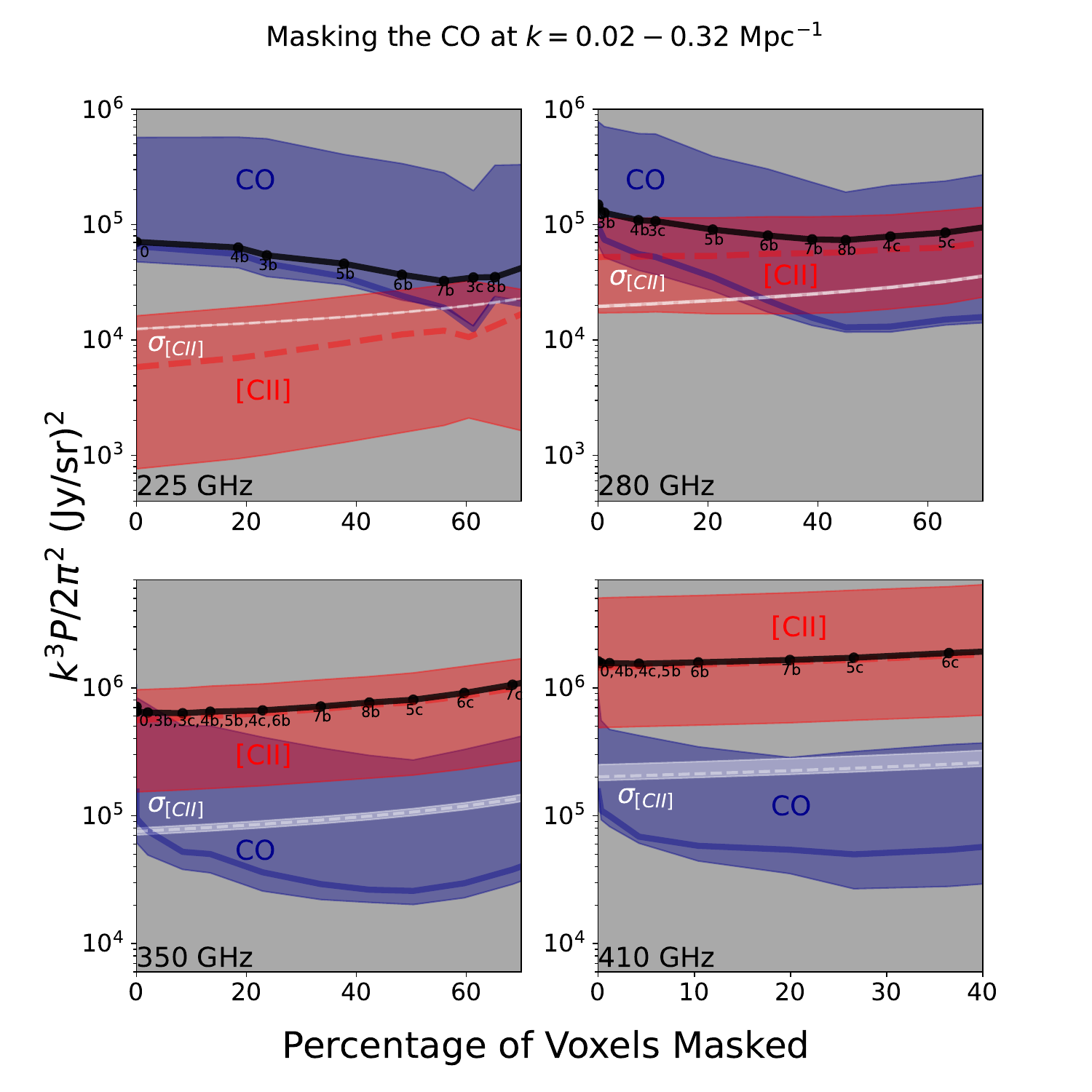}
  \caption{Same plot as Fig. \ref{fig:coimpact} but with the masking bias of [CII] taken into account. We also show the [CII] uncertainty ($\sigma_{ \rm [CII]}$, white line) and the total line (CO+[CII]) PS for the fiducial models (black line)}
  \label{fig:allimpact}
\end{figure*}

We now turn our attention to assessing the impact of masking on the CO-contaminated [CII] PS, $P_{\rm (CO+[CII])}$, specifically within the $k=0.02-0.32 \ \text{Mpc}^{-1}$ range (Fig. \ref{fig:allimpact}). The masks are applied with the optimal sequence inferred using the CO-only tomographies shown in Fig.~\ref{fig:coimpact} (Sect.~\ref{Subsubsection:MaskCO}). Building upon the analysis presented in Sect.~\ref{Subsubsection:MaskCII}, we include the bias introduced by masking on the [CII] signal, which translates into a gradual increase in the [CII] PS as masking depth increases. Finally, as a last piece of information, the uncertainty associated with the [CII] PS for a 2000-hour FYST survey, denoted as $\sigma_{\rm [CII]}$, is plotted in Fig.~\ref{fig:allimpact} as well. This uncertainty, encompassing both instrumental white noise and sample variance, is defined in Paper I by the equation:
\begin{equation} \label{signaltonoise} \sigma _{ \rm [CII]} =\frac{P _ { \mathrm { [CII] } } ( k , z ) + P _ { \mathrm { N } }}{\ \sqrt { N _ {\rm{m} } ( k ) }}. \end{equation}

Assuming that the distribution of masked voxels is uniform, this uncertainty increases with the masking depth proportionally to the decrease in the number of $k$-modes across all $k$-bins. A Monte Carlo algorithm validated this uniformity assumption, establishing a consistent relationship across all scales we examined:

\begin{equation} \label{maskuncertain} \frac{\sigma {\rm [CII]}}{\sigma{\rm [CII],masked}}= \sqrt{\frac{N_{k, \ \rm{not \ masked}}}{ N_k}}=\sqrt{\frac{N_{ \rm{voxels}, \ \rm{not \ masked}}} {N_{\rm{voxels}}}}. \end{equation}

By integrating all the elements presented in Fig.~\ref{fig:allimpact}, we can assess whether the objective of masking has been achieved, that is, to reduce the amplitude of the CO PS below the measurement uncertainty of the [CII] PS ($\sigma_{\rm[CII]}$), while ensuring that this uncertainty remains below the amplitude of the [CII] PS. Two distinct cases are evident in Fig.~\ref{fig:allimpact}. Firstly, in the higher frequency bands ($350 \pm 20$ and $410 \pm 20$ GHz), $P_{\rm(CO+[CII])}$ initially undergoes a slight decrease at masking depths of $<10\%$, attributed to the masking of the rare but bright $J_{\rm up} = 3$ and $J_{\rm up} = 4$ sources, immediately followed by a gradual increase, mainly due to the [CII] PS masking bias (the CO PS masking bias only playing a role at masking depths $>20\%$). As already mentioned, since the [CII] emission is invariably uncorrelated with the mask, such an increase in PS from the onset of the masking procedure is characteristic of the dominance of the [CII] signal in the unmasked tomography. At these high-frequency and at the optimal masking depth of $<10\%$, the [CII] signal dominates and, more importantly, it is well above $\sigma_{\rm CII}$. This guarantees accurate detection of the [CII] PS by the FYST LIM survey in these bands. Secondly, in the lower frequency bands ($225 \pm 20$ and $280 \pm 20$ GHz), $P_{\rm (CO+CII)}$ decreases progressively with masking depth, reaches a minimum at the optimal masking depth of $\approx 40-60 \%$ and finally increases due to a combination of CO and [CII] masking biases. At $280 \pm 20$ GHz, the decrease in $P_{\rm (CO+CII)}$ is less pronounced than at $225 \pm 20$ GHz, and in the former case, the masked tomography effectively switches from CO-dominant to [CII]-dominant, whereas in the latter case, the masked tomography remains CO-dominant. At $280 \pm 20$ GHz, and at the optimal masking depth of 40\%, the [CII] signal dominates and lies above $\sigma_{\rm CII}$. FYST LIM will thus detect, albeit marginally, the [CII] PS in this band (signal-to-noise ratio of two and CO contamination of around 15\%). In contrast, at $225 \pm 20$ GHz and at optimal masking depth of 60\%, the CO signal still dominates. In this band, an alternative masking approach will be required.

\begin{figure*}[ht]
     \centering
     \begin{subfigure}[b]{7cm}
         \centering
         \includegraphics[width=\textwidth]{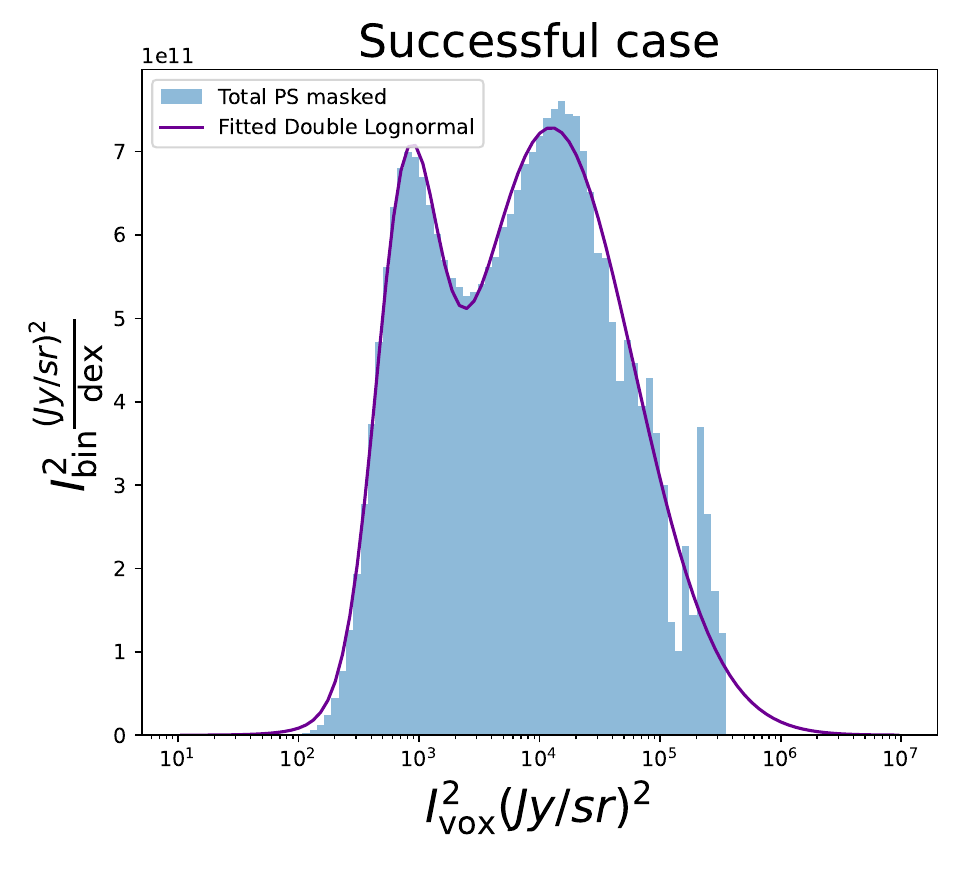}
     \end{subfigure}
     %\hfill
     \begin{subfigure}[b]{7cm}
         \centering
         \includegraphics[width=\textwidth]{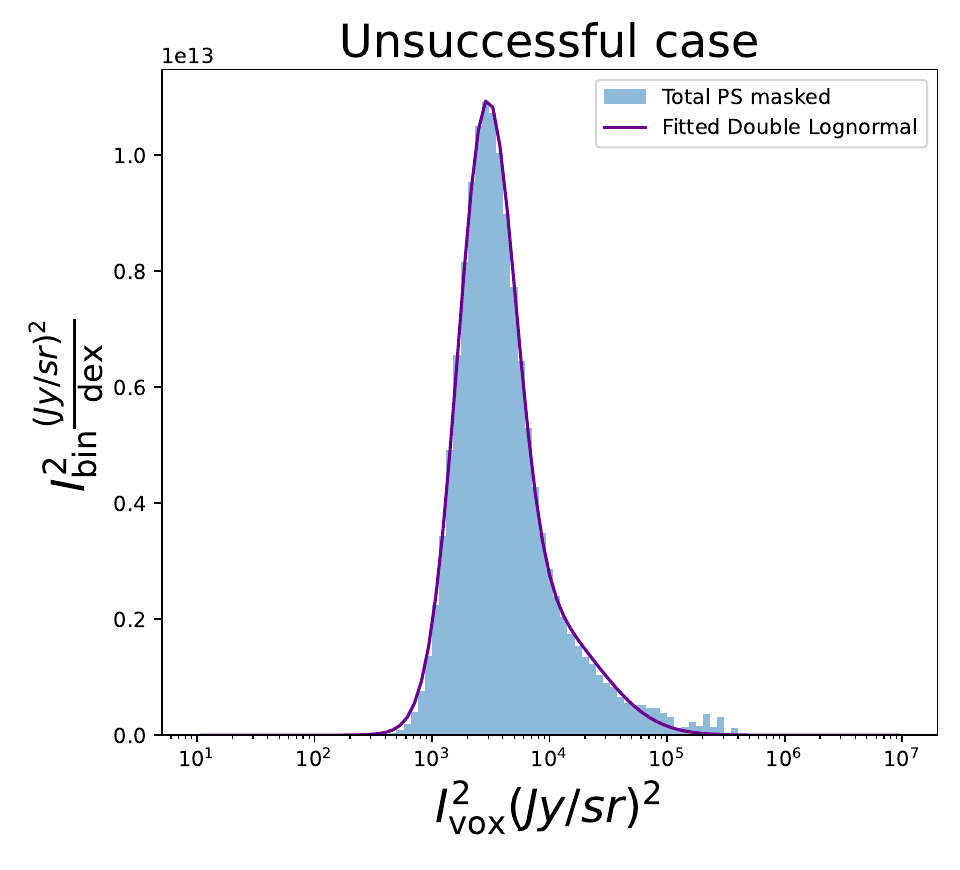}
     \end{subfigure}
    %\hfill
     \begin{subfigure}[b]{7cm}
         \centering
         \includegraphics[width=\textwidth]{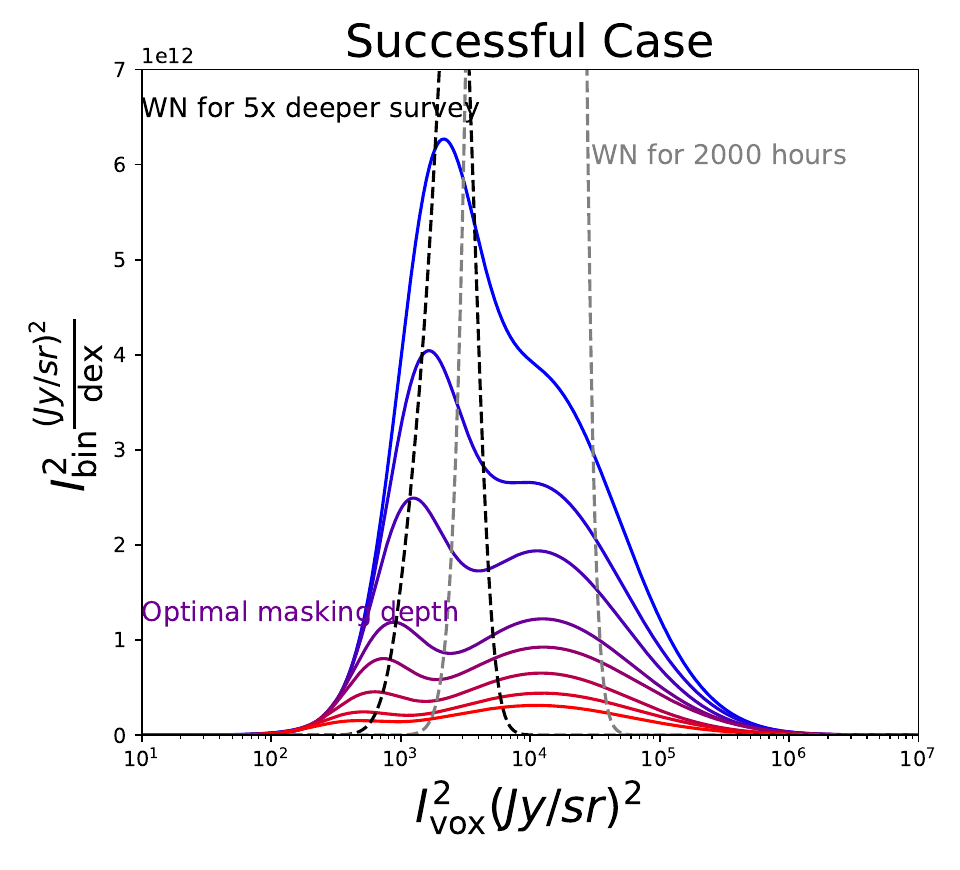}
     \end{subfigure}
     %\hfill
     \begin{subfigure}[b]{7cm}
         \centering
         \includegraphics[width=\textwidth]{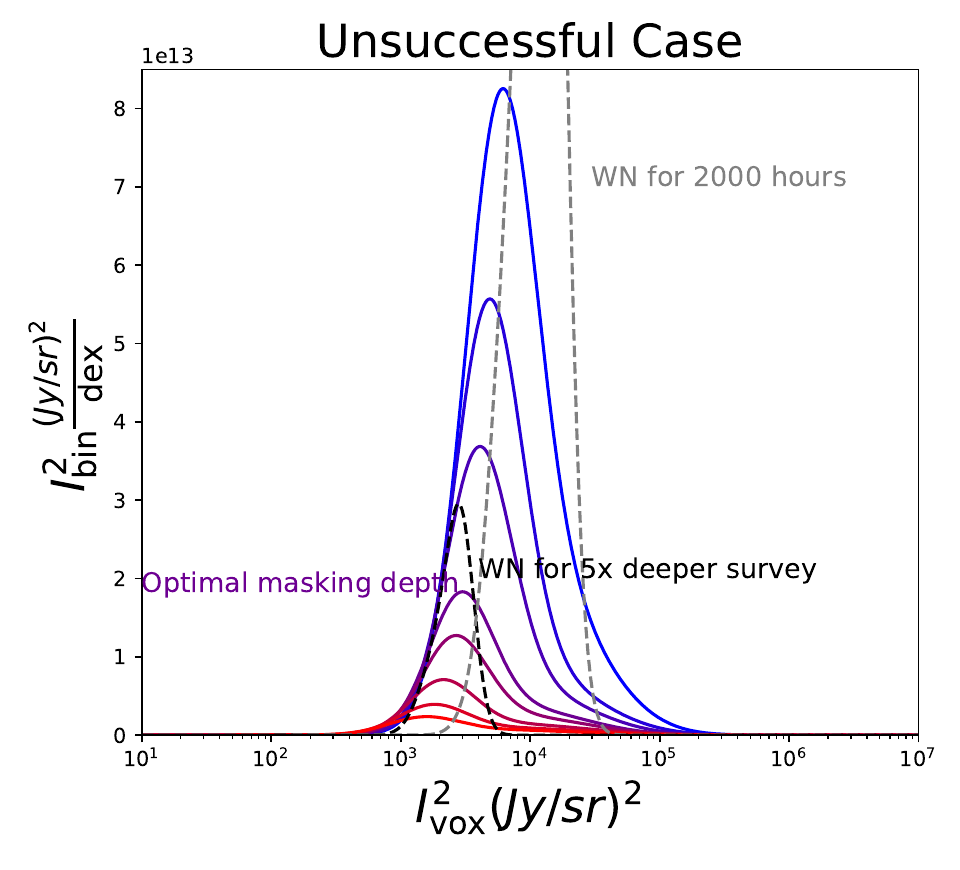}
     \end{subfigure}
 \caption{Pixel distribution histograms weighted with the voxel intensity squared, illustrating the distinction between successful (left; fiducial CO) and unsuccessful (right; pessimistic CO) cases of CO masking of the FYST $280 \pm 20$ GHz tomography. The top panels show the weighted voxel histograms of these $280 \pm 20$ GHz tomography masked to the optimal depth and fitted double lognormal distributions. The lower panels detail the progressive masking depth, each curve representing the lognormal fits of the weighted histogram at increasing depth. This highlights the emergence of a bimodal distribution in the successful case versus a single peak distribution in the unsuccessful case. The lognormal fit to the weighted histogram of a white noise tomography equivalent to the upcoming 2000 hours FYST LIM (black dashed line) and a fit to the white noise tomography of a survey 5 times deeper (gray dashed line) are superimposed.}
\label{fig:weight}
\end{figure*}

From the observational point of view, the results of Fig.~\ref{fig:allimpact} make clear that for a real dataset, the evolution of the PS as function of the masking depth and the location of the optimal masking depth provide key insights into the initial contamination levels of the tomography. A very low value for the optimal masking depth ($\lesssim10 \%$), associated with an amplified PS as we surpass this threshold, unambiguously indicates the predominance of [CII] in the initial tomography. In contrast, a value of the optimal masking depth greater than $\gtrsim10\%$ indicates the dominance by CO foregrounds in the initial tomography. In this case, accurately assessing the extent of CO contamination at the optimal masking depth is challenging in a real dataset as no clear signatures are visible in the PS and as the [CII] and CO emitters share similar spatial frequency characteristics, complicating their differentiation in Fourier space.

Separation between the [CII] and CO emitters becomes clearer when analyzing how these populations contribute to the average total power of the signal. Specifically, this can be achieved by examining their contributions to the zero displacement of the PS from constructing a histogram that represents the distribution of squared voxel intensities, each weighted by its own squared intensity (Fig.~\ref{fig:weight}). It is crucial to note, however, that in order to create this histogram, knowledge of the absolute intensity of the tomography is needed, while this information is partially lost by our atmospheric and astrophysical continuum foregrounds removal. To estimate a posteriori the magnitude of the mean intensity, we used the PS value at large scales, since we demonstrated in Paper I that
$ \ k^{3} P_{\rm clustering} ( k \approx 10^{-2} \ {\rm Mpc}^{-1}) \approx I_{\rm mean} ^2. $

Equipped with this proxy for the mean intensity, we can generate a weighted histogram of the tomography, masked to the optimal depth. When this histogram displays two peaks instead of one, it signifies a transition from a CO-dominated to a [CII]-dominated tomography, achieved by eliminating the brighter CO sources. Conversely, the presence of a single peak signifies that the tomography remains CO-dominated, primarily by the less intense CO sources that were not masked in earlier iterations. Fig. \ref{fig:weight} illustrates these two cases for the $280 \pm 20$ GHz tomography. For the first case we considered our fiducial [CII] and CO PS whereas for the second we combined our fiducial [CII] with our most pessimistic CO PS predictions. For the first case of the fiducial CO PS, as we increase the masking depth, a secondary peak emerges at the brighter end of the histogram, corresponding to the population of [CII] emitters. Notably, at the optimal masking depth, these two peaks have similar heights, marking the transition from CO dominance to [CII] dominance in the PS of this masked tomography. This emergence and predominance of this secondary peak thus signifies the success of the masking process at the optimal masking depth. On the contrary, in the second case, even at the optimal masking depth, the [CII] peak remains hidden in the bright tail of the CO peak. This single-peak histogram, which moves only towards lower voxel intensities as the masking depth increases, signifies the failure of the masking process at the optimal masking depth.

Unfortunately, this rather simple method to distinguish CO-dominated from CII-dominated tomography at the optimal masking depth is strongly affected by the presence of white noise. For example, considering the white noise of the plane 2000 hours of the FYST LIM survey, the peak(s) of the weighted histogram is entirely hidden by it (Fig. \ref{fig:weight}). Nevertheless, with five times deeper observations, the white noise distribution moves enough towards fainter voxels ($I_{\rm vox} ^{2} < 10^{4} \ (\rm{Jy/sr})^{2}$) allowing us to identify the emergence of the [CII]-related secondary peak at $I_{\rm vox} ^{2} > 10^{4} \ (\rm{Jy/sr})^{2}$. This white noise distribution effectively acts as a threshold on the $x$-axis, below which observations become indiscernible. 

In essence, our technique unfolds in two steps. We begin by incrementally masking the LIM tomography until we identify the point where the PS at $k=0.02-0.32 \ {\rm Mpc}^1 $ is minimized, revealing the optimal depth for masking. Following this, we examine the weighted voxel histogram of the tomography masked to this depth. In this histogram --which is calibrated using the PS measurement and weighted by the square of the voxel intensities-- we look for a bimodal distribution as confirmation of successful [CII] signal isolation.

\section{Discussion}

\begin{figure}[ht] % Use the figure* environment for a two-column wide figure
    \centering
    \includegraphics[width=0.45 \textwidth]{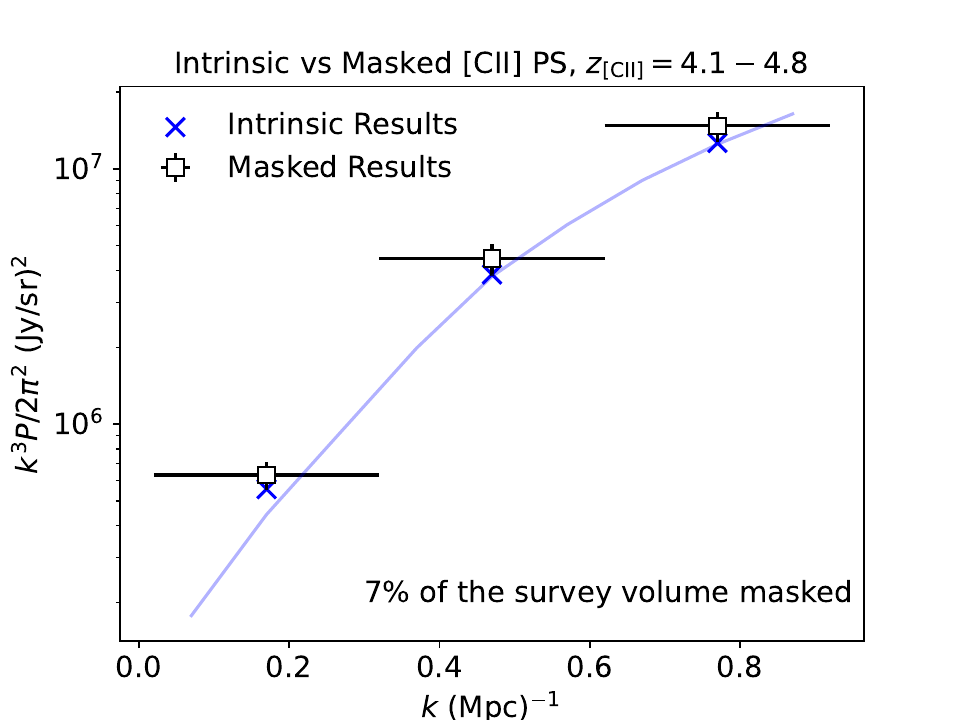}
    \caption{Comparison of the [CII] PS for [CII] observations in the redshift range $z_{\rm [CII]} = 4.1-4.8$. The squares denote the combined CO+[CII] PS after masking CO bright voxels, resulting in a $7\%$ reduction in survey volume. The associated error bars on the squares capture the uncertainty in [CII] measurements due to white instrumental noise and sample variance. The blue crosses depict the intrinsic [CII] PS results, binned at the same $\Delta k$. The continuous line represents the intrinsic [CII] PS results but with a finer binning.}
    \label{fig:masktransfer2}
\end{figure}

This study aimed to quantify the contamination of the \mbox{(post-)EoR} [CII] LIM signal by foreground CO emissions for the upcoming FYST LIM survey, utilizing post-processing of the TNG300 hydrodynamical simulation. The strength of our approach lies in its capacity to generate a range of predictions that account for the inherent uncertainties in CO luminosity functions \citep{Riechers2019, Decarli2020}, and for variations in the CO SLED of galaxies. Based on these predictions, we assessed the effectiveness of a masking technique that leverages an external catalog to identify and subsequently mask bright CO emitters. This led to the identification of a method to determine the optimal masking depth and a novel indicator for successful CO foreground removal. 

Our analysis revealed significant variations in CO contamination across different frequency bands of the EoR-Spec, with notably lower contamination at higher frequencies. This variation significantly impacts the efficiency of our masking technique, which shows greater effectiveness at frequencies where initial contamination is lower. Specifically, we found that the targeted [CII] signal could be detected with minimal masking and thus minimal masking bias at 410 GHz and 350 GHz (Fig. \ref{fig:masktransfer2}). In contrast, at 280 GHz and particularly at 225 GHz, our simple CO masking technique faced challenges due to the high level of foreground CO emissions, underscoring the need for alternative strategies at these frequencies.

Our work aligns with the efforts of \cite{VanCuyck} performed in the framework of the CONCERTO experiment, which similarly aimed to test via simulation the extraction of the [CII] signal from the (post-)EoR universe using a comparable masking strategy. Despite the challenges in making direct comparisons due to differences in angular and spectral resolution, both studies highlight the limitations of masking strategies at high redshifts, especially beyond $z \geq 6.5$. This consensus emphasizes the complexities of masking strategies and their variable effectiveness across redshift. Diverging from \cite{VanCuyck}, we chose to exclude the [CI] line from our simulations due to its highly uncertain intensity values, aligning with methodologies like those suggested by \cite{Yue:Ferrara2019}. The absence of [CI] contamination in our simulations is probably largely compensated by the very wide range of CO models, including very pessimistic ones, taken into account in our study. The precise impact of [CI] lines, along with other foregrounds, is expected to be clarified with the advent of first light science observations.

A notable limitation of our study, also highlighted by \cite{Gkogou2022} and \cite{VanCuyck}, concerns the impact of sample variance, of both the CO and [CII] emitters. Our reliance on the single realization of the TNG300 simulation introduces a potential limitation in terms of the accuracy of our results. Nonetheless, given the wide range of models considered in our simulation, we assume that the influence of sample variance on the effectiveness of our masking technique is relatively minor. Nevertheless, recognizing that sample variance could pose a significant challenge in the context of future forward modeling of the [CII] LIM observations, we will explore the development of multiple realizations of wide observational cones by employing machine learning techniques \citep{Garcia}.

Our decontamination method itself comes with its limitations. Despite its reliability, informed masking is not the most efficient technique, particularly in terms of survey volume loss, at low frequencies. The challenge remains to develop more refined techniques, that de-confuse the line PS without the cost of losing survey volume. There are several mathematically elegant techniques that could achieve this goal. One such approach would be utilizing the wide frequency coverage of the EoR-Spec to measure the CO PS indirectly from the measurements of the cross-correlation of three lines coming from the same galaxies, leveraging the statistical independence of the rest of the line-components. Another approach involves exploiting the anisotropic nature of the 2D PS of interloper lines \citep{Cheng:Chang:Bock:2016}. Upon projection on the frame of reference of the targeted line, the PS of these interloper lines exhibits distinct anisotropies that can be used to differentiate between the target signal and the foreground contamination. Both approaches are promising, yet they share the limitation of failing to recover the valuable Fourier phase information, which is essential for analyzing histograms related to voxel intensity distribution. A third approach that recaptures the Fourier phase information is using conditional generative adversarial networks to de-confuse the signals at the map level rather than at the PS level. \cite{Moriwaki2020} focused on de-confusing two lines (H$\alpha$ and OIII), using a large set of simulated data for training their models. Although this approach is expected to be very effective, it is challenging to estimate the uncertainties involved in the de-confusion method of the measurements. Finally, it is worth noting the existence of a fourth approach that is more robust against the model uncertainty and still recaptures the phase information and that relies on the identification and extraction of foreground sources that are characterized by multiple interloping emission lines. This can be done by fitting multiple SLED templates to each line-of-sight of the tomography using the sparse approximation \citep{Cheng2020}. Given the wide frequency range of the EoR-Spec, we consider this method to be the most promising alternative approach beyond targeted masking. This approach will be tested and evaluated in the context of the FYST LIM survey on the third part of our paper series.

While line foregrounds present significant challenges, the importance of atmospheric and astrophysical continuum foregrounds should not be overlooked. Although \cite{VanCuyck} successfully applied the asymmetric re-weighted penalized least-squares technique to remove contamination from the cosmic infrared background, reinforcing our focus on line interlopers as the primary challenge at low frequencies ($225 \pm 20$ GHz, $280 \pm 20$ GHz), the issue of atmospheric foregrounds remains unresolved and may only be addressed through the use of mock detector data streams. Unfortunately, it is at the high frequencies ($350 \pm 20$ GHz, $410 \pm 20$ GHz), where CO contamination is not problematic that we expect atmospheric contamination to play a critical role. This juxtaposition makes atmospheric interference potentially the primary barrier to detection at the two higher frequency bands of FYST LIM survey.

\section{Conclusion}
\label{sec:Conclusion}

Our study evaluated the level of contamination from CO rotational line emission when observing the (post-)EoR [CII] PS using the upcoming FYST LIM survey and presented a method for mitigating it. We created mock tomographies of the CO line emitters by post-processing the TNG300 simulation, accounting for the inherent uncertainties in the observed CO luminosity functions and the observed CO SLEDs. The resulting range of CO predictions is in good agreement with the ASPECS observations. 

Our analysis revealed that the total mean intensity of CO emissions remains relatively constant, at approximately 1000~Jy/sr, across the observed frequency spectrum. This emission predominantly originates from low-J CO transitions ($J_{\rm up}<8$) in galaxies at lower redshifts ($z<2$). In contrast, the [CII] mean intensity exhibits a significant upward trend, increasing from about 10~Jy/sr at 225 GHz to 2000~Jy/sr at 410 GHz, making the [CII] the dominant emission line at frequencies above 364 GHz.

Based on these predictions, we assessed a foreground removal strategy that involves identifying and masking voxels containing bright CO emitters, utilizing mock external catalogs that exhibit realistic features. A key aspect of our analysis was the determination of mask effectiveness, which we defined as the ability of a mask to reduce the standard deviation of the intensity of the signal, normalized by the number of voxels masked. Through iterative application and reassessment of these masks on CO-only tomography, we were able to refine our masking sequence to achieve optimal foreground removal. The optimal sequence for CO masks found here is broadly applicable to real observations, as we see very little variation into this sequence across the broad range of CO model considered here.

Our analysis also highlighted the frequency dependency of the optimal masking depth, which aligns with the point where the PS of the masked tomography reaches its minimum. While less than 10\% masking depth is required at higher frequencies ($350 \pm 20$ GHz and $410 \pm 20$ GHz), a deeper masking of more than 40\% is necessary at lower frequencies ($280 \pm 20$ GHz and $225 \pm 20$ GHz). 

At these masking depths, according to our fiducial models, [CII] emission dominates and significantly exceeds the typical instrumental white noise levels at higher frequencies, ensuring a robust detection of the [CII] PS at $350 \pm 20$ GHz and $410 \pm 20$ GHz. A tentative detection is also expected at $280 \pm 20$ GHz, where [CII] marginally prevails over CO contamination. However, at $225 \pm 20$ GHz, despite deep masking, CO contamination continues to overshadow [CII] emission, highlighting the urgent need for advanced decontamination techniques.

While estimating the optimal masking depth for decontaminating a real LIM tomography is relatively straightforward, discerning the levels of decontamination achieved at this depth presents a more complex challenge. This complexity arises because CO and [CII] emitters share similar spatial frequency characteristics, making it difficult to distinguish between the two solely based on their distribution in Fourier space. A practical solution involves creating a weighted voxel histogram, where each voxel is weighted by the square of its intensity and the histogram is calibrated using the PS measurement of the tomography masked at the optimal depth. Detailed examination of this histogram for signs of bimodality can reveal a successful transition from a CO-dominated to a [CII]-dominated state. This method will, however, only be applicable to data with very low instrumental white noise, about five times lower than that anticipated for 2000 hours of observation with the EoR-Spec.

Our study lays the groundwork for future LIM studies into the high-redshift universe. The development of decontamination techniques capable of extracting the [CII] high-redshift signal as well as the detailed modeling of the astrophysical and atmospheric foregrounds suitable for forward modeling are the necessary next steps.

\begin{acknowledgements}
We thank Dongwoo Chung, Patrick Breysse, Yoko Okada, Anirban Roy, Steve Choi, Eiichiro Komatsu, Thoma Nikola, Reinhold Schaaf, Ralf Antonius Timmermann, Kaustuv moni Basu, Vyoma Muralidhara, Maude Charmetant, Nicholas Battaglia, Gordon Stacey and other members of the CCAT science working group for detailed discussions about EoR-Spec; We are grateful to Matthieu Bethermin, Athanasia Gkogkou, Mathilde Van Cuyck, Desika Narayanan, Anthony Pullen, Sylvia Adscheid, Elena Marcuzzo, Prachi Khatri, Benedetta Spina, and Cristiano Porciani for the helpful discussions. This research was carried out within the Collaborative Research Center 956 (project ID 184018867), sub-projects A1 and C4 and within the Collaborative Research Center 1601 (project ID 500700252), sub-projects C3 and C6, funded by the Deutsche Forschungsgemeinschaft (DFG). This research made use of NASA’s Astrophysics Data System Bibliographic Services; Matplotlib \citep{Hunter:2007}; Astropy, a community-developed core Python package for astronomy \citep{Astropy}; PlotDigitizer (http://plotdigitizer.sourceforge.net/).
\end{acknowledgements}

\bibliographystyle{aa} %style aa.bst
\bibliography{sorted_unique}

\begin{thebibliography}{101}
\expandafter\ifx\csname natexlab\endcsname\relax\def\natexlab#1{#1}\fi

\bibitem[{{Aihara} {et~al.}(2018){Aihara}, {Armstrong}, {Bickerton}, {Bosch}, {Coupon}, {Furusawa}, {Hayashi}, {Ikeda}, {Kamata}, {Karoji}, {Kawanomoto}, {Koike}, {Komiyama}, {Lang}, {Lupton}, {Mineo}, {Miyatake}, {Miyazaki}, {Morokuma}, {Obuchi}, {Oishi}, {Okura}, {Price}, {Takata}, {Tanaka}, {Tanaka}, {Tanaka}, {Uchida}, {Uraguchi}, {Utsumi}, {Wang}, {Yamada}, {Yamanoi}, {Yasuda}, {Arimoto}, {Chiba}, {Finet}, {Fujimori}, {Fujimoto}, {Furusawa}, {Goto}, {Goulding}, {Gunn}, {Harikane}, {Hattori}, {Hayashi}, {He{\l}miniak}, {Higuchi}, {Hikage}, {Ho}, {Hsieh}, {Huang}, {Huang}, {Imanishi}, {Iwata}, {Jaelani}, {Jian}, {Kashikawa}, {Katayama}, {Kojima}, {Konno}, {Koshida}, {Kusakabe}, {Leauthaud}, {Lee}, {Lin}, {Lin}, {Mandelbaum}, {Matsuoka}, {Medezinski}, {Miyama}, {Momose}, {More}, {More}, {Mukae}, {Murata}, {Murayama}, {Nagao}, {Nakata}, {Niida}, {Niikura}, {Nishizawa}, {Oguri}, {Okabe}, {Ono}, {Onodera}, {Onoue}, {Ouchi}, {Pyo}, {Shibuya}, {Shimasaku}, {Simet}, {Speagle}, {Spergel}, {Strauss}, {Sugahara},
  {Sugiyama}, {Suto}, {Suzuki}, {Tait}, {Takada}, {Terai}, {Toba}, {Turner}, {Uchiyama}, {Umetsu}, {Urata}, {Usuda}, {Yeh}, \& {Yuma}}]{Aihara}
{Aihara}, H., {Armstrong}, R., {Bickerton}, S., {et~al.} 2018, \pasj, 70, S8

\bibitem[{{Aravena} {et~al.}(2016){Aravena}, {Spilker}, {Bethermin}, {Bothwell}, {Chapman}, {de Breuck}, {Furstenau}, {G{\'o}nzalez-L{\'o}pez}, {Greve}, {Litke}, {Ma}, {Malkan}, {Marrone}, {Murphy}, {Stark}, {Strandet}, {Vieira}, {Weiss}, {Welikala}, {Wong}, \& {Collier}}]{Aravena2016}
{Aravena}, M., {Spilker}, J.~S., {Bethermin}, M., {et~al.} 2016, \mnras, 457, 4406

\bibitem[{{Astropy Collaboration} {et~al.}(2013){Astropy Collaboration}, {Robitaille}, {Tollerud}, {Greenfield}, {Droettboom}, {Bray}, {Aldcroft}, {Davis}, {Ginsburg}, {Price-Whelan}, {Kerzendorf}, {Conley}, {Crighton}, {Barbary}, {Muna}, {Ferguson}, {Grollier}, {Parikh}, {Nair}, {Unther}, {Deil}, {Woillez}, {Conseil}, {Kramer}, {Turner}, {Singer}, {Fox}, {Weaver}, {Zabalza}, {Edwards}, {Azalee Bostroem}, {Burke}, {Casey}, {Crawford}, {Dencheva}, {Ely}, {Jenness}, {Labrie}, {Lim}, {Pierfederici}, {Pontzen}, {Ptak}, {Refsdal}, {Servillat}, \& {Streicher}}]{Astropy}
{Astropy Collaboration}, {Robitaille}, T.~P., {Tollerud}, E.~J., {et~al.} 2013, \aap, 558, A33

\bibitem[{{Barkana} \& {Loeb}(2001)}]{Barkana2001}
{Barkana}, R. \& {Loeb}, A. 2001, \physrep, 349, 125

\bibitem[{{Becker} {et~al.}(2015){Becker}, {Bolton}, \& {Lidz}}]{Becker2015}
{Becker}, G.~D., {Bolton}, J.~S., \& {Lidz}, A. 2015, \pasa, 32, e045

\bibitem[{{Becker} {et~al.}(2001){Becker}, {Fan}, {White}, {Strauss}, {Narayanan}, {Lupton}, {Gunn}, {Annis}, {Bahcall}, {Brinkmann}, {Connolly}, {Csabai}, {Czarapata}, {Doi}, {Heckman}, {Hennessy}, {Ivezi{\'c}}, {Knapp}, {Lamb}, {McKay}, {Munn}, {Nash}, {Nichol}, {Pier}, {Richards}, {Schneider}, {Stoughton}, {Szalay}, {Thakar}, \& {York}}]{Becker2001}
{Becker}, R.~H., {Fan}, X., {White}, R.~L., {et~al.} 2001, \aj, 122, 2850

\bibitem[{{Bernal} \& {Kovetz}(2022)}]{Bernal2022}
{Bernal}, J.~L. \& {Kovetz}, E.~D. 2022, \aapr, 30, 5

\bibitem[{{B{\'e}thermin} {et~al.}(2020){B{\'e}thermin}, {Fudamoto}, {Ginolfi}, {Loiacono}, {Khusanova}, {Capak}, {Cassata}, {Faisst}, {Le F{\`e}vre}, {Schaerer}, {Silverman}, {Yan}, {Amorin}, {Bardelli}, {Boquien}, {Cimatti}, {Davidzon}, {Dessauges-Zavadsky}, {Fujimoto}, {Gruppioni}, {Hathi}, {Ibar}, {Jones}, {Koekemoer}, {Lagache}, {Lemaux}, {Moreau}, {Oesch}, {Pozzi}, {Riechers}, {Talia}, {Toft}, {Vallini}, {Vergani}, {Zamorani}, \& {Zucca}}]{BetALP2020}
{B{\'e}thermin}, M., {Fudamoto}, Y., {Ginolfi}, M., {et~al.} 2020, \aap, 643, A2

\bibitem[{{B{\'e}thermin} {et~al.}(2022){B{\'e}thermin}, {Gkogkou}, {Van Cuyck}, {Lagache}, {Beelen}, {Aravena}, {Benoit}, {Bounmy}, {Calvo}, {Catalano}, {de Batz de Trenquelleon}, {De Breuck}, {Fasano}, {Ferrara}, {Goupy}, {Hoarau}, {Horellou}, {Hu}, {Julia}, {Knudsen}, {Lambert}, {Macias-Perez}, {Marpaud}, {Monfardini}, {Pallottini}, {Ponthieu}, {Roehlly}, {Vallini}, {Walter}, \& {Weiss}}]{Bethermin2022}
{B{\'e}thermin}, M., {Gkogkou}, A., {Van Cuyck}, M., {et~al.} 2022, \aap, 667, A156

\bibitem[{{Boogaard} {et~al.}(2023){Boogaard}, {Decarli}, {Walter}, {Wei{\ss}}, {Popping}, {Neri}, {Aravena}, {Riechers}, {Ellis}, {Carilli}, {Cox}, \& {Pety}}]{Boogaard2023}
{Boogaard}, L.~A., {Decarli}, R., {Walter}, F., {et~al.} 2023, \apj, 945, 111

\bibitem[{{Boogaard} {et~al.}(2020){Boogaard}, {van der Werf}, {Weiss}, {Popping}, {Decarli}, {Walter}, {Aravena}, {Bouwens}, {Riechers}, {Gonz{\'a}lez-L{\'o}pez}, {Smail}, {Carilli}, {Kaasinen}, {Daddi}, {Cox}, {D{\'\i}az-Santos}, {Inami}, {Cortes}, \& {Wagg}}]{Boogaard2020}
{Boogaard}, L.~A., {van der Werf}, P., {Weiss}, A., {et~al.} 2020, \apj, 902, 109

\bibitem[{{Bouwens} {et~al.}(2015){Bouwens}, {Illingworth}, {Oesch}, {Trenti}, {Labb{\'e}}, {Bradley}, {Carollo}, {van Dokkum}, {Gonzalez}, {Holwerda}, {Franx}, {Spitler}, {Smit}, \& {Magee}}]{Bouwens:Illingworth:2015}
{Bouwens}, R.~J., {Illingworth}, G.~D., {Oesch}, P.~A., {et~al.} 2015, \apj, 803, 34

\bibitem[{{Bouwens} {et~al.}(2016){Bouwens}, {Smit}, {Labb{\'e}}, {Franx}, {Caruana}, {Oesch}, {Stefanon}, \& {Rasappu}}]{Bouwens2016}
{Bouwens}, R.~J., {Smit}, R., {Labb{\'e}}, I., {et~al.} 2016, \apj, 831, 176

\bibitem[{{Bouwens} {et~al.}(2022){Bouwens}, {Smit}, {Schouws}, {Stefanon}, {Bowler}, {Endsley}, {Gonzalez}, {Inami}, {Stark}, {Oesch}, {Hodge}, {Aravena}, {da Cunha}, {Dayal}, {de Looze}, {Ferrara}, {Fudamoto}, {Graziani}, {Li}, {Nanayakkara}, {Pallottini}, {Schneider}, {Sommovigo}, {Topping}, {van der Werf}, {Algera}, {Barrufet}, {Hygate}, {Labb{\'e}}, {Riechers}, \& {Witstok}}]{Bowens2022b}
{Bouwens}, R.~J., {Smit}, R., {Schouws}, S., {et~al.} 2022, \apj, 931, 160

\bibitem[{{Boylan-Kolchin} {et~al.}(2015){Boylan-Kolchin}, {Weisz}, {Johnson}, {Bullock}, {Conroy}, \& {Fitts}}]{Boylan2015}
{Boylan-Kolchin}, M., {Weisz}, D.~R., {Johnson}, B.~D., {et~al.} 2015, \mnras, 453, 1503

\bibitem[{{Breysse} {et~al.}(2015){Breysse}, {Kovetz}, \& {Kamionkowski}}]{Breysse2015}
{Breysse}, P.~C., {Kovetz}, E.~D., \& {Kamionkowski}, M. 2015, \mnras, 452, 3408

\bibitem[{{Carilli} \& {Walter}(2013)}]{CW13}
{Carilli}, C.~L. \& {Walter}, F. 2013, \araa, 51, 105

\bibitem[{{Cassata} {et~al.}(2020){Cassata}, {Liu}, {Groves}, {Schinnerer}, {Ibar}, {Sargent}, {Karim}, {Talia}, {F{\`e}vre}, {Tasca}, {Lemaux}, {Ribeiro}, {Fiore}, {Romano}, {Mancini}, {Morselli}, {Rodighiero}, {Rodr{\'\i}guez-Mu{\~n}oz}, {Enia}, \& {Smolcic}}]{Cassata2020}
{Cassata}, P., {Liu}, D., {Groves}, B., {et~al.} 2020, \apj, 891, 83

\bibitem[{{CCAT-Prime Collaboration} {et~al.}(2023){CCAT-Prime Collaboration}, {Aravena}, {Austermann}, {Basu}, {Battaglia}, {Beringue}, {Bertoldi}, {Bigiel}, {Bond}, {Breysse}, {Broughton}, {Bustos}, {Chapman}, {Charmetant}, {Choi}, {Chung}, {Clark}, {Cothard}, {Crites}, {Dev}, {Douglas}, {Duell}, {D{\"u}nner}, {Ebina}, {Erler}, {Fich}, {Fissel}, {Foreman}, {Freundt}, {Gallardo}, {Gao}, {Garc{\'\i}a}, {Giovanelli}, {Golec}, {Groppi}, {Haynes}, {Henke}, {Hensley}, {Herter}, {Higgins}, {Hlo{\v{z}}ek}, {Huber}, {Huber}, {Hubmayr}, {Jackson}, {Johnstone}, {Karoumpis}, {Keating}, {Komatsu}, {Li}, {Magnelli}, {Matthews}, {Mauskopf}, {McMahon}, {Meerburg}, {Meyers}, {Muralidhara}, {Murray}, {Niemack}, {Nikola}, {Okada}, {Puddu}, {Riechers}, {Rosolowsky}, {Rossi}, {Rotermund}, z, {Sadavoy}, {Schaaf}, {Schilke}, {Scott}, {Simon}, {Sinclair}, {Sivakoff}, {Stacey}, {Stutz}, {Stutzki}, {Tahani}, {Thanjavur}, {Timmermann}, {Ullom}, {Engelen}, {Vavagiakis}, {Vissers}, {Wheeler}, {White}, {Zhu}, \& {Zou}}]{FYST}
{CCAT-Prime Collaboration}, {Aravena}, M., {Austermann}, J.~E., {et~al.} 2023, \apjs, 264, 7

\bibitem[{{Cheng} {et~al.}(2016){Cheng}, {Chang}, {Bock}, {Bradford}, \& {Cooray}}]{Cheng:Chang:Bock:2016}
{Cheng}, Y.-T., {Chang}, T.-C., {Bock}, J., {Bradford}, C.~M., \& {Cooray}, A. 2016, \apj, 832, 165

\bibitem[{{Cheng} {et~al.}(2020){Cheng}, {Chang}, \& {Bock}}]{Cheng2020}
{Cheng}, Y.-T., {Chang}, T.-C., \& {Bock}, J.~J. 2020, \apj, 901, 142

\bibitem[{{Chung} {et~al.}(2022){Chung}, {Breysse}, {Cleary}, {Ihle}, {Padmanabhan}, {Silva}, {Richard Bond}, {Borowska}, {Catha}, {Church}, {Dunne}, {Kristian Eriksen}, {Kristine Foss}, {Gaier}, {Ott Gundersen}, {Harper}, {Harris}, {Hensley}, {Hobbs}, {Keating}, {Kim}, {Lamb}, {Lawrence}, {Gahr Sturtzel Lunde}, {Murray}, {Pearson}, {Philip}, {Rasmussen}, {Readhead}, {Rennie}, {Stutzer}, {Uzgil}, {Viero}, {Watts}, {Wechsler}, {Kathrine Wehus}, {Woody}, \& {Comap Collaboration}}]{Chung2022}
{Chung}, D.~T., {Breysse}, P.~C., {Cleary}, K.~A., {et~al.} 2022, \apj, 933, 186

\bibitem[{{Clarke} {et~al.}(2024){Clarke}, {Karoumpis}, {Riechers}, {Magnelli}, {Okada}, {Dev}, {Nikola}, \& {Bertoldi}}]{Clarke2024}
{Clarke}, J., {Karoumpis}, C., {Riechers}, D., {et~al.} 2024, \aap, 689, A101

\bibitem[{{CONCERTO Collaboration} {et~al.}(2020){CONCERTO Collaboration}, {Ade}, {Aravena}, {Barria}, {Beelen}, {Benoit}, {B{\'e}thermin}, {Bounmy}, {Bourrion}, {Bres}, {De Breuck}, {Calvo}, {Cao}, {Catalano}, {D{\'e}sert}, {Dur{\'a}n}, {Fasano}, {Fenouillet}, {Garcia}, {Garde}, {Goupy}, {Groppi}, {Hoarau}, {Lagache}, {Lambert}, {Leggeri}, {Levy-Bertrand}, {Mac{\'\i}as-P{\'e}rez}, {Mani}, {Marpaud}, {Mauskopf}, {Monfardini}, {Pisano}, {Ponthieu}, {Prieur}, {Roni}, {Roudier}, {Tourres}, \& {Tucker}}]{CONCERTO2020}
{CONCERTO Collaboration}, {Ade}, P., {Aravena}, M., {et~al.} 2020, \aap, 642, A60

\bibitem[{{Conselice} {et~al.}(2016){Conselice}, {Wilkinson}, {Duncan}, \& {Mortlock}}]{Conselice}
{Conselice}, C.~J., {Wilkinson}, A., {Duncan}, K., \& {Mortlock}, A. 2016, \apj, 830, 83

\bibitem[{{Daddi} {et~al.}(2015){Daddi}, {Dannerbauer}, {Liu}, {Aravena}, {Bournaud}, {Walter}, {Riechers}, {Magdis}, {Sargent}, {B{\'e}thermin}, {Carilli}, {Cibinel}, {Dickinson}, {Elbaz}, {Gao}, {Gobat}, {Hodge}, \& {Krips}}]{Daddi15}
{Daddi}, E., {Dannerbauer}, H., {Liu}, D., {et~al.} 2015, \aap, 577, A46

\bibitem[{{Dav{\'e}} {et~al.}(2020){Dav{\'e}}, {Crain}, {Stevens}, {Narayanan}, {Saintonge}, {Catinella}, \& {Cortese}}]{Dave2020}
{Dav{\'e}}, R., {Crain}, R.~A., {Stevens}, A. R.~H., {et~al.} 2020, \mnras, 497, 146

\bibitem[{{De Looze} {et~al.}(2014){De Looze}, {Cormier}, {Lebouteiller}, {Madden}, {Baes}, {Bendo}, {Boquien}, {Boselli}, {Clements}, {Cortese}, {Cooray}, {Galametz}, {Galliano}, {Graci{\'a}-Carpio}, {Isaak}, {Karczewski}, {Parkin}, {Pellegrini}, {R{\'e}my-Ruyer}, {Spinoglio}, {Smith}, \& {Sturm}}]{DeLooze2014}
{De Looze}, I., {Cormier}, D., {Lebouteiller}, V., {et~al.} 2014, \aap, 568, A62

\bibitem[{{Decarli} {et~al.}(2020){Decarli}, {Aravena}, {Boogaard}, {Carilli}, {Gonz{\'a}lez-L{\'o}pez}, {Walter}, {Cortes}, {Cox}, {da Cunha}, {Daddi}, {D{\'\i}az-Santos}, {Hodge}, {Inami}, {Neeleman}, {Novak}, {Oesch}, {Popping}, {Riechers}, {Smail}, {Uzgil}, {van der Werf}, {Wagg}, \& {Weiss}}]{Decarli2020}
{Decarli}, R., {Aravena}, M., {Boogaard}, L., {et~al.} 2020, \apj, 902, 110

\bibitem[{{Decarli} {et~al.}(2016){Decarli}, {Walter}, {Aravena}, {Carilli}, {Bouwens}, {da Cunha}, {Daddi}, {Elbaz}, {Riechers}, {Smail}, {Swinbank}, {Weiss}, {Bacon}, {Bauer}, {Bell}, {Bertoldi}, {Chapman}, {Colina}, {Cortes}, {Cox}, {G{\'o}nzalez-L{\'o}pez}, {Inami}, {Ivison}, {Hodge}, {Karim}, {Magnelli}, {Ota}, {Popping}, {Rix}, {Sargent}, {van der Wel}, \& {van der Werf}}]{Decarli2016}
{Decarli}, R., {Walter}, F., {Aravena}, M., {et~al.} 2016, \apj, 833, 70

\bibitem[{{den Brok} {et~al.}(2021){den Brok}, {Chatzigiannakis}, {Bigiel}, {Puschnig}, {Barnes}, {Leroy}, {Jim{\'e}nez-Donaire}, {Usero}, {Schinnerer}, {Rosolowsky}, {Faesi}, {Grasha}, {Hughes}, {Kruijssen}, {Liu}, {Neumann}, {Pety}, {Querejeta}, {Saito}, {Schruba}, \& {Stuber}}]{denBrok}
{den Brok}, J.~S., {Chatzigiannakis}, D., {Bigiel}, F., {et~al.} 2021, \mnras, 504, 3221

\bibitem[{{Downes} \& {Solomon}(1998)}]{Downes1998}
{Downes}, D. \& {Solomon}, P.~M. 1998, \apj, 507, 615

\bibitem[{{Dragone}(1978)}]{Dragone1978}
{Dragone}, C. 1978, AT T Technical Journal, 57, 2663

\bibitem[{{Elbaz} {et~al.}(2007){Elbaz}, {Daddi}, {Le Borgne}, {Dickinson}, {Alexander}, {Chary}, {Starck}, {Brandt}, {Kitzbichler}, {MacDonald}, {Nonino}, {Popesso}, {Stern}, \& {Vanzella}}]{Elbaz07}
{Elbaz}, D., {Daddi}, E., {Le Borgne}, D., {et~al.} 2007, \aap, 468, 33

\bibitem[{{Euclid Collaboration} {et~al.}(2020){Euclid Collaboration}, {Desprez}, {Paltani}, {Coupon}, {Almosallam}, {Alvarez-Ayllon}, {Amaro}, {Brescia}, {Brodwin}, {Cavuoti}, {De Vicente-Albendea}, {Fotopoulou}, {Hatfield}, {Hartley}, {Ilbert}, {Jarvis}, {Longo}, {Rau}, {Saha}, {Speagle}, {Tramacere}, {Castellano}, {Dubath}, {Galametz}, {Kuemmel}, {Laigle}, {Merlin}, {Mohr}, {Pilo}, {Salvato}, {Andreon}, {Auricchio}, {Baccigalupi}, {Balaguera-Antol{\'\i}nez}, {Baldi}, {Bardelli}, {Bender}, {Biviano}, {Bodendorf}, {Bonino}, {Bozzo}, {Branchini}, {Brinchmann}, {Burigana}, {Cabanac}, {Camera}, {Capobianco}, {Cappi}, {Carbone}, {Carretero}, {Carvalho}, {Casas}, {Casas}, {Castander}, {Castignani}, {Cimatti}, {Cledassou}, {Colodro-Conde}, {Congedo}, {Conselice}, {Conversi}, {Copin}, {Corcione}, {Courtois}, {Cuby}, {Da Silva}, {de la Torre}, {Degaudenzi}, {Di Ferdinando}, {Douspis}, {Duncan}, {Dupac}, {Ealet}, {Fabbian}, {Fabricius}, {Farrens}, {Ferreira}, {Finelli}, {Fosalba}, {Fourmanoit}, {Frailis},
  {Franceschi}, {Fumana}, {Galeotta}, {Garilli}, {Gillard}, {Gillis}, {Giocoli}, {Gozaliasl}, {Graci{\'a}-Carpio}, {Grupp}, {Guzzo}, {Hailey}, {Haugan}, {Holmes}, {Hormuth}, {Humphrey}, {Jahnke}, {Keihanen}, {Kermiche}, {Kilbinger}, {Kirkpatrick}, {Kitching}, {Kohley}, {Kubik}, {Kunz}, {Kurki-Suonio}, {Ligori}, {Lilje}, {Lloro}, {Maino}, {Maiorano}, {Marggraf}, {Markovic}, {Martinet}, {Marulli}, {Massey}, {Maturi}, {Mauri}, {Maurogordato}, {Medinaceli}, {Mei}, {Meneghetti}, {Metcalf}, {Meylan}, {Moresco}, {Moscardini}, {Munari}, {Niemi}, {Padilla}, {Pasian}, {Patrizii}, {Pettorino}, {Pires}, {Polenta}, {Poncet}, {Popa}, {Potter}, {Pozzetti}, {Raison}, {Renzi}, {Rhodes}, {Riccio}, {Rossetti}, {Saglia}, {Sapone}, {Schneider}, {Scottez}, {Secroun}, {Serrano}, {Sirignano}, {Sirri}, {Stanco}, {Stern}, {Sureau}, {Tallada Cresp{\'\i}}, {Tavagnacco}, {Taylor}, {Tenti}, {Tereno}, {Toledo-Moreo}, {Torradeflot}, {Valenziano}, {Valiviita}, {Vassallo}, {Viel}, {Wang}, {Welikala}, {Whittaker}, {Zacchei}, {Zamorani},
  {Zoubian}, \& {Zucca}}]{Euclid2020}
{Euclid Collaboration}, {Desprez}, G., {Paltani}, S., {et~al.} 2020, \aap, 644, A31

\bibitem[{{Euclid Collaboration} {et~al.}(2021){Euclid Collaboration}, {Ilbert}, {de la Torre}, {Martinet}, {Wright}, {Paltani}, {Laigle}, {Davidzon}, {Jullo}, {Hildebrandt}, {Masters}, {Amara}, {Conselice}, {Andreon}, {Auricchio}, {Azzollini}, {Baccigalupi}, {Balaguera-Antol{\'\i}nez}, {Baldi}, {Balestra}, {Bardelli}, {Bender}, {Biviano}, {Bodendorf}, {Bonino}, {Borgani}, {Boucaud}, {Bozzo}, {Branchini}, {Brescia}, {Burigana}, {Cabanac}, {Camera}, {Capobianco}, {Cappi}, {Carbone}, {Carretero}, {Carvalho}, {Casas}, {Castander}, {Castellano}, {Castignani}, {Cavuoti}, {Cimatti}, {Cledassou}, {Colodro-Conde}, {Congedo}, {Conversi}, {Copin}, {Corcione}, {Costille}, {Coupon}, {Courtois}, {Cropper}, {Cuby}, {Da Silva}, {Degaudenzi}, {Di Ferdinando}, {Dubath}, {Duncan}, {Dupac}, {Dusini}, {Ealet}, {Fabricius}, {Farrens}, {Ferreira}, {Finelli}, {Fosalba}, {Fotopoulou}, {Franceschi}, {Franzetti}, {Galeotta}, {Garilli}, {Gillard}, {Gillis}, {Giocoli}, {Gozaliasl}, {Graci{\'a}-Carpio}, {Grupp}, {Guzzo}, {Haugan},
  {Holmes}, {Hormuth}, {Jahnke}, {Keihanen}, {Kermiche}, {Kiessling}, {Kirkpatrick}, {Kunz}, {Kurki-Suonio}, {Ligori}, {Lilje}, {Lloro}, {Maino}, {Maiorano}, {Marggraf}, {Markovic}, {Marulli}, {Massey}, {Maturi}, {Mauri}, {Maurogordato}, {McCracken}, {Medinaceli}, {Mei}, {Metcalf}, {Moresco}, {Morin}, {Moscardini}, {Munari}, {Nakajima}, {Neissner}, {Niemi}, {Nightingale}, {Padilla}, {Pasian}, {Patrizii}, {Pedersen}, {Pello}, {Pettorino}, {Pires}, {Polenta}, {Poncet}, {Popa}, {Potter}, {Pozzetti}, {Raison}, {Renzi}, {Rhodes}, {Riccio}, {Romelli}, {Roncarelli}, {Rossetti}, {Saglia}, {S{\'a}nchez}, {Sapone}, {Schneider}, {Schrabback}, {Scottez}, {Secroun}, {Seidel}, {Serrano}, {Sirignano}, {Sirri}, {Stanco}, {Sureau}, {Tallada Cresp{\'a}}, {Tenti}, {Teplitz}, {Tereno}, {Toledo-Moreo}, {Torradeflot}, {Tramacere}, {Valentijn}, {Valenziano}, {Valiviita}, {Vassallo}, {Wang}, {Welikala}, {Weller}, {Whittaker}, {Zacchei}, {Zamorani}, {Zoubian}, \& {Zucca}}]{Ilbert2021}
{Euclid Collaboration}, {Ilbert}, O., {de la Torre}, S., {et~al.} 2021, \aap, 647, A117

\bibitem[{{Euclid Collaboration} {et~al.}(2022){Euclid Collaboration}, {Moneti}, {McCracken}, {Shuntov}, {Kauffmann}, {Capak}, {Davidzon}, {Ilbert}, {Scarlata}, {Toft}, {Weaver}, {Chary}, {Cuby}, {Faisst}, {Masters}, {McPartland}, {Mobasher}, {Sanders}, {Scaramella}, {Stern}, {Szapudi}, {Teplitz}, {Zalesky}, {Amara}, {Auricchio}, {Bodendorf}, {Bonino}, {Branchini}, {Brau-Nogue}, {Brescia}, {Brinchmann}, {Capobianco}, {Carbone}, {Carretero}, {Castander}, {Castellano}, {Cavuoti}, {Cimatti}, {Cledassou}, {Congedo}, {Conselice}, {Conversi}, {Copin}, {Corcione}, {Costille}, {Cropper}, {Da Silva}, {Degaudenzi}, {Douspis}, {Dubath}, {Duncan}, {Dupac}, {Dusini}, {Farrens}, {Ferriol}, {Fosalba}, {Frailis}, {Franceschi}, {Fumana}, {Garilli}, {Gillis}, {Giocoli}, {Granett}, {Grazian}, {Grupp}, {Haugan}, {Hoekstra}, {Holmes}, {Hormuth}, {Hudelot}, {Jahnke}, {Kermiche}, {Kiessling}, {Kilbinger}, {Kitching}, {Kohley}, {K{\"u}mmel}, {Kunz}, {Kurki-Suonio}, {Ligori}, {Lilje}, {Lloro}, {Maiorano}, {Mansutti}, {Marggraf},
  {Markovic}, {Marulli}, {Massey}, {Maurogordato}, {Meneghetti}, {Merlin}, {Meylan}, {Moresco}, {Moscardini}, {Munari}, {Niemi}, {Padilla}, {Paltani}, {Pasian}, {Pedersen}, {Pires}, {Poncet}, {Popa}, {Pozzetti}, {Raison}, {Rebolo}, {Rhodes}, {Rix}, {Roncarelli}, {Rossetti}, {Saglia}, {Schneider}, {Secroun}, {Seidel}, {Serrano}, {Sirignano}, {Sirri}, {Stanco}, {Tallada-Cresp{\'\i}}, {Taylor}, {Tereno}, {Toledo-Moreo}, {Torradeflot}, {Wang}, {Welikala}, {Weller}, {Zamorani}, {Zoubian}, {Andreon}, {Bardelli}, {Camera}, {Graci{\'a}-Carpio}, {Medinaceli}, {Mei}, {Polenta}, {Romelli}, {Sureau}, {Tenti}, {Vassallo}, {Zacchei}, {Zucca}, {Baccigalupi}, {Balaguera-Antol{\'\i}nez}, {Bernardeau}, {Biviano}, {Bolzonella}, {Bozzo}, {Burigana}, {Cabanac}, {Cappi}, {Carvalho}, {Casas}, {Castignani}, {Colodro-Conde}, {Coupon}, {Courtois}, {Di Ferdinando}, {Farina}, {Finelli}, {Flose-Reimberg}, {Fotopoulou}, {Galeotta}, {Ganga}, {Garcia-Bellido}, {Gaztanaga}, {Gozaliasl}, {Hook}, {Joachimi}, {Kansal}, {Keihanen},
  {Kirkpatrick}, {Lindholm}, {Mainetti}, {Maino}, {Maoli}, {Martinelli}, {Martinet}, {Maturi}, {Metcalf}, {Morgante}, {Morisset}, {Nucita}, {Patrizii}, {Potter}, {Renzi}, {Riccio}, {S{\'a}nchez}, {Sapone}, {Schirmer}, {Schultheis}, {Scottez}, {Sefusatti}, {Teyssier}, {Tubio}, {Tutusaus}, {Valiviita}, {Viel}, \& {Hildebrandt}}]{Euclid2022}
{Euclid Collaboration}, {Moneti}, A., {McCracken}, H.~J., {et~al.} 2022, \aap, 658, A126

\bibitem[{{Fan} {et~al.}(2006){Fan}, {Strauss}, {Richards}, {Hennawi}, {Becker}, {White}, {Diamond-Stanic}, {Donley}, {Jiang}, {Kim}, {Vestergaard}, {Young}, {Gunn}, {Lupton}, {Knapp}, {Schneider}, {Brandt}, {Bahcall}, {Barentine}, {Brinkmann}, {Brewington}, {Fukugita}, {Harvanek}, {Kleinman}, {Krzesinski}, {Long}, {Neilsen}, {Nitta}, {Snedden}, \& {Voges}}]{Fan2006}
{Fan}, X., {Strauss}, M.~A., {Richards}, G.~T., {et~al.} 2006, \aj, 131, 1203

\bibitem[{{Garcia} {et~al.}(2023){Garcia}, {Narayanan}, {Popping}, {Anirudh}, {Sutherland}, \& {Kaasinen}}]{Garcia}
{Garcia}, K., {Narayanan}, D., {Popping}, G., {et~al.} 2023, arXiv e-prints; \apj, in press, arXiv:2311.01508

\bibitem[{{Gkogkou} {et~al.}(2023){Gkogkou}, {B{\'e}thermin}, {Lagache}, {Van Cuyck}, {Jullo}, {Aravena}, {Beelen}, {Benoit}, {Bounmy}, {Calvo}, {Catalano}, {Cora}, {Croton}, {de la Torre}, {Fasano}, {Ferrara}, {Goupy}, {Hoarau}, {Hu}, {Ishiyama}, {Knudsen}, {Lambert}, {Mac{\'\i}as-P{\'e}rez}, {Marpaud}, {Mellema}, {Monfardini}, {Pallottini}, {Ponthieu}, {Prada}, {Roehlly}, {Vallini}, \& {Walter}}]{Gkogou2022}
{Gkogkou}, A., {B{\'e}thermin}, M., {Lagache}, G., {et~al.} 2023, \aap, 670, A16

\bibitem[{{Harikane} {et~al.}(2022){Harikane}, {Inoue}, {Mawatari}, {Hashimoto}, {Yamanaka}, {Fudamoto}, {Matsuo}, {Tamura}, {Dayal}, {Yung}, {Hutter}, {Pacucci}, {Sugahara}, \& {Koekemoer}}]{Harikane2022}
{Harikane}, Y., {Inoue}, A.~K., {Mawatari}, K., {et~al.} 2022, \apj, 929, 1

\bibitem[{{Huber} {et~al.}(2022){Huber}, {Choi}, {Duell}, {Freundt}, {Gallardo}, {Keller}, {Li}, {Lin}, {Niemack}, {Nikola}, {Reichers}, {Stacey}, {Vavagiakis}, \& {Zou}}]{Huber22}
{Huber}, Z.~B., {Choi}, S.~K., {Duell}, C.~J., {et~al.} 2022, in Millimeter, Submillimeter, and Far-Infrared Detectors and Instrumentation for Astronomy XI, Vol. 12190, 121902H

\bibitem[{Hunter(2007)}]{Hunter:2007}
Hunter, J.~D. 2007, Computing in Science \& Engineering, 9, 90

\bibitem[{{Kaasinen} {et~al.}(2019){Kaasinen}, {Scoville}, {Walter}, {Da Cunha}, {Popping}, {Pavesi}, {Darvish}, {Casey}, {Riechers}, \& {Glover}}]{Kaasinen2019}
{Kaasinen}, M., {Scoville}, N., {Walter}, F., {et~al.} 2019, \apj, 880, 15

\bibitem[{{Kamenetzky} {et~al.}(2017){Kamenetzky}, {Rangwala}, \& {Glenn}}]{Kamenetzky2017}
{Kamenetzky}, J., {Rangwala}, N., \& {Glenn}, J. 2017, \mnras, 471, 2917

\bibitem[{{Karoumpis} {et~al.}(2022){Karoumpis}, {Magnelli}, {Romano-D{\'\i}az}, {Haslbauer}, \& {Bertoldi}}]{Karoumpis}
{Karoumpis}, C., {Magnelli}, B., {Romano-D{\'\i}az}, E., {Haslbauer}, M., \& {Bertoldi}, F. 2022, \aap, 659, A12

\bibitem[{{Kennicutt}(1998)}]{Kennicutt98}
{Kennicutt}, Robert~C., J. 1998, \araa, 36, 189

\bibitem[{{Kim}(2011)}]{Kim2011}
{Kim}, J. 2011, \aap, 531, A32

\bibitem[{{Kogut} {et~al.}(2015){Kogut}, {Dwek}, \& {Moseley}}]{Kogut2015}
{Kogut}, A., {Dwek}, E., \& {Moseley}, S.~H. 2015, \apj, 806, 234

\bibitem[{{Kovetz} {et~al.}(2019){Kovetz}, {Breysse}, {Lidz}, {Bock}, {Bradford}, {Chang}, {Foreman}, {Padmanabhan}, {Pullen}, {Riechers}, {Silva}, \& {Switzer}}]{Kovetz2019}
{Kovetz}, E., {Breysse}, P.~C., {Lidz}, A., {et~al.} 2019, \baas, 51, 101

\bibitem[{{Lagache} {et~al.}(2018){Lagache}, {Cousin}, \& {Chatzikos}}]{Lagache2018}
{Lagache}, G., {Cousin}, M., \& {Chatzikos}, M. 2018, \aap, 609, A130

\bibitem[{{Laigle} {et~al.}(2016){Laigle}, {McCracken}, {Ilbert}, {Hsieh}, {Davidzon}, {Capak}, {Hasinger}, {Silverman}, {Pichon}, {Coupon}, {Aussel}, {Le Borgne}, {Caputi}, {Cassata}, {Chang}, {Civano}, {Dunlop}, {Fynbo}, {Kartaltepe}, {Koekemoer}, {Le F{\`e}vre}, {Le Floc'h}, {Leauthaud}, {Lilly}, {Lin}, {Marchesi}, {Milvang-Jensen}, {Salvato}, {Sanders}, {Scoville}, {Smolcic}, {Stockmann}, {Taniguchi}, {Tasca}, {Toft}, {Vaccari}, \& {Zabl}}]{Laigle2016}
{Laigle}, C., {McCracken}, H.~J., {Ilbert}, O., {et~al.} 2016, \apjs, 224, 24

\bibitem[{{Le F{\`e}vre} {et~al.}(2020){Le F{\`e}vre}, {B{\'e}thermin}, {Faisst}, {Jones}, {Capak}, {Cassata}, {Silverman}, {Schaerer}, {Yan}, {Amorin}, {Bardelli}, {Boquien}, {Cimatti}, {Dessauges-Zavadsky}, {Giavalisco}, {Hathi}, {Fudamoto}, {Fujimoto}, {Ginolfi}, {Gruppioni}, {Hemmati}, {Ibar}, {Koekemoer}, {Khusanova}, {Lagache}, {Lemaux}, {Loiacono}, {Maiolino}, {Mancini}, {Narayanan}, {Morselli}, {M{\'e}ndez-Hern{\`a}ndez}, {Oesch}, {Pozzi}, {Romano}, {Riechers}, {Scoville}, {Talia}, {Tasca}, {Thomas}, {Toft}, {Vallini}, {Vergani}, {Walter}, {Zamorani}, \& {Zucca}}]{LeFevreALPINE}
{Le F{\`e}vre}, O., {B{\'e}thermin}, M., {Faisst}, A., {et~al.} 2020, \aap, 643, A1

\bibitem[{{Lehmer} {et~al.}(2005){Lehmer}, {Brandt}, {Alexander}, {Bauer}, {Schneider}, {Tozzi}, {Bergeron}, {Garmire}, {Giacconi}, {Gilli}, {Hasinger}, {Hornschemeier}, {Koekemoer}, {Mainieri}, {Miyaji}, {Nonino}, {Rosati}, {Silverman}, {Szokoly}, \& {Vignali}}]{Lehmer2005}
{Lehmer}, B.~D., {Brandt}, W.~N., {Alexander}, D.~M., {et~al.} 2005, \apjs, 161, 21

\bibitem[{{Lenki{\'c}} {et~al.}(2020){Lenki{\'c}}, {Bolatto}, {F{\"o}rster Schreiber}, {Tacconi}, {Neri}, {Combes}, {Walter}, {Garc{\'\i}a-Burillo}, {Genzel}, {Lutz}, \& {Cooper}}]{Lenkic2020}
{Lenki{\'c}}, L., {Bolatto}, A.~D., {F{\"o}rster Schreiber}, N.~M., {et~al.} 2020, \aj, 159, 190

\bibitem[{{Madau} \& {Dickinson}(2014)}]{Madau2014}
{Madau}, P. \& {Dickinson}, M. 2014, \araa, 52, 415

\bibitem[{{Magdis} {et~al.}(2021){Magdis}, {Gobat}, {Valentino}, {Daddi}, {Zanella}, {Kokorev}, {Toft}, {Jin}, \& {Whitaker}}]{Magdis2021}
{Magdis}, G.~E., {Gobat}, R., {Valentino}, F., {et~al.} 2021, \aap, 647, A33

\bibitem[{{Magnelli} {et~al.}(2020){Magnelli}, {Boogaard}, {Decarli}, {G{\'o}nzalez-L{\'o}pez}, {Novak}, {Popping}, {Smail}, {Walter}, {Aravena}, {Assef}, {Bauer}, {Bertoldi}, {Carilli}, {Cortes}, {Cunha}, {Daddi}, {D{\'\i}az-Santos}, {Inami}, {Ivison}, {F{\`e}vre}, {Oesch}, {Riechers}, {Rix}, {Sargent}, {Werf}, {Wagg}, \& {Weiss}}]{Magnelli2020}
{Magnelli}, B., {Boogaard}, L., {Decarli}, R., {et~al.} 2020, \apj, 892, 66

\bibitem[{{Mashian} {et~al.}(2015){Mashian}, {Sturm}, {Sternberg}, {Janssen}, {Hailey-Dunsheath}, {Fischer}, {Contursi}, {Gonz{\'a}lez-Alfonso}, {Graci{\'a}-Carpio}, {Poglitsch}, {Veilleux}, {Davies}, {Genzel}, {Lutz}, {Tacconi}, {Verma}, {Wei{\ss}}, {Polisensky}, \& {Nikola}}]{Mashian15}
{Mashian}, N., {Sturm}, E., {Sternberg}, A., {et~al.} 2015, \apj, 802, 81

\bibitem[{{Moriwaki} {et~al.}(2020){Moriwaki}, {Filippova}, {Shirasaki}, \& {Yoshida}}]{Moriwaki2020}
{Moriwaki}, K., {Filippova}, N., {Shirasaki}, M., \& {Yoshida}, N. 2020, \mnras, 496, L54

\bibitem[{{Moriwaki} \& {Yoshida}(2021)}]{Moriwaki2021}
{Moriwaki}, K. \& {Yoshida}, N. 2021, \apjl, 923, L7

\bibitem[{{Narayanan} \& {Krumholz}(2014)}]{Narayanan2014}
{Narayanan}, D. \& {Krumholz}, M.~R. 2014, \mnras, 442, 1411

\bibitem[{{Parshley} {et~al.}(2018){Parshley}, {Niemack}, {Hills}, {Dicker}, {D{\"u}nner}, {Erler}, {Gallardo}, {Gudmundsson}, {Herter}, {Koopman}, {Limon}, {Matsuda}, {Mauskopf}, {Riechers}, {Stacey}, \& {Vavagiakis}}]{Parshley2018}
{Parshley}, S.~C., {Niemack}, M., {Hills}, R., {et~al.} 2018, in Ground-based and Airborne Telescopes VII, Vol. 10700, 1070041

\bibitem[{{Peebles}(1968)}]{Peebles1968}
{Peebles}, P.~J.~E. 1968, \apj, 153, 1

\bibitem[{{Perot} \& {Fabry}(1899)}]{FP1899}
{Perot}, A. \& {Fabry}, C. 1899, \apj, 9, 87

\bibitem[{{Pillepich} {et~al.}(2019){Pillepich}, {Nelson}, {Springel}, {Pakmor}, {Torrey}, {Weinberger}, {Vogelsberger}, {Marinacci}, {Genel}, {van der Wel}, \& {Hernquist}}]{Pille19}
{Pillepich}, A., {Nelson}, D., {Springel}, V., {et~al.} 2019, \mnras, 490, 3196

\bibitem[{{Pillepich} {et~al.}(2018){Pillepich}, {Springel}, {Nelson}, {Genel}, {Naiman}, {Pakmor}, {Hernquist}, {Torrey}, {Vogelsberger}, {Weinberger}, \& {Marinacci}}]{Pillepich:Springel:Nelson2017}
{Pillepich}, A., {Springel}, V., {Nelson}, D., {et~al.} 2018, \mnras, 473, 4077

\bibitem[{{Ponthieu} {et~al.}(2011){Ponthieu}, {Grain}, \& {Lagache}}]{Ponthieu:Grain2011}
{Ponthieu}, N., {Grain}, J., \& {Lagache}, G. 2011, \aap, 535, A90

\bibitem[{{Popping} {et~al.}(2019){Popping}, {Pillepich}, {Somerville}, {Decarli}, {Walter}, {Aravena}, {Carilli}, {Cox}, {Nelson}, {Riechers}, {Weiss}, {Boogaard}, {Bouwens}, {Contini}, {Cortes}, {da Cunha}, {Daddi}, {D{\'\i}az-Santos}, {Diemer}, {Gonz{\'a}lez-L{\'o}pez}, {Hernquist}, {Ivison}, {Le F{\`e}vre}, {Marinacci}, {Rix}, {Swinbank}, {Vogelsberger}, {van der Werf}, {Wagg}, \& {Yung}}]{Popping2019}
{Popping}, G., {Pillepich}, A., {Somerville}, R.~S., {et~al.} 2019, \apj, 882, 137

\bibitem[{{Pullen} {et~al.}(2023){Pullen}, {Breysse}, {Oxholm}, {Switzer}, {Anderson}, {Barrentine}, {Bolatto}, {Cataldo}, {Essinger-Hileman}, {Maniyar}, {Stevenson}, {Somerville}, {Volpert}, {Wollack}, {Yang}, {Yung}, \& {Zhou}}]{Pullen2022}
{Pullen}, A.~R., {Breysse}, P.~C., {Oxholm}, T., {et~al.} 2023, \mnras, 521, 6124

\bibitem[{{Riechers} {et~al.}(2020){Riechers}, {Boogaard}, {Decarli}, {Gonz{\'a}lez-L{\'o}pez}, {Smail}, {Walter}, {Aravena}, {Carilli}, {Cortes}, {Cox}, {D{\'\i}az-Santos}, {Hodge}, {Inami}, {Ivison}, {Kaasinen}, {Wagg}, {Wei{\ss}}, \& {van der Werf}}]{Riechers2020}
{Riechers}, D.~A., {Boogaard}, L.~A., {Decarli}, R., {et~al.} 2020, \apjl, 896, L21

\bibitem[{{Riechers} {et~al.}(2019){Riechers}, {Pavesi}, {Sharon}, {Hodge}, {Decarli}, {Walter}, {Carilli}, {Aravena}, {da Cunha}, {Daddi}, {Dickinson}, {Smail}, {Capak}, {Ivison}, {Sargent}, {Scoville}, \& {Wagg}}]{Riechers2019}
{Riechers}, D.~A., {Pavesi}, R., {Sharon}, C.~E., {et~al.} 2019, \apj, 872, 7

\bibitem[{{Roy} {et~al.}(2023){Roy}, {Valent{\'\i}n-Mart{\'\i}nez}, {Wang}, {Battaglia}, \& {van Engelen}}]{Roy2023}
{Roy}, A., {Valent{\'\i}n-Mart{\'\i}nez}, D., {Wang}, K., {Battaglia}, N., \& {van Engelen}, A. 2023, \apj, 957, 87

\bibitem[{{Schaerer} {et~al.}(2020{\natexlab{a}}){Schaerer}, {Ginolfi}, {B{\'e}thermin}, {Fudamoto}, {Oesch}, {Le F{\`e}vre}, {Faisst}, {Capak}, {Cassata}, {Silverman}, {Yan}, {Jones}, {Amorin}, {Bardelli}, {Boquien}, {Cimatti}, {Dessauges-Zavadsky}, {Giavalisco}, {Hathi}, {Fujimoto}, {Ibar}, {Koekemoer}, {Lagache}, {Lemaux}, {Loiacono}, {Maiolino}, {Narayanan}, {Morselli}, {M{\'e}ndez-Hern{\`a}ndez}, {Pozzi}, {Riechers}, {Talia}, {Toft}, {Vallini}, {Vergani}, {Zamorani}, \& {Zucca}}]{Schaerer2020}
{Schaerer}, D., {Ginolfi}, M., {B{\'e}thermin}, M., {et~al.} 2020{\natexlab{a}}, \aap, 643, A3

\bibitem[{{Schaerer} {et~al.}(2020{\natexlab{b}}){Schaerer}, {Ginolfi}, {B{\'e}thermin}, {Fudamoto}, {Oesch}, {Le F{\`e}vre}, {Faisst}, {Capak}, {Cassata}, {Silverman}, {Yan}, {Jones}, {Amorin}, {Bardelli}, {Boquien}, {Cimatti}, {Dessauges-Zavadsky}, {Giavalisco}, {Hathi}, {Fujimoto}, {Ibar}, {Koekemoer}, {Lagache}, {Lemaux}, {Loiacono}, {Maiolino}, {Narayanan}, {Morselli}, {M{\'e}ndez-Hern{\`a}ndez}, {Pozzi}, {Riechers}, {Talia}, {Toft}, {Vallini}, {Vergani}, {Zamorani}, \& {Zucca}}]{ALPINE}
{Schaerer}, D., {Ginolfi}, M., {B{\'e}thermin}, M., {et~al.} 2020{\natexlab{b}}, \aap, 643, A3

\bibitem[{{Scoville} {et~al.}(2017){Scoville}, {Lee}, {Vanden Bout}, {Diaz-Santos}, {Sanders}, {Darvish}, {Bongiorno}, {Casey}, {Murchikova}, {Koda}, {Capak}, {Vlahakis}, {Ilbert}, {Sheth}, {Morokuma-Matsui}, {Ivison}, {Aussel}, {Laigle}, {McCracken}, {Armus}, {Pope}, {Toft}, \& {Masters}}]{Scoville2017}
{Scoville}, N., {Lee}, N., {Vanden Bout}, P., {et~al.} 2017, \apj, 837, 150

\bibitem[{{Seager} {et~al.}(2000){Seager}, {Sasselov}, \& {Scott}}]{Seager2000}
{Seager}, S., {Sasselov}, D.~D., \& {Scott}, D. 2000, \apjs, 128, 407

\bibitem[{Shannon(1949)}]{Shanon49}
Shannon, C. 1949, Proceedings of the IRE, 37, 10

\bibitem[{{Silich} {et~al.}(2009){Silich}, {Tenorio-Tagle}, {Torres-Campos}, {Mu{\~n}oz-Tu{\~n}{\'o}n}, {Monreal-Ibero}, \& {Melo}}]{Silich2009}
{Silich}, S., {Tenorio-Tagle}, G., {Torres-Campos}, A., {et~al.} 2009, \apj, 700, 931

\bibitem[{{Silva} {et~al.}(2015){Silva}, {Santos}, {Cooray}, \& {Gong}}]{Silva2015}
{Silva}, M., {Santos}, M.~G., {Cooray}, A., \& {Gong}, Y. 2015, \apj, 806, 209

\bibitem[{{Solomon} \& {Vanden Bout}(2005)}]{Solomon2005}
{Solomon}, P.~M. \& {Vanden Bout}, P.~A. 2005, \araa, 43, 677

\bibitem[{{Stacey} {et~al.}(1991){Stacey}, {Geis}, {Genzel}, {Lugten}, {Poglitsch}, {Sternberg}, \& {Townes}}]{Stacey1991}
{Stacey}, G.~J., {Geis}, N., {Genzel}, R., {et~al.} 1991, \apj, 373, 423

\bibitem[{{Sun} {et~al.}(2021){Sun}, {Chang}, {Uzgil}, {Bock}, {Bradford}, {Butler}, {Caze-Cortes}, {Cheng}, {Cooray}, {Crites}, {Hailey-Dunsheath}, {Emerson}, {Frez}, {Hoscheit}, {Hunacek}, {Keenan}, {Li}, {Madonia}, {Marrone}, {Moncelsi}, {Shiu}, {Trumper}, {Turner}, {Weber}, {Wei}, \& {Zemcov}}]{TIMENEW}
{Sun}, G., {Chang}, T.~C., {Uzgil}, B.~D., {et~al.} 2021, \apj, 915, 33

\bibitem[{{Sun} {et~al.}(2018){Sun}, {Moncelsi}, {Viero}, {Silva}, {Bock}, {Bradford}, {Chang}, {Cheng}, {Cooray}, {Crites}, {Hailey-Dunsheath}, {Uzgil}, {Hunacek}, \& {Zemcov}}]{Sun2018}
{Sun}, G., {Moncelsi}, L., {Viero}, M.~P., {et~al.} 2018, \apj, 856, 107

\bibitem[{{Tacconi} {et~al.}(2018){Tacconi}, {Genzel}, {Saintonge}, {Combes}, {Garc{\'\i}a-Burillo}, {Neri}, {Bolatto}, {Contini}, {F{\"o}rster Schreiber}, {Lilly}, {Lutz}, {Wuyts}, {Accurso}, {Boissier}, {Boone}, {Bouch{\'e}}, {Bournaud}, {Burkert}, {Carollo}, {Cooper}, {Cox}, {Feruglio}, {Freundlich}, {Herrera-Camus}, {Juneau}, {Lippa}, {Naab}, {Renzini}, {Salome}, {Sternberg}, {Tadaki}, {{\"U}bler}, {Walter}, {Weiner}, \& {Weiss}}]{Tacconi2018}
{Tacconi}, L.~J., {Genzel}, R., {Saintonge}, A., {et~al.} 2018, \apj, 853, 179

\bibitem[{{Valentino} {et~al.}(2020){Valentino}, {Daddi}, {Puglisi}, {Magdis}, {Liu}, {Kokorev}, {Cortzen}, {Madden}, {Aravena}, {G{\'o}mez-Guijarro}, {Lee}, {Le Floc'h}, {Gao}, {Gobat}, {Bournaud}, {Dannerbauer}, {Jin}, {Dickinson}, {Kartaltepe}, \& {Sanders}}]{Valentino2020}
{Valentino}, F., {Daddi}, E., {Puglisi}, A., {et~al.} 2020, \aap, 641, A155

\bibitem[{{Vallini} {et~al.}(2015){Vallini}, {Gallerani}, {Ferrara}, {Pallottini}, \& {Yue}}]{Vallini2015}
{Vallini}, L., {Gallerani}, S., {Ferrara}, A., {Pallottini}, A., \& {Yue}, B. 2015, \apj, 813, 36

\bibitem[{{Van Cuyck} {et~al.}(2023){Van Cuyck}, {Ponthieu}, {Lagache}, {Beelen}, {B{\'e}thermin}, {Gkogkou}, {Aravena}, {Benoit}, {Bounmy}, {Calvo}, {Catalano}, {D{\'e}sert}, {Dup{\'e}}, {Fasano}, {Ferrara}, {Goupy}, {Hoarau}, {Hu}, {Lambert}, {Mac{\'\i}as-P{\'e}rez}, {Marpaud}, {Mellema}, {Monfardini}, \& {Pallottini}}]{VanCuyck}
{Van Cuyck}, M., {Ponthieu}, N., {Lagache}, G., {et~al.} 2023, \aap, 676, A62

\bibitem[{Vavagiakis {et~al.}(2018)Vavagiakis, Ahmed, Ali, Basu, Battaglia, Bertoldi, Bond, Bustos, Chapman, Chung, Coppi, Cothard, Dicker, Duell, Duff, Erler, Fich, Galitzki, Gallardo, Henderson, Herter, Hilton, Hubmayr, Irwin, Koopman, McMahon, Murray, Niemack, Nikola, Nolta, Orlowski-Scherer, Parshley, Riechers, Rossi, Scott, Sierra, Silva-Feaver, Simon, Stacey, Stevens, Ullom, Vissers, Walker, Wollack, Xu, \& Zhu}]{Vava18}
Vavagiakis, E.~M., Ahmed, Z., Ali, A., {et~al.} 2018, in Millimeter, Submillimeter, and Far-Infrared Detectors and Instrumentation for Astronomy IX, Vol. 10708 (SPIE), 107081U

\bibitem[{{Venemans} {et~al.}(2013){Venemans}, {Findlay}, {Sutherland}, {De Rosa}, {McMahon}, {Simcoe}, {Gonz{\'a}lez-Solares}, {Kuijken}, \& {Lewis}}]{Veneman2013}
{Venemans}, B.~P., {Findlay}, J.~R., {Sutherland}, W.~J., {et~al.} 2013, \apj, 779, 24

\bibitem[{{Vieira} {et~al.}(2020){Vieira}, {Aguirre}, {Bradford}, {Filippini}, {Groppi}, {Marrone}, {Bethermin}, {Chang}, {Devlin}, {Dore}, {Fu}, {Hailey Dunsheath}, {Holder}, {Keating}, {Keenan}, {Kovetz}, {Lagache}, {Mauskopf}, {Narayanan}, {Popping}, {Shirokoff}, {Somerville}, {Trumper}, {Uzgil}, \& {Zmuidzinas}}]{TIM2020}
{Vieira}, J., {Aguirre}, J., {Bradford}, C.~M., {et~al.} 2020, arXiv e-prints, arXiv:2009.14340

\bibitem[{{Wang} \& {Hwang}(2020)}]{Wang2020}
{Wang}, T.-M. \& {Hwang}, C.-Y. 2020, \aap, 641, A24

\bibitem[{{Wang} {et~al.}(2022){Wang}, {Magnelli}, {Schinnerer}, {Liu}, {Modak}, {Jim{\'e}nez-Andrade}, {Karoumpis}, {Kokorev}, \& {Bertoldi}}]{Wang2022}
{Wang}, T.-M., {Magnelli}, B., {Schinnerer}, E., {et~al.} 2022, \aap, 660, A142

\bibitem[{{Weaver} {et~al.}(2022){Weaver}, {Kauffmann}, {Ilbert}, {McCracken}, {Moneti}, {Toft}, {Brammer}, {Shuntov}, {Davidzon}, {Hsieh}, {Laigle}, {Anastasiou}, {Jespersen}, {Vinther}, {Capak}, {Casey}, {McPartland}, {Milvang-Jensen}, {Mobasher}, {Sanders}, {Zalesky}, {Arnouts}, {Aussel}, {Dunlop}, {Faisst}, {Franx}, {Furtak}, {Fynbo}, {Gould}, {Greve}, {Gwyn}, {Kartaltepe}, {Kashino}, {Koekemoer}, {Kokorev}, {Le F{\`e}vre}, {Lilly}, {Masters}, {Magdis}, {Mehta}, {Peng}, {Riechers}, {Salvato}, {Sawicki}, {Scarlata}, {Scoville}, {Shirley}, {Silverman}, {Sneppen}, {Smolc̆i{\'c}}, {Steinhardt}, {Stern}, {Tanaka}, {Taniguchi}, {Teplitz}, {Vaccari}, {Wang}, \& {Zamorani}}]{Weaver2022}
{Weaver}, J.~R., {Kauffmann}, O.~B., {Ilbert}, O., {et~al.} 2022, \apjs, 258, 11

\bibitem[{{Wilson} {et~al.}(2017){Wilson}, {Cooray}, {Nayyeri}, {Bonato}, {Bradford}, {Clements}, {De Zotti}, {D{\'\i}az-Santos}, {Farrah}, {Magdis}, {Micha{\l}owski}, {Pearson}, {Rigopoulou}, {Valtchanov}, {Wang}, \& {Wardlow}}]{Wilson17}
{Wilson}, D., {Cooray}, A., {Nayyeri}, H., {et~al.} 2017, \apj, 848, 30

\bibitem[{{Yue} \& {Ferrara}(2019)}]{Yue:Ferrara2019}
{Yue}, B. \& {Ferrara}, A. 2019, \mnras, 490, 1928

\bibitem[{{Yue} {et~al.}(2015){Yue}, {Ferrara}, {Pallottini}, {Gallerani}, \& {Vallini}}]{Yue:Ferrara2015}
{Yue}, B., {Ferrara}, A., {Pallottini}, A., {Gallerani}, S., \& {Vallini}, L. 2015, \mnras, 450, 3829

\bibitem[{{Zaroubi}(2013)}]{Zaroubi2013}
{Zaroubi}, S. 2013, Astrophysics and Space Science Library, Vol. 396, {The Epoch of Reionization}, ed. T.~{Wiklind}, B.~{Mobasher}, \& V.~{Bromm}, 45

\bibitem[{{Zel'dovich} {et~al.}(1969){Zel'dovich}, {Kurt}, \& {Syunyaev}}]{Zeldovic1969}
{Zel'dovich}, Y.~B., {Kurt}, V.~G., \& {Syunyaev}, R.~A. 1969, Soviet Journal of Experimental and Theoretical Physics, 28, 146

\bibitem[{{Zhou} {et~al.}(2023){Zhou}, {Gong}, {Deng}, {Zhang}, {Yue}, \& {Chen}}]{Zhou2023}
{Zhou}, X., {Gong}, Y., {Deng}, F., {et~al.} 2023, \mnras, 521, 278

\bibitem[{{Zou} {et~al.}(2022){Zou}, {Choi}, {Cothard}, {Freundt}, {Huber}, {Li}, {Niemack}, {Nikola}, {Riechers}, {Rossi}, {Stacey}, \& {Vavagiakis}}]{Zou2022}
{Zou}, B., {Choi}, S.~K., {Cothard}, N.~F., {et~al.} 2022, in Millimeter, Submillimeter, and Far-Infrared Detectors and Instrumentation for Astronomy XI, Vol. 12190, 121902B

\end{thebibliography}

\newpage
% Start of the Appendix
\begin{appendices}
\renewcommand{\thefigure}{A\arabic{figure}}
\setcounter{figure}{0}

\section*{Appendix A. Impact of masking on the PS of \\ individual CO lines}
\label{sec:individualCO}
%\renewcommand{\thesection}{Appendix \Alph{section}}
%\label{sec:individualCO}

\begin{figure*}[h!]
    \centering
    \includegraphics[width=0.8\textwidth]{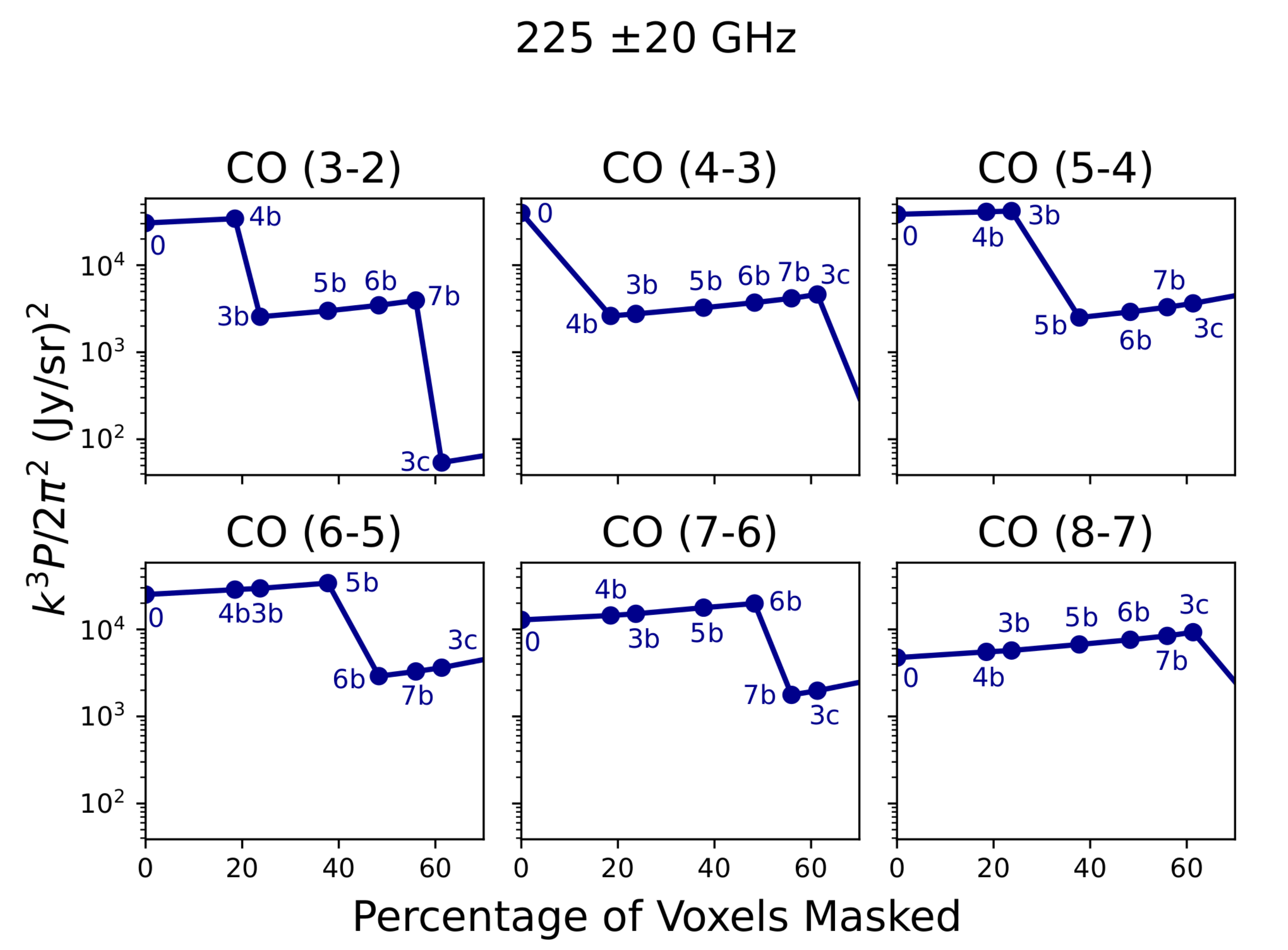}
    \vspace{0.1cm} % 
    \includegraphics[width=0.8\textwidth]{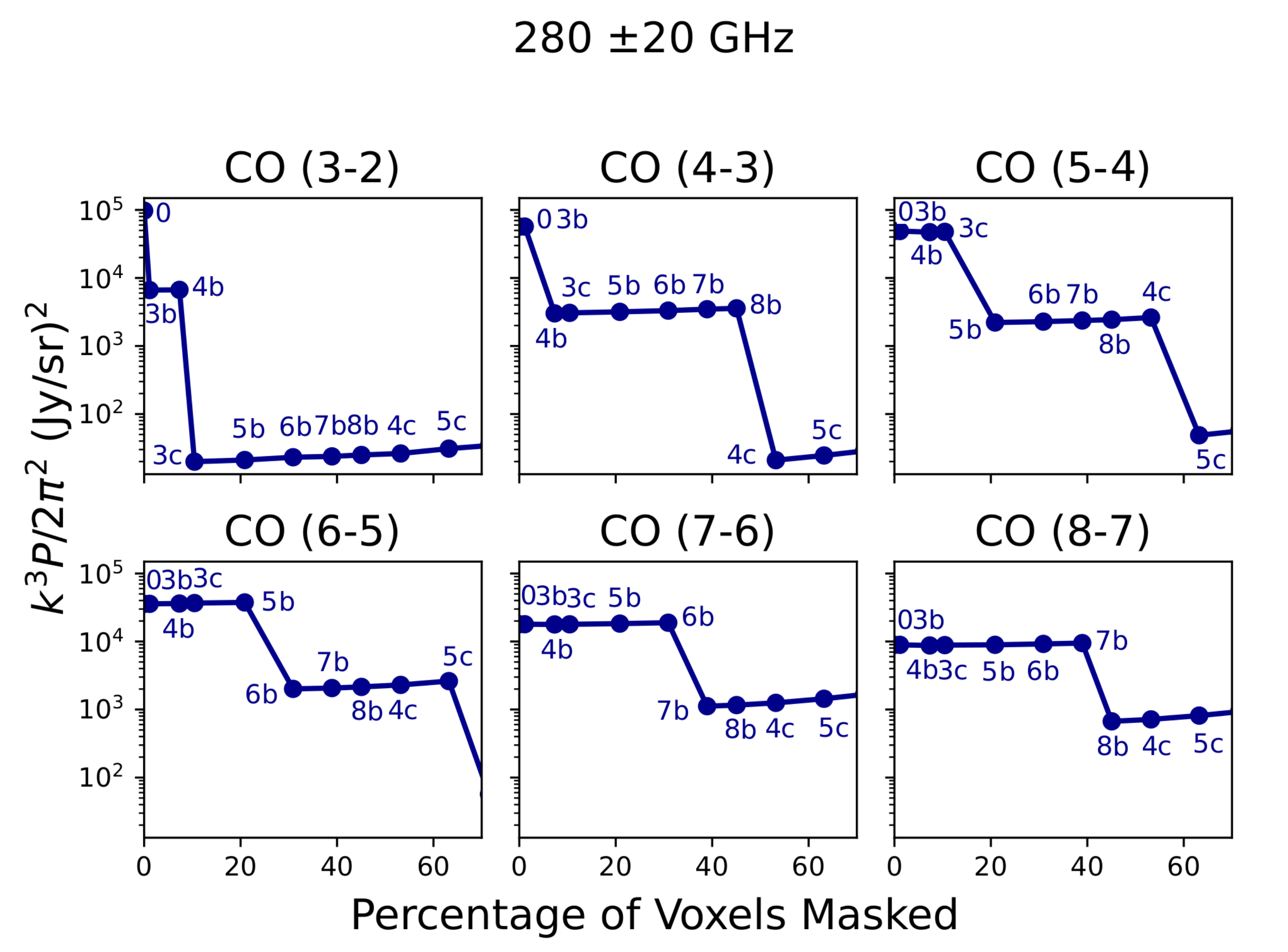}
    \caption{Effect of masking on the PS of the brightest individual CO lines ($J_{\rm up}=3-8$). The lines represent our fiducial model, with the same optimal masking sequence applied as in Fig.~\ref{fig:coimpact} and Fig.~\ref{fig:allimpact}. The results are plotted for the frequency ranges $225 \pm 40$~GHz (corresponding to $z_{\rm [CII]}=7.4$) and $280 \pm 40$~GHz ($z_{\rm [CII]}=5.8$)}
    \label{fig:FigureA}

\end{figure*}

\begin{figure*}[h!]
    \centering
    \includegraphics[width=0.8\textwidth]{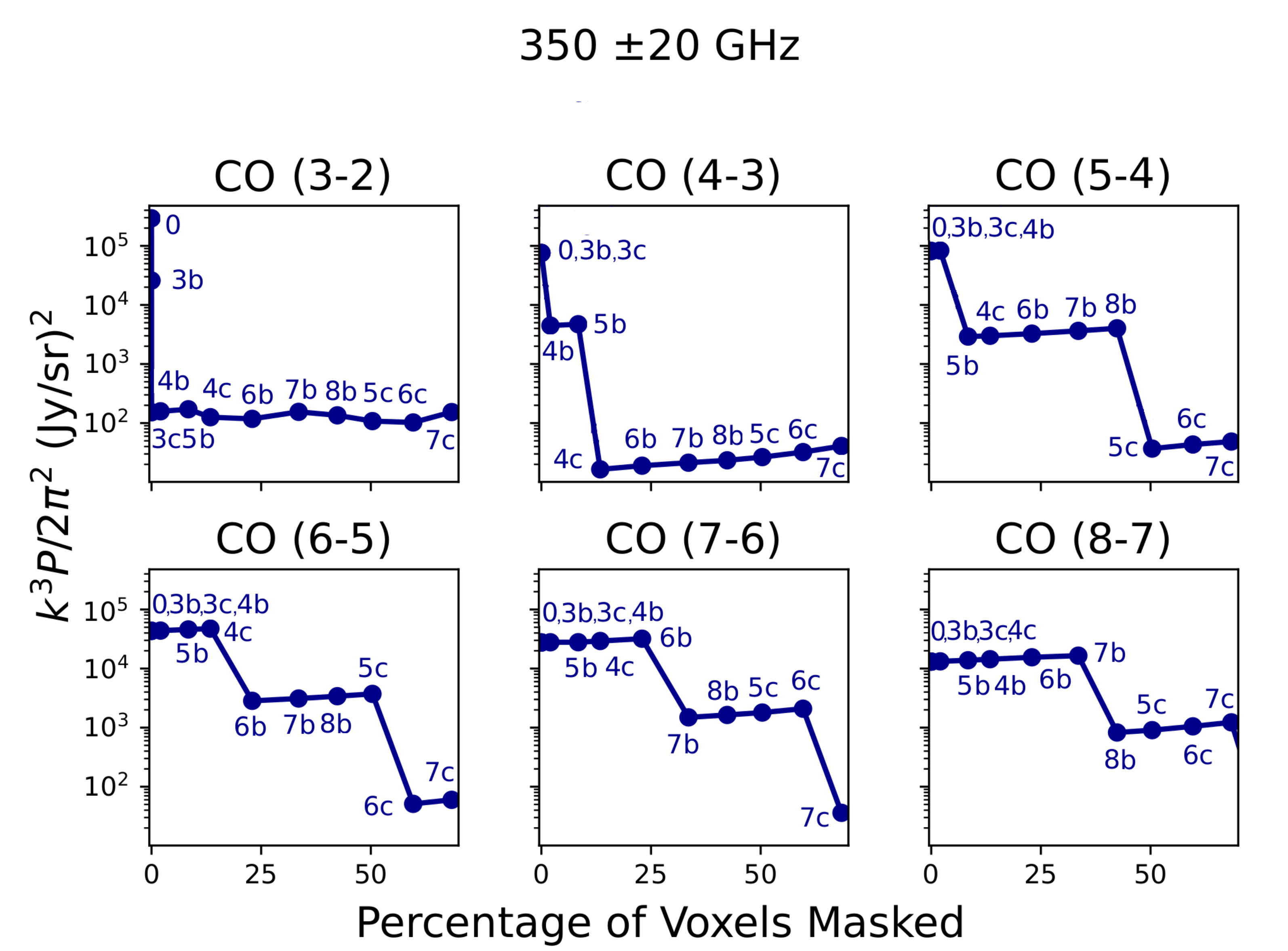}
    \vspace{0.8cm} 
    \includegraphics[width=0.8\textwidth]{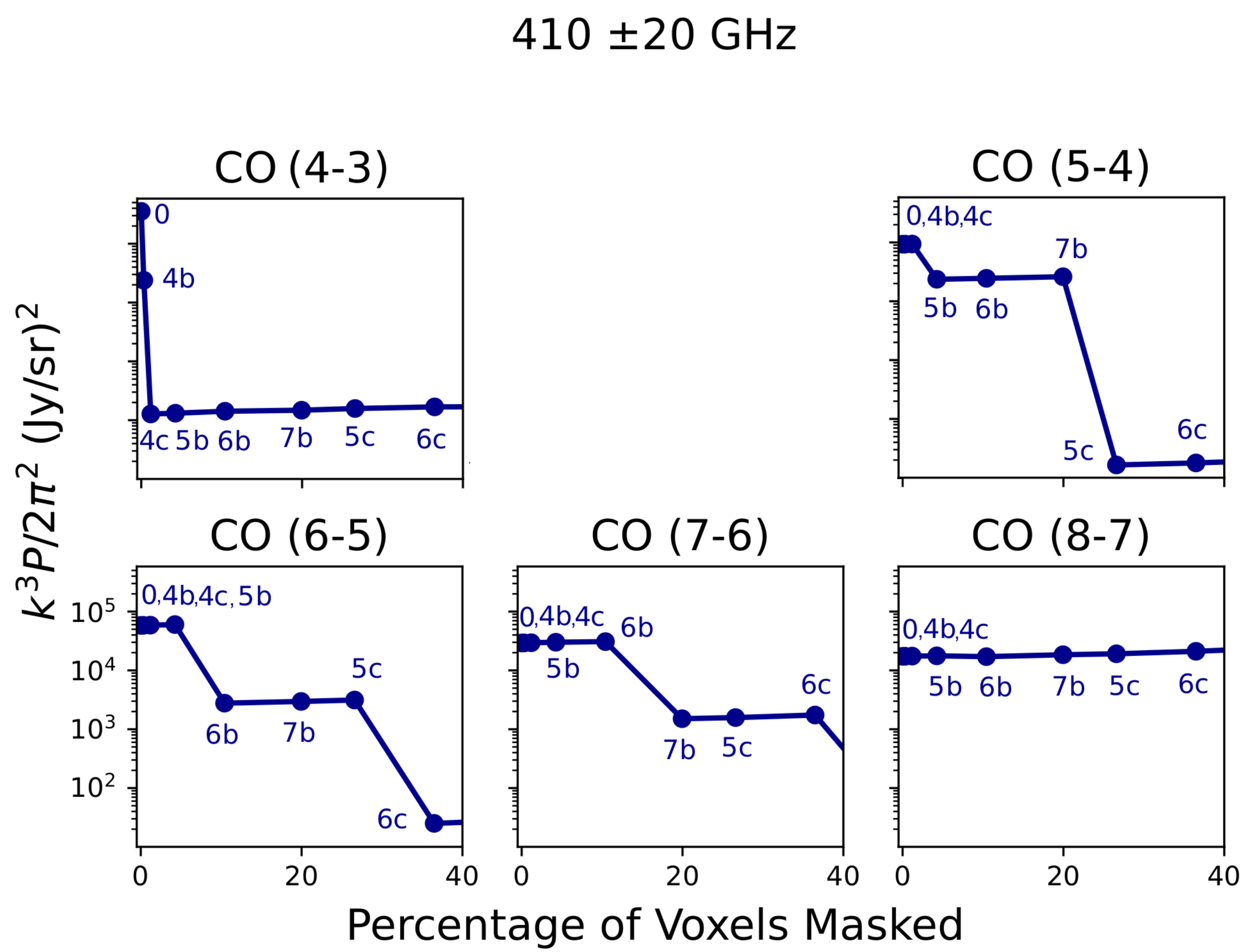}
    \caption{Same as Fig.~\ref{fig:FigureA}, but for the frequency ranges $350 \pm 40$~GHz ($z_{\rm [CII]}=4.3$) and $410 \pm 40$~GHz ($z_{\rm [CII]}=3.7$).}
    \label{fig:FigureB}

\end{figure*}

This appendix presents the PS of individual CO ($J_{\rm up}=3-8$) spectral lines in the frame of reference of the [CII] line at scales of $k = 0.02-0.32$ Mpc$^{-1}$ as a function of the percentage of the survey volume masked (Fig. \ref{fig:FigureA} and \ref{fig:FigureB}) following the optimal masking sequence described in Sect. \ref{Subsection:Evaluating}. For each CO line PS, two distinct behaviors are observed during the masking process. When the masking step specifically targets the spectral line in question, there is a sharp decrease in the PS. Conversely, when the masking step targets a different spectral line, a slow increase in the PS is observed due to the masking bias (see Sect. \ref{Subsubsection:MaskCII}).

The behavior of the PS during masking is influenced by how the galaxies emitting the lines are distributed within the 3D tomography. Lines from galaxies that are highly localized within the survey volume, such as CO(3-2) at 350 GHz, exhibit a sharp decrease in their PS when masking targets them. Additionally, there is almost no increase in PS when masking targets other lines due to the minimal residuals left after masking. In contrast, lines from galaxies more evenly distributed throughout the survey volume, such as CO(4-3) at 225 GHz, show a less pronounced decrease in PS when masked, but a stronger increase when other lines are masked, because there remains a substantial amount of residual emission that gets amplified. A key indicator of how uniformly the intensity of each line is distributed in the tomography is the CV of its intensity, as shown in Fig.~\ref{fig:VAR} and described in detail in Sect.~\ref{Subsection:Intensities}.

\end{appendices}

\end{document}